\RequirePackage[2020-02-02]{latexrelease}
\documentclass[amsmath, aps, prf]{revtex4-2}

\usepackage{bm}
\usepackage[font=footnotesize,labelfont=bf]{caption}
\usepackage[font=footnotesize,labelfont=bf]{subcaption}
\usepackage{multirow}

\usepackage{graphicx}
\usepackage{bm}
\usepackage[utf8]{inputenc}

\usepackage{booktabs}
\usepackage{float}
\usepackage{xcolor}

\newcommand{\refeq}[1]{Eq.\,(\ref{#1})}

\newcommand{\reffig}[1]{Fig.\,\ref{#1}}
\newcommand{\Reffig}[1]{Figure \ref{#1}}
\newcommand{\reffigs}[1]{Fig.\,\ref{#1}}

\newcommand{\refsec}[1]{Section \ref{#1}}

\begin{document}

\title{Motion and deformation of capsules flowing through a corner in the inertial and non-inertial regimes}

\author{Damien P. Huet}
\affiliation{Department of Mathematics, University of British Columbia, 1984 Mathematics Road, Vancouver BC V6T 1Z2, Canada}
\author{Antoine Morente}
\affiliation{Department of Mathematics, University of British Columbia, 1984 Mathematics Road, Vancouver BC V6T 1Z2, Canada}
\author{Guodong Gai}
\affiliation{Department of Mathematics, University of British Columbia, 1984 Mathematics Road, Vancouver BC V6T 1Z2, Canada}
\author{Anthony Wachs}
\email[]{wachs@math.ubc.ca}
\affiliation{Department of Mathematics, University of British Columbia, 1984 Mathematics Road, Vancouver BC V6T 1Z2, Canada}
\affiliation{Department of Chemical and Biological Engineering, University of British Columbia, 2360 E Mall, Vancouver BC V6T 1Z3, Canada}

\date{\today}


\begin{abstract}
We investigate the inertial and non-inertial dynamics of three-dimensional elastic capsules flowing through a square channel presenting a sharp corner. Our study analyzes the trajectory, surface area, velocity and membrane stress of the capsules in the case of a single capsule, a system of two interacting capsules and a train of ten capsules released upstream of the corner. The channel Reynolds number $Re$ ranges from 0.01 to 50 and the Capillary number $Ca$, which measures the ratio of the viscous and elastic stresses, ranges from 0.075 to 0.35. We find that in the inertial regime, the membrane stretch and stress increase dramatically as compared to the non-inertial case, and that the velocity overshoot inside the corner is also enhanced. The maximum capsule deformation is observed to depend nearly linearly on $Ca$ and $Re$. Additionally, we report a repelling mechanism between two confined capsules when their initial interspacing distance $d$ is smaller than a critical value $d_c$. The deformation of the leading capsule is found to be mitigated by the presence of the following capsule. In the case of multiple capsules flowing through the corner, we observe that the increase in the maximum surface area of the trailing capsules eventually saturates at the tail of the train. Moreover, we find that the corner tends to separate the capsules regardless of their upstream interspacing distances $d$. This study contributes to the elaboration of practical guidelines for controlling capsule breakup and predicting throughput in both inertial and non-inertial microfluidic experiments.
\end{abstract}

\pacs{}
\maketitle

\section{Introduction\label{sec:corner_intro}}
Membrane-enclosed fluid objects, or capsules, are everywhere in natural and industrial processes, from red blood cells (RBCs), circulating tumor cells (CTCs) or flowing eggs in biology to encapsulated substances in the pharmaceutical, cosmetic and food industries \cite{barthes2016motion}. The study of microcapsules in particular is of primary importance in a variety of biological applications, such as sorting and enriching solutions of biological microcapsules, e.g. to segregate RBCs or CTCs, as well as efficiently manufacturing capsules enclosing an active substance in the field of targeted drug delivery \cite{bae2011targeted, kumari2016nanocarriers}.
In the past decade, microfluidic devices have been shown to accomplish a variety of tasks including cell segregation based on size and deformability \cite{zhu2014microfluidic, haner2021sorting, haner2020deformation, fang2022efficient}, concentration enrichment \cite{wang2016motion, wang2018path, lu2021path} and cell characterization \cite{gubspun2016characterization, lin2021high, wang2023computational}.
Moreover, the increase in computing power has recently allowed numerical studies to contribute to the design of microfluidic devices. For example, Zhu et al. \cite{zhu2014microfluidic} numerically investigated an original microchannel geometry consisting of a semi-circular pillar located at the center of a microchannel: their study showed that this design can efficiently segregate cells based on membrane deformability. Recently, experiments were conducted using their microfluidic design and concluded that it can indeed sort cells based solely on membrane stiffness, with relatively high efficacy \cite{haner2021sorting}. With regards to cell characterization, Gubspun et al. \cite{gubspun2016characterization} proposed a method to determine capsule properties such as the membrane shear modulus by comparing the experimental and numerical ``parachute" shape of capsules in a straight microchannel. While the majority of microfluidic investigations operate in Stokes conditions, in recent years the design and study of inertial microfluidic devices has risen due to their ability to accurately segregate capsules by size and to extract them from their solvant \cite{fang2022efficient, zhu2021polymer}.
Inertial focusing in microfluidic devices typically relies on a spiral-shaped channel concentrating heavier capsules to the outer, lower-curvature edge of the channel, while lighter capsules concentrate closer to the inner, higher-curvature edge. A smooth geometry such as a spiral-shaped channel usually does not induce a high strain nor stress on a suspended capsule even in inertial regimes, however little is known about the strains and stresses induced by commonly encountered sharp geometries such as forks and corners on a capsule flowing in the presence of inertia. Moreover, the effect of such sharp geometries on the hydrodynamic interactions of a train of several capsules in inertial regimes is also an open question. More insight in these directions is of practical interest in the design and operation of inertial microfluidic devices because (i) the devices should not compromise the mechanical integrity of the capsules, i.e. it is critical to avoid capsule breakup, and (ii) cell-sorting processes typically operate in very dilute regimes to avoid capsule interactions, while a better understanding of such interactions would allow to operate these devices at a moderate to high concentration optimizing efficacy and throughput.

In the past four decades, a significant research effort has been invested into the modeling and the study of capsule deformations in non-inertial regimes, primarily because this regime is encountered in microcirculation such as capillary vessels and in traditional microfluidic devices. Using formalism from the thin-shell theory \cite{green1960large}, Barthès-Biesel \& Rallison first published an analytical solution for the time-dependant deformation of an elastic capsule in an unbounded, creeping shear flow in the limit of small deformations \cite{barthes1981time}. Over a decade later, Pozrikidis was able to go beyond the assumption of small deformations using a Boundary Integral Method (BIM) \cite{pozrikidis1995finite}. The same method was used to consider finite deformations of sheared capsules which inner and outer fluid viscosities differ \cite{ramanujan1998deformation}, as well as to study the contribution of bending stresses \cite{pozrikidis2001effect}, allowing to consider RBCs suspended in an unbounded shear flow \cite{pozrikidis2003numerical}. Besides unbounded geometries, Zhao et al. \cite{zhao2010spectral} simulated RBCs in straight and constricted channels using a spectral BIM.
A similar method was later used by Hu et al. \cite{hu2013characterizing} to consider an initially spherical capsule flowing through a square channel of width similar to the capsule diameter: the originality of their work is that they performed experiments and showed remarkable agreement between the measured and the computed capsule shape. Concomitantly, Park and Dimitrakopoulos \cite{park2013transient} studied the deformation of a capsule with non-unity viscosity ratio flowing through a sharp constriction.
More recently, Balogh \& Bagchi \cite{balogh2017computational, balogh2018analysis, balogh2019cell} used a Front-Tracking Method (FTM) to analyze the motion and deformation of RBCs through complex geometries resembling capillary vessels found in human microcirculation: their simulations exhibited in particular the well-known cell-free layer observed experimentally between the RBCs and the vessel walls \cite{bugliarello1970velocity, namgung2011effect}.

Regarding the study of flowing capsules in the presence of inertia, the aforementioned analytical theory for small deformations as well as the popular BIM both fall short of accounting for the convective term in the fluid momentum equation. Doddi \& Bagchi \cite{doddi2008effect} first studied inertial capsules in the context of two interacting capsules in a shear flow using the FTM. They showed in particular that the two capsules engage in spiralling motions at sufficiently high inertia.
The inertial motion of a deformable capsule was then studied in straight microchannels \cite{kilimnik2011inertial, raffiee2017elasto}, where several equilibrium positions are found away from the channel centerline, along the cross-section diagonals.
With regards to curved channels, Ebrahimi \& Bagchi \cite{ebrahimi2021inertial} recently investigated the migration of a single capsule over an impressive amount of varying parameters: the channel Reynolds number, the capsule deformability, as well as the aspect ratio and curvature of the channel were all varied independently. Their study shows that for sufficiently high inertia, exactly two focusing locations appear near the centers of the vortices of the secondary flow, known as Dean's vortices.
However no mention of the membrane internal strains and stresses is found in their work, as their goal was not to investigate the capsule integrity in such flows.

While straight and curved microchannels are essential components of microfluidic devices, such simple geometries do not account for the numerous junctions, corners and coils commonly found in these devices. To bridge this gap, Zhu \& Brandt \cite{zhu2015motion} investigated the non-inertial motion and the deformation of a single elastic capsule in a sharp corner. They showed that the capsule follows the streamlines of the undisturbed flow regardless of membrane deformability. Due to lubrication forces, the capsule velocity decreases when approaching the corner, reaches a minimum along the corner diagonal, and rises back to its steady state with an overshoot increasing with deformability. Similarly, the surface area of the capsule reaches a maximum inside the corner and reaches its steady value with an undershoot more pronounced as deformability is increased. Also reported in their study is the maximum stress in the capsule membrane, which can be used to assess mechanical integrity and characterize the cell mechanical properties. They find that the maximum stress deviation increases and shifts from the front to the top of the capsule with increasing deformability.
Wang et al. \cite{wang2016motion, wang2018path} later considered the inertial and non-inertial path selection of a single capsule through Y- and T-junctions, both typically encountered in microfluidic geometries. They observe that at high inertia, the capsule does not necessarily favor the daughter branch with the largest flow rate, and that this effect is more pronounced for stiff membranes (corresponding to a low capillary number).
Recently, Lu et al. \cite{lu2021path} investigated the interaction and path selection of capsules in a T-junction at moderate inertia, with the goal of enriching capsule solutions. When considering a pair of capsules, they show that the leading capsule is weakly affected by the presence of a trailing capsule, but that the reverse is not true. They find that the trailing capsule enters a different branch depending on the initial interspacing distance and on the flow rate split ratio between the two daughter branches of the T-junction. They then consider a train of capsules and find two distinct regimes: (i) the interspacing distance is low and the capsule interaction is high, resulting in an unsteady regime and affecting the trajectories of the capsules, and (ii) the interspacing distance is large and the capsule interaction is low, leaving the capsule trajectories identical to that of a single capsule. Interestingly, they report that the critical interspacing distance between two capsules plotted against the flow rate split ratio of the daughter branches results in a master curve independent of membrane deformability, capsule size, and Reynolds number.

In this study, we investigate the inertial and non-inertial motion and the interaction of deformable capsules flowing through a sharp corner, which is a very common geometry in microfluidic devices. As the efficiency of these devices is defined in terms of the capsules throughput, which can be optimized by increasing the flow rate as well as the concentration of capsules, our objective is two-fold: first, we aim to quantify the effet of inertia on the deformation of a single capsule in a microfluidic-relevant geometry, second, we seek to describe the hydrodynamic interactions and deformation differences between leading and trailing capsules when a pair and a train of capsules are considered.
The rest of this paper is organized as follows. In \refsec{sec:corner_equations}, we describe the governing equations as well as the flow configuration and the considered parameter space. In \refsec{sec:corner_method_validation}, we give an overview of our numerical method and we investigate the impact of the inlet length. We analyze the motion of a single capsule in \refsec{sec:corner_single}, both in the non-inertial and in the inertial regimes.  \refsec{sec:corner_double} is devoted to the analysis of binary interactions of a pair of capsules, where the influence of the initial interspacing distance is investigated. In \refsec{sec:corner_train}, we consider a train of ten capsules flowing through the corner and we discuss the velocity and deformation discrepancies between the leading and trailing capsules. Finally, we conclude in \refsec{sec:corner_conclusion}.

The documented source code allowing to reproduce all of the simulations and figures presented in this study is freely available online \cite{huet_sandbox}.

\section{Governing equations and problem statement\label{sec:corner_equations}}
The capsule membrane $\Gamma$ is assumed infinitely thin and is surrounded by an incompressible, Newtonian fluid of constant viscosity and density. In all of this study, the capsule inner and outer fluids are assumed identical: in particular their viscosity ratio is unity. The fluid is described by the mass and momentum conservation equations:
\begin{equation}
    \nabla \cdot \bm{\tilde{u}} = 0
    \label{eq:mass_conservation}
\end{equation}
\begin{equation}
    \frac{\partial \bm{\tilde{u}}}{\partial \tilde{t}} + \bm{\tilde{u}} \cdot \nabla \bm{\tilde{u}} = \frac{1}{\tilde{\rho}} \nabla \tilde{p} + \tilde{\nu} \Delta \bm{\tilde{u}} + \frac{1}{\tilde{\rho}} \bm{\tilde{f_b}}
    \label{eq:momentum_conservation}
\end{equation}
where $\bm{\tilde{u}}$ is the velocity field, $\tilde{p}$ is the pressure field, $\tilde{\rho}$ is the density, $\tilde{\nu} = \tilde{\mu}/\tilde{\rho}$ is the kinematic viscosity, $\tilde{\mu}$ is the dynamic viscosity and $\bm{\tilde{f_b}}$ is a body term accounting for the action of the membrane on its surrounding fluid. The dimensional quantities are denoted by the $\sim$ symbol. The membrane exhibits elasticity and bending resistance, and its action on the fluid is local, resulting in the following expression for $\bm{\tilde{f_b}}$:
\begin{equation}
    \bm{\tilde{f_b}} = \left( \bm{\tilde{f}_\text{elastic}} + \bm{\tilde{f}_\text{bending}} \right) \tilde{\delta}(\bm{\tilde{x}} - \bm{\tilde{x}_\Gamma}),
    \label{eq:body_force}
\end{equation}
where $\tilde{\delta}(\bm{\tilde{x}} - \bm{\tilde{x}_\Gamma})$ is a Dirac distribution that is non-zero on the surface of the membrane.

The shear and area-dilatation membrane stresses are described using the thin-shell theory, and are briefly summarized here. The interested reader is referred to Green \& Adkins \cite{green1960large} as well as to the analytical study of Barthès-Biesel \& Rallison \cite{barthes1981time} for more details. We adopt a neo-Hookean law \cite{green1960large}, which surface strain-energy function is expressed as:
\begin{equation}
    \tilde{W_s}^{NH} = \frac{\tilde{E_s}}{2} \left( \lambda_1^2 \lambda_2^2  + \frac{1}{\lambda_1^2 \lambda_2^2}\right),
\end{equation}
where $\lambda_{1,2}$ are the principal stretches in the two tangential directions, and $\tilde{E_s}$ is a shear modulus. The principal stresses $\tilde{\sigma}_{1,2}$ are given by:
\begin{equation}
    \tilde{\sigma}_{i} = \frac{1}{\lambda_j} \frac{\partial \tilde{W_s}^{NH}}{\partial \lambda_i}, \qquad i, j \in \{1,2\}, \quad i \neq j.
    \label{eq:principal_stress}
\end{equation}

The bending stresses for biological membrane are governed by the Helfrich's bending energy $\mathcal{E}_b$ \cite{helfrich1973elastic, zhong1989bending}:
\begin{equation}
    \tilde{\mathcal{E}}_b = \frac{\tilde{E_b}}{2} \int_\Gamma \left( 2\tilde{\kappa} - \tilde{\kappa}_0 \right)^2 dS,
    \label{eq:bending_energy}
\end{equation}
where $\tilde{E_b}$ is the bending modulus, $\tilde{\kappa}$ is the mean curvature and $\tilde{\kappa}_0$ is a reference curvature. Taking the variational formulation of \refeq{eq:bending_energy} leads to the bending force per unit area $\tilde{A}$:
\begin{equation}
    \bm{\tilde{f}_\text{bending}}/\tilde{A} = -2 \tilde{E}_b ( \Delta_s(\tilde{\kappa}) + 2 (\tilde{\kappa} - \tilde{\kappa}_0)( \tilde{\kappa}^2 - \tilde{\kappa}_g + \tilde{\kappa}_0\tilde{\kappa}) ) \bm{n},
  \label{eq:bending_force}
\end{equation}
where $\tilde{\kappa}_g$ is the Gaussian curvature and $\bm{n}$ is the outer normal vector.

\begin{figure}
    \centering
    \begin{subfigure}{.39\textwidth}
        \centering
        \includegraphics[width=\columnwidth, trim=140 20 150 0, clip]{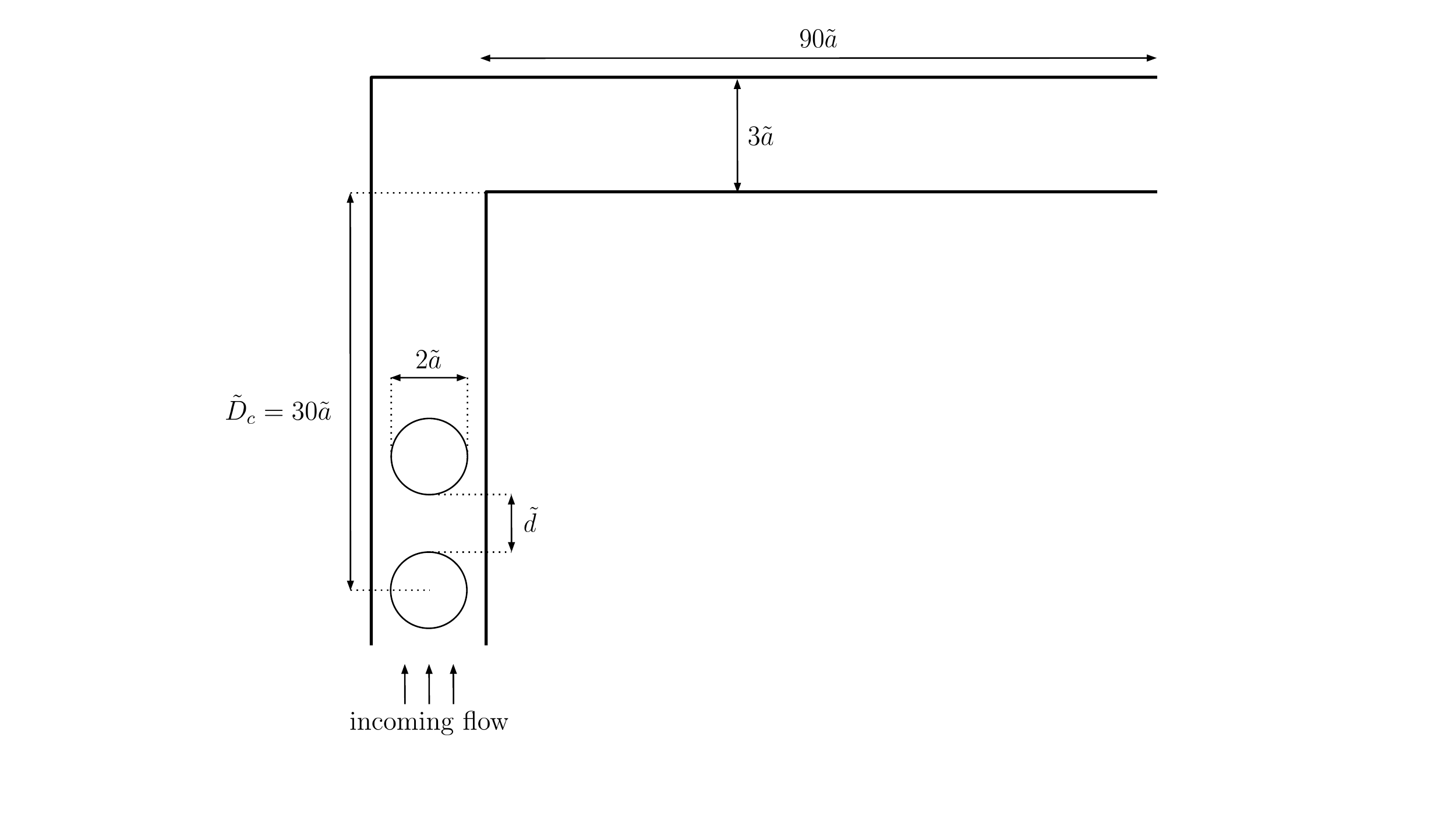}
        \caption{}
    \end{subfigure}
    \hspace{4em}
    \begin{subfigure}{.29\textwidth}
        \centering
        \includegraphics[width=\columnwidth]{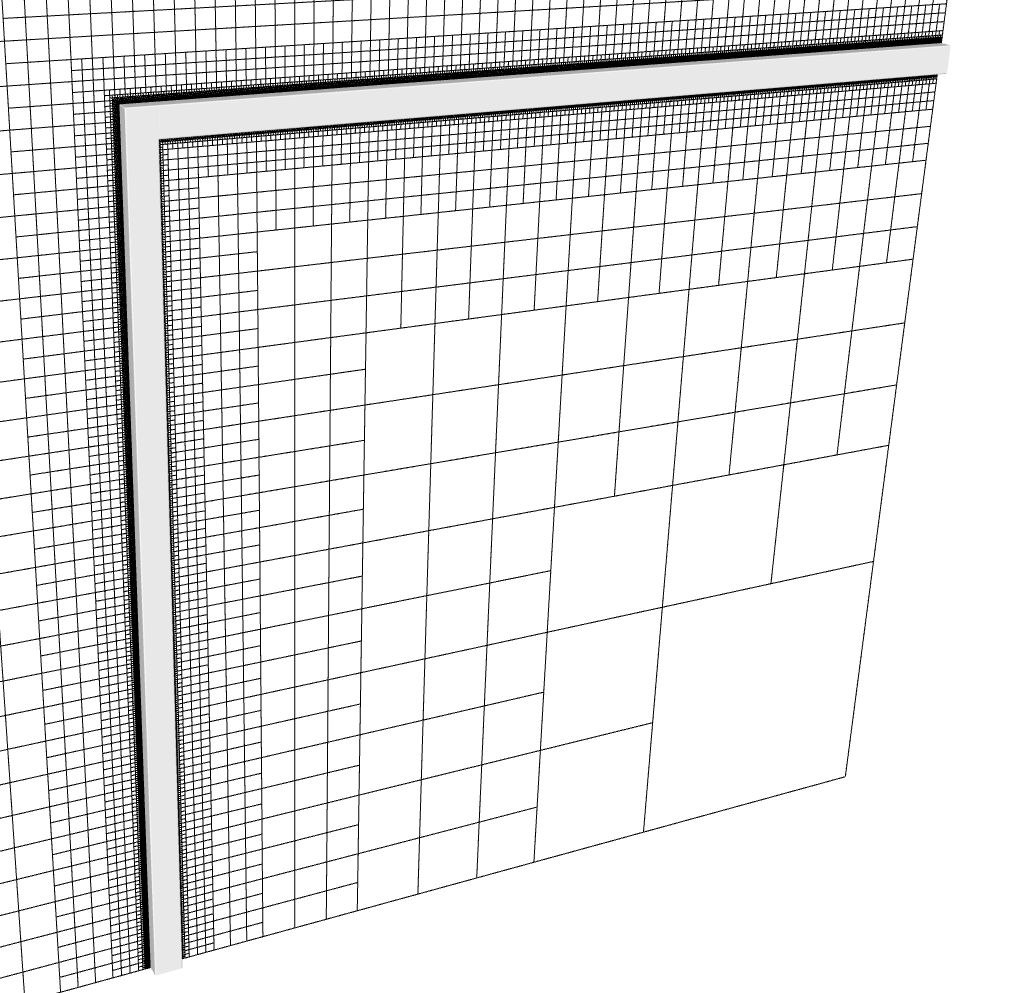}
        \caption{}
    \end{subfigure}
    \caption{(a) Schematic of the geometry of the fluid domain. The channel has a square cross-section of side length $3\tilde{a}$. (b) Visualization of the full channel and the computational grid over the symmetry plane of the channel.}
    \label{fig:corner_geometry}
\end{figure}

At $t=0$, an initially spherical capsule of radius $\tilde{a}$ is placed in a square channel of width $\tilde{W} = 3\tilde{a}$ at a distance $\tilde{h}_0 = 30\tilde{a}$ from a sharp corner, as represented in \reffig{fig:corner_geometry}. An average cross-section velocity $\tilde{U_0}$ is imposed at the inlet boundary, while the outflow boundary condition $\partial \bm{\tilde{u}}_n / \partial \bm{n} = 0$ is imposed at the outlet boundary. When several capsules are considered, we use the same initial conditions as Lu et al. \cite{lu2021path}: a trailing capsule is inserted in the simulation only after the centroid of its preceding capsule has advanced by a distance $\tilde{d}$.
Our problem is governed by the trailing dimensionless numbers:

\begin{enumerate}
    \item The channel Reynolds number $Re = \tilde{\rho} \tilde{U_0} \tilde{W}/\tilde{\mu}$,
    \item The Capillary number $Ca = \tilde{\mu} \tilde{U_0} \tilde{a}/\tilde{E_s}$, representing the ratio of viscous stresses over elastic stresses,
    \item The reduced bending stiffness coefficient $E_b = \tilde{E_b}/(\tilde{E_s} a^2)$,
    \item The confinement ratio $\beta = 2\tilde{a}/\tilde{W}$,
    \item The reduced initial gap between capsules $d_0 = \tilde{d}/2\tilde{a} - 1$.
\end{enumerate}
In this study, the Reynolds number $Re$ ranges from 0.01 to 50, the Capillary number $Ca$ varies from $0.075$ to $0.35$, and the reduced initial gap $d_0$ is chosen from 0.125 to 1. The reduced bending stiffness $E_b$ and the confinement ratio $\beta$ are both kept constant, with $\beta = 2/3$ and $E_b = 5 \cdot 10^{-3}$ as proposed by Pozrikidis \cite{pozrikidis2010computational}. The reference curvature $\tilde{\kappa_0}$ is equal to $-2.09/\tilde{a}$ in this study, as is common for some biological membranes such as RBC membranes \cite{pozrikidis2005resting, yazdani2011phase}.
In the rest of this study, we use the capsule radius $\tilde{a}$ as the characteristic length scale, and we define the characteristic time scale as the radio of the capsule radius over the average cross-section velocity, i.e. $t = \tilde{a}/\tilde{U_0}$.

\section{Numerical method and validations\label{sec:corner_method_validation}}
We use our adaptive Front-Tracking Method (FTM) to solve the above equations: we provide below a brief overview of the numerical method, while an in-depth description is available in \cite{huet2022cartesian}.
\refeq{eq:mass_conservation} and \refeq{eq:momentum_conservation} are solved using the Finite Volume method on an adaptive octree grid using the open-source software Basilisk \cite{Popinet2015}. The membrane is discretized using an unstructured triangulation and \refeq{eq:principal_stress} is solved using a linear Finite Element Method, while \refeq{eq:bending_force} is solved using a paraboloid-fitting method. The membrane triangulation and the octree grid communicate by means of the immersed boundary method \cite{Peskin1977, Peskin2002}, where the Dirac distribution in \refeq{eq:body_force} is regularized using a cosine-based formulation:
\begin{equation}
  \tilde{\delta}(\bm{x_0} - \bm{x}) =
  \begin{cases}
  \begin{aligned}
    & \frac{1}{64 \tilde{\Delta}^3} \prod_{i = 1}^3 \left( 1 + \cos\left( \frac{\pi}{2\tilde{\Delta}} (x_{0, i} - x_i) \right) \right) \quad \text{if} \quad |x_{0,i} - x_i| < 2 \tilde{\Delta}\\
    & 0 \quad \text{otherwise}
  \end{aligned}
\end{cases},
\label{eq:dirac-delta}
\end{equation}
where $\bm{x_0} = [x_{0,1} \; x_{0,2} \; x_{0,3}]$ is the location of a Lagrangian node on the surface discretization of the membrane, and $\tilde{\Delta}$ is the local mesh size of the Eulerian octree grid. Extensive validation of the present numerical method was the focus of our previous study \cite{huet2022cartesian} and is therefore not presented here. Nonetheless, the convergence with respect to the Eulerian grid as well as the release distance of the capsule from the corner are investigated below.

\begin{figure}
    \centering
    \begin{subfigure}{.48\textwidth}
        \centering
        \includegraphics[width=.9\textwidth]{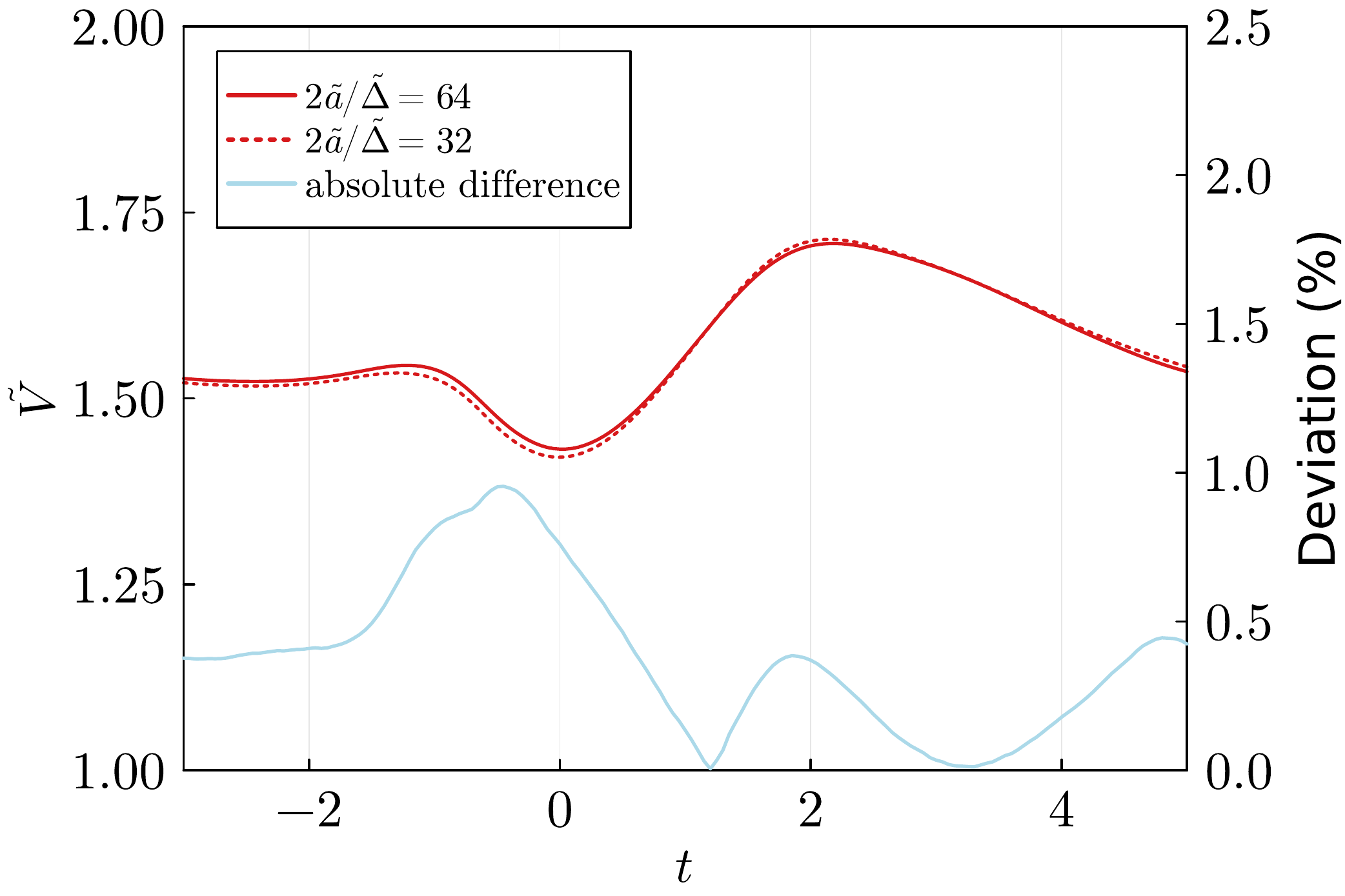}
        \caption{}
    \end{subfigure}
    \hspace{4em}
    \begin{subfigure}{.3\textwidth}
        \includegraphics[width=.7\textwidth, trim=340 830 800 250, clip]{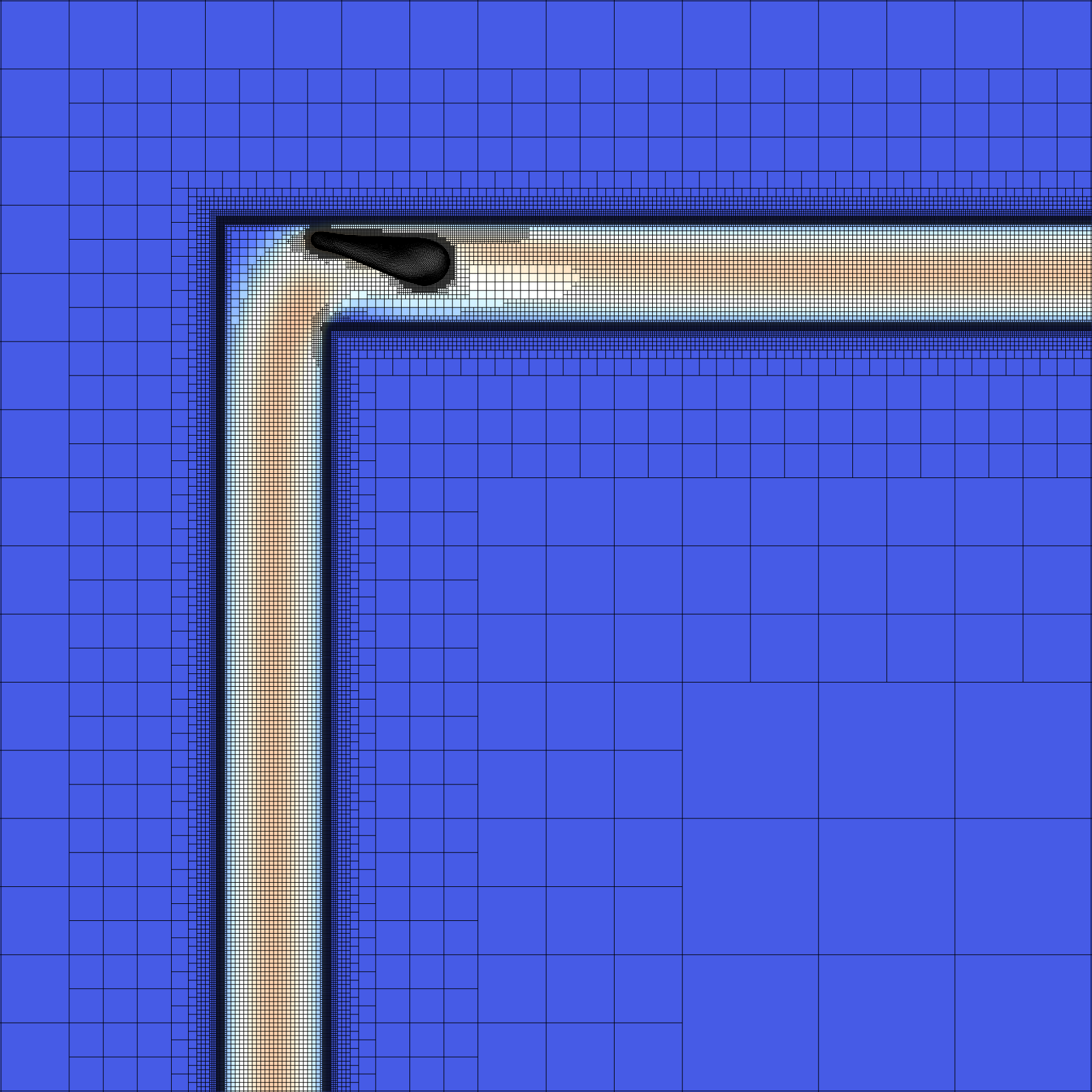}
        \vspace{3em}
        \caption{}
    \end{subfigure}
    \caption{(a) Centroid velocity of a capsule at $Ca = 0.35$ and $Re=50$ for two grid resolutions: 32 grid cells per initial diameter (red dotted line) and 64 grid cells per initial diameter (red solid line). The blue curve denotes the deviation in the centroid velocities for these two grid resolutions. (b) Corresponding shape and grid resolutions of the capsule and the flow field: blue means zero velocity and red means large velocity.}
    \label{fig:conv_vel}
\end{figure}

In the immersed boundary method, it is well known that the support of the regularized Dirac distribution may extend outside of the fluid domain if the immersed object of interest becomes very close to the domain walls \cite{Uhlmann2005, lu2021path, wang2018path}. In order to avoid unphysical loss of momentum for the specific membrane nodes close to the wall, it is important to ensure that none of the supports of the regularized Dirac distribution extend outside of the fluid domain, i.e. that there always exist more than two grid cells between membrane nodes and the domain boundaries. As such, we simulate the dynamics of a capsule for two different grid resolutions in the configuration where it is most deformed and is the closest to the channel wall, as shown in \reffig{fig:conv_vel}b. \Reffig{fig:conv_vel}a shows the velocity of the capsule $\tilde{V}$ inside and downstream of the corner for Eulerian resolutions equivalent to 32 and 64 grid cells per initial capsule diameter, as well as the deviation of the velocities in these two configurations. Excellent agreement is found between the velocities computed using the two grid resolutions, with the maximum discrepancy lower than 1\% and the average discrepancy over the considered time range of about 0.5\%. Moreover, in both configurations it was found that more than 3 grid cells are present in the lubrication layer between the capsule tail and the upper corner wall. These results indicate that an equivalent grid resolution of 32 grid cells per capsule initial diameter is sufficient to obtain converged solutions, and that the present simulations do not suffer from immersed boundary stencils extending outside of the fluid domain.

\begin{figure}
    \centering
    \begin{subfigure}{.49\textwidth}
        \centering
        \includegraphics[width=.8\columnwidth]{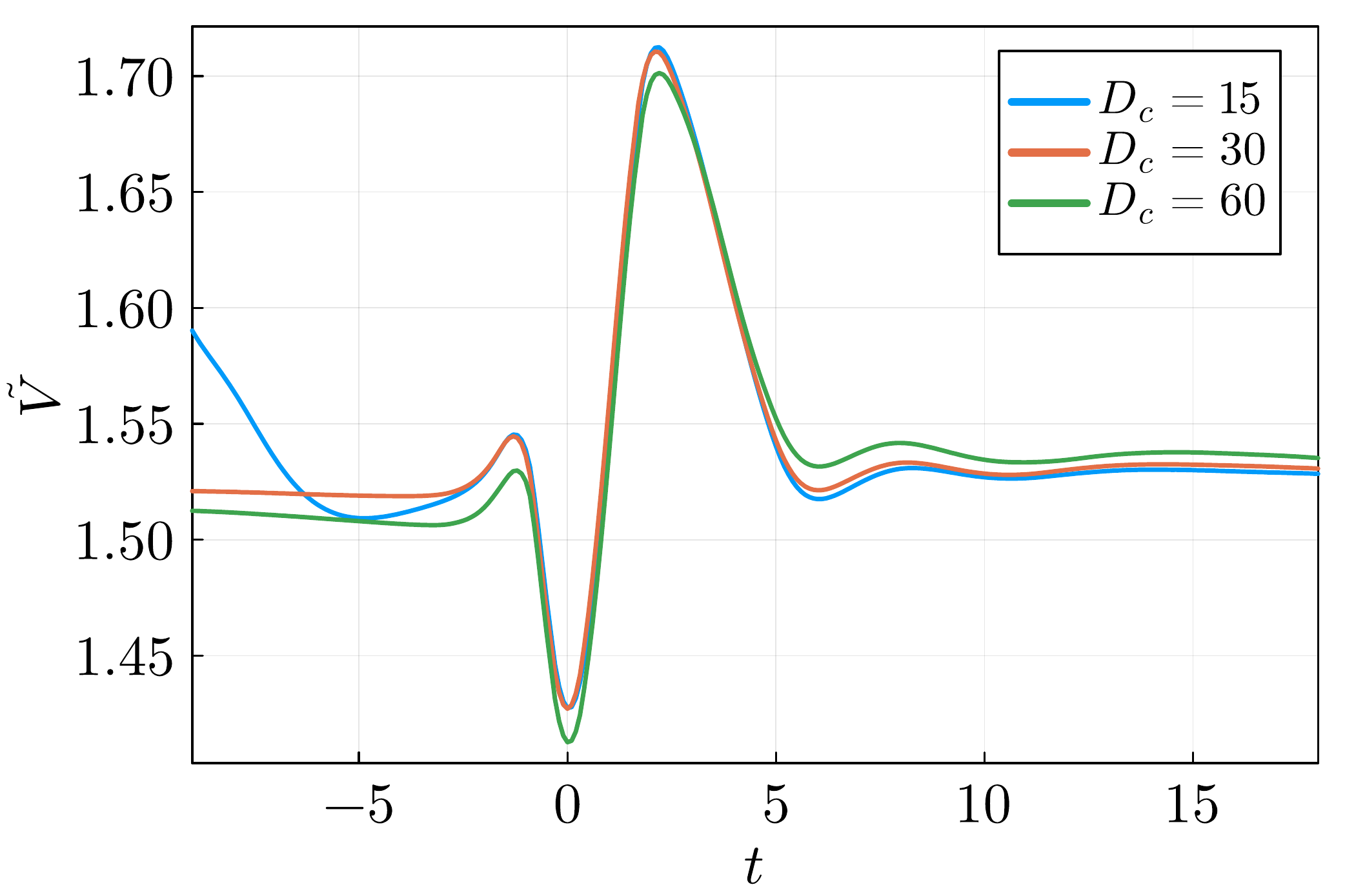}
        \caption{}
        \label{fig:inlet_length_vel}
    \end{subfigure}
    \begin{subfigure}{.49\textwidth}
        \centering
        \includegraphics[width=.8\columnwidth]{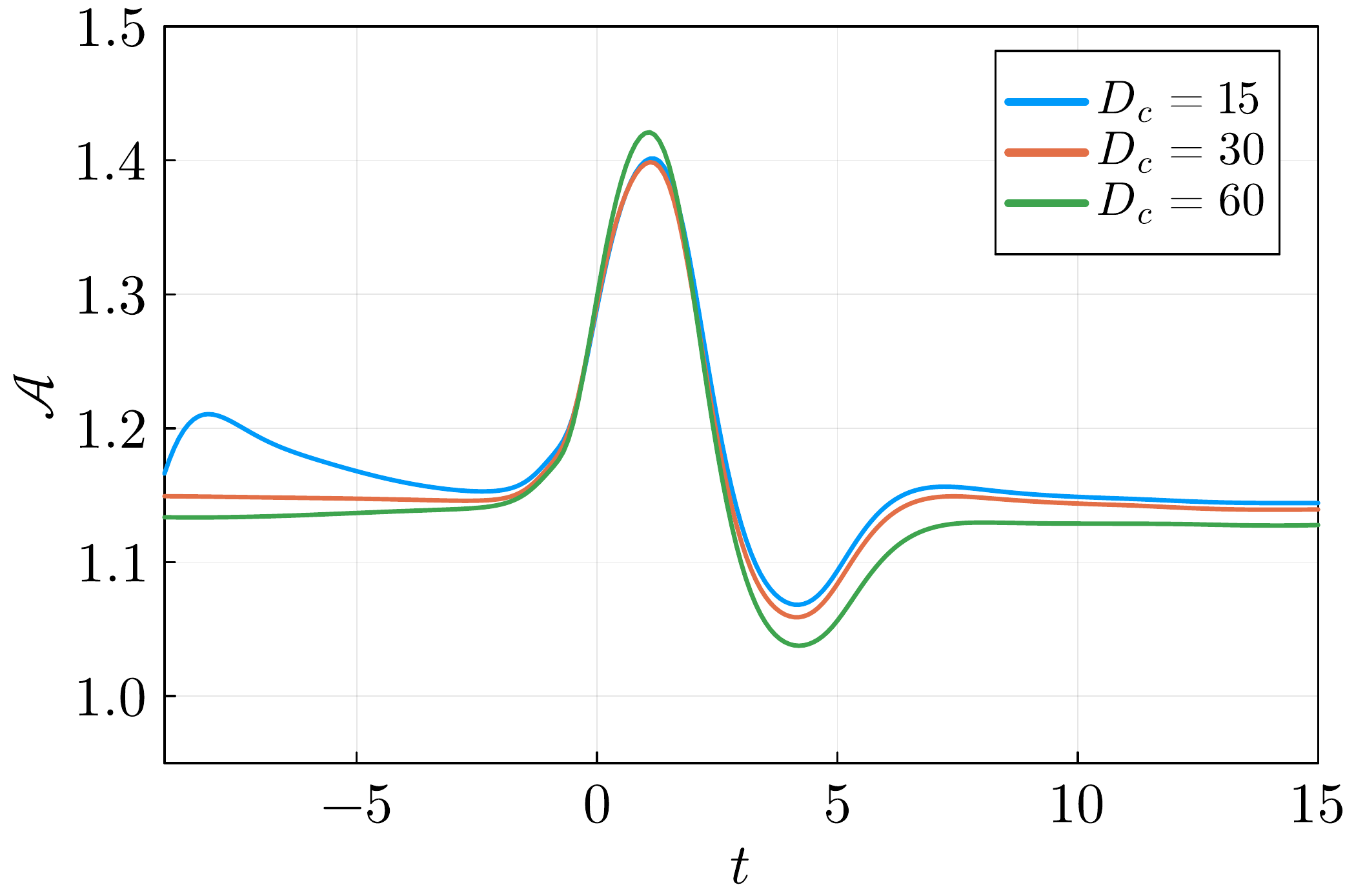}
        \caption{}
        \label{fig:inlet_length_area}
    \end{subfigure}
    \caption{Centroid velocity (a) and normalized surface area (b) of a capsule flowing through a corner from three distinct normalized release distances $D_c=15$, 30 and 60, at $Re = 50$ and $Ca = 0.35$.}
    \label{fig:inlet_length}
\end{figure}

Next we investigate the influence of the normalized release distance $D_{c}$ between the initial position of the capsule centroid and the corner. Indeed, after its release the capsule relaxes from a spherical to an equilibrium steady shape and it is important that this steady state is reached before the capsule enters the corner. As such, we consider three initial distances $D_c = 15$, 30 and 60 in the most challenging configuration at $Re = 50$ and $Ca = 0.35$, i.e. the capsule is highly deformable and placed in a highly inertial flow. The inlet boundary is located at a distance of 90$a$ away from the corner and is therefore sufficiently far away from the capsule to not alter its response. The norm of the capsule centroid velocity $\tilde{V}$ and the reduced capsule surface area $\mathcal{A} = \tilde{\mathcal{A}}/4\pi a^2$ are shown in \reffig{fig:inlet_length}, where the origin of the reduced time $t$ is chosen at the time the capsule reaches a minimum velocity $\tilde{V}_{min}$. In \reffig{fig:inlet_length_vel} we remark that the capsule velocity $\tilde{V}$ at $D_c=15$ decreases significantly prior to entering the corner: this is because the initially spherical capsule is located farther away from the channel walls and is therefore advected faster than when it has reached a steady shape. We observe that neither the capsule velocity shown in \reffig{fig:inlet_length_vel} nor the normalized surface area shown in \reffig{fig:inlet_length_area} present a steady state before the capsule enters the corner in the case $D_c = 15$. Therefore a larger initial distance $D_c$ should be used. When considering $D_c = 30$, both the velocity and the normalized surface area present steady values before the corner. Interestingly, inside and after the corner the capsule velocity and surface area almost overlap when the capsule is released 15 and 30 initial radii away from the corner, suggesting that the corner resets the dynamics of the capsule regardless of its previous state. The fact that steady values for the velocity and the surface area of the capsule are reached before the corner for $D_c = 30$ suggests that this initial release distance is suitable for the rest of this study. Interestingly, releasing the capsule at $D_c = 60$ leads to an unexpected result: the capsule seems to no longer be in a steady motion as its velocity (respectively its normalized surface area) is slightly decreasing (respectively slightly increasing) prior to entering the corner. This suggests that in this challenging configuration, the relaxation of the capsule from a fixed spherical shape to a steady ``parachute" shape occurs over very long time scales. However, the magnitude of the deviations between the capsule velocity and surface area in the cases $D_c = 30$ and 60 is at most 3\%. As the capsule has already reached a pseudo steady state by the time it reaches the corner in the case of $D_c = 30$, and as the aforementioned discrepancies are small, we choose $D_c = 30$ in the rest of this study. Again, this short study of the impact of the initial release distance on the capsule dynamics was performed in our most challenging configuration as we considered our highest Reynolds number and highest Capillary number. The discrepancy between the cases $D_c = 30$ and 60 is less pronounced $-$ sometimes nonexistent $-$ for less deformable membranes and less inertial flows.




\section{Motion and deformation of a single capsule\label{sec:corner_single}}

We consider the motion of a single capsule through a square duct at $Ca=0.075, 0.15, 0.25, 0.35$
and $Re=0.01, 1, 25, 50$, extending the investigation carried out in a non-inertial framework by Zhu \& Brandt \cite{zhu2015motion}. In order to establish the influence of the increasing effect of inertia on the motion and the deformation of a single capsule, we first recall the overall dynamics of a capsule moving through a duct corner in the Stokes regime, as detailed in \cite{zhu2015motion}. The capsule once released from its initial position moves along the center of the channel due to the symmetry of the flow far from the corner. While approaching the corner, the capsule velocity decreases until reaching a minimum in the corner region. The capsule experiences moderate to high deformation (depending on the Capillary number considered) due to the flow acceleration, and its velocity strongly increases; this phenomenon being referred to as the overshoot of velocity. Further away from the corner, the capsule moves in the downstream branch of the duct, relaxing to a steady state (shape and velocity), and moving along the center of the duct.

We investigate the influence of the Reynolds number $Re$ and the Capillary number $Ca$ on the dynamics and the deformation of the capsule, reporting the time evolution of its surface area $\mathcal{A}$ scaled by the initial surface area of the capsule $\mathcal{A}_\text{sphere} = 4\pi \tilde{a}^2$, as well as the velocity $V$ of the capsule centroid scaled by its equilibrium velocity $V_{eq}$ before the capsule enters the corner region. In the remainder of this study and unless otherwise stated, the time origin is chosen such that $t = 0$ when capsule velocity reaches a global minimum, i.e. $V_{min } = V(t = 0)$. We borrow this convention from Zhu \& Brandt \cite{zhu2015motion}, as it corresponds to setting the time origin when the capsule is located at the heart of the corner.

\subsection{Influence of the Reynolds and Capillary numbers\label{sec:corner_single_re}}

\begin{figure}
    \centering
 \begin{subfigure}{.49\textwidth}
        \centering
        \includegraphics[width=.8\columnwidth, trim=0 0 0 0, clip]{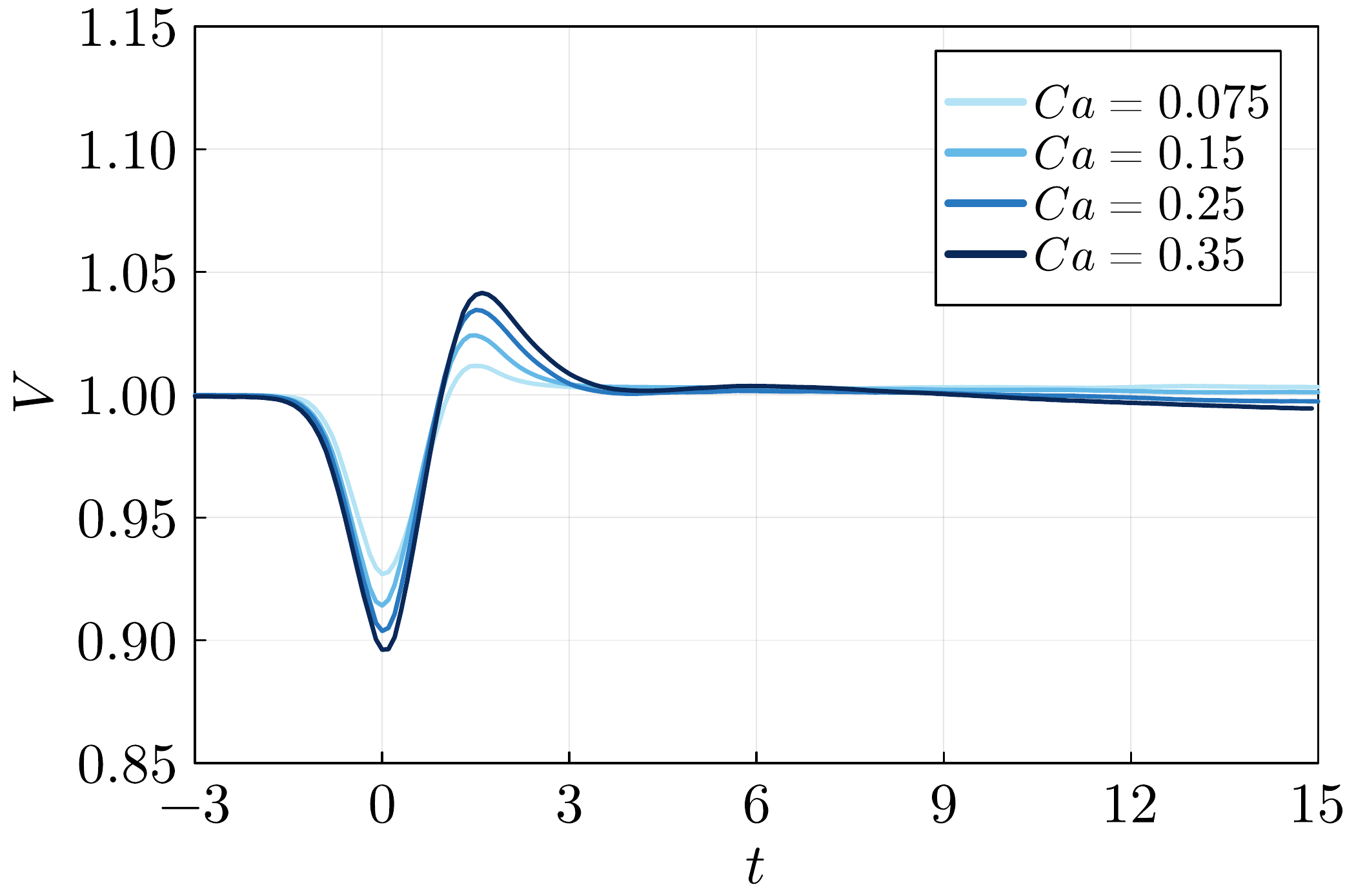}
        \caption{$Re=0.01$}
    \end{subfigure}
    \begin{subfigure}{.49\textwidth}
        \centering
        \includegraphics[width=.8\columnwidth, trim=0 0 0 0, clip]{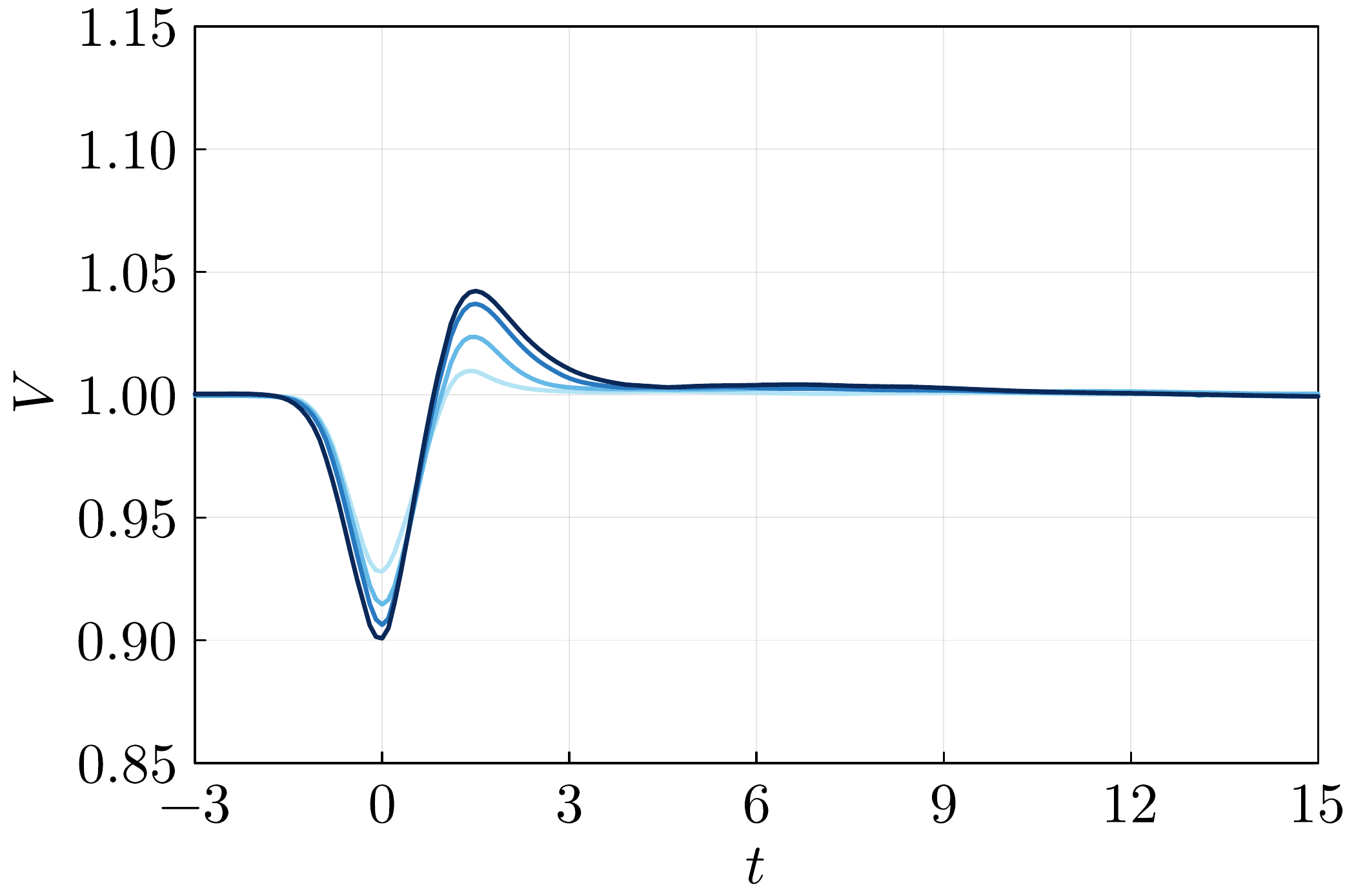}
        \caption{$Re=1$}
    \end{subfigure}\\

    \begin{subfigure}{.49\textwidth}
        \centering
        \includegraphics[width=.8\columnwidth, trim=0 0 0 0, clip]{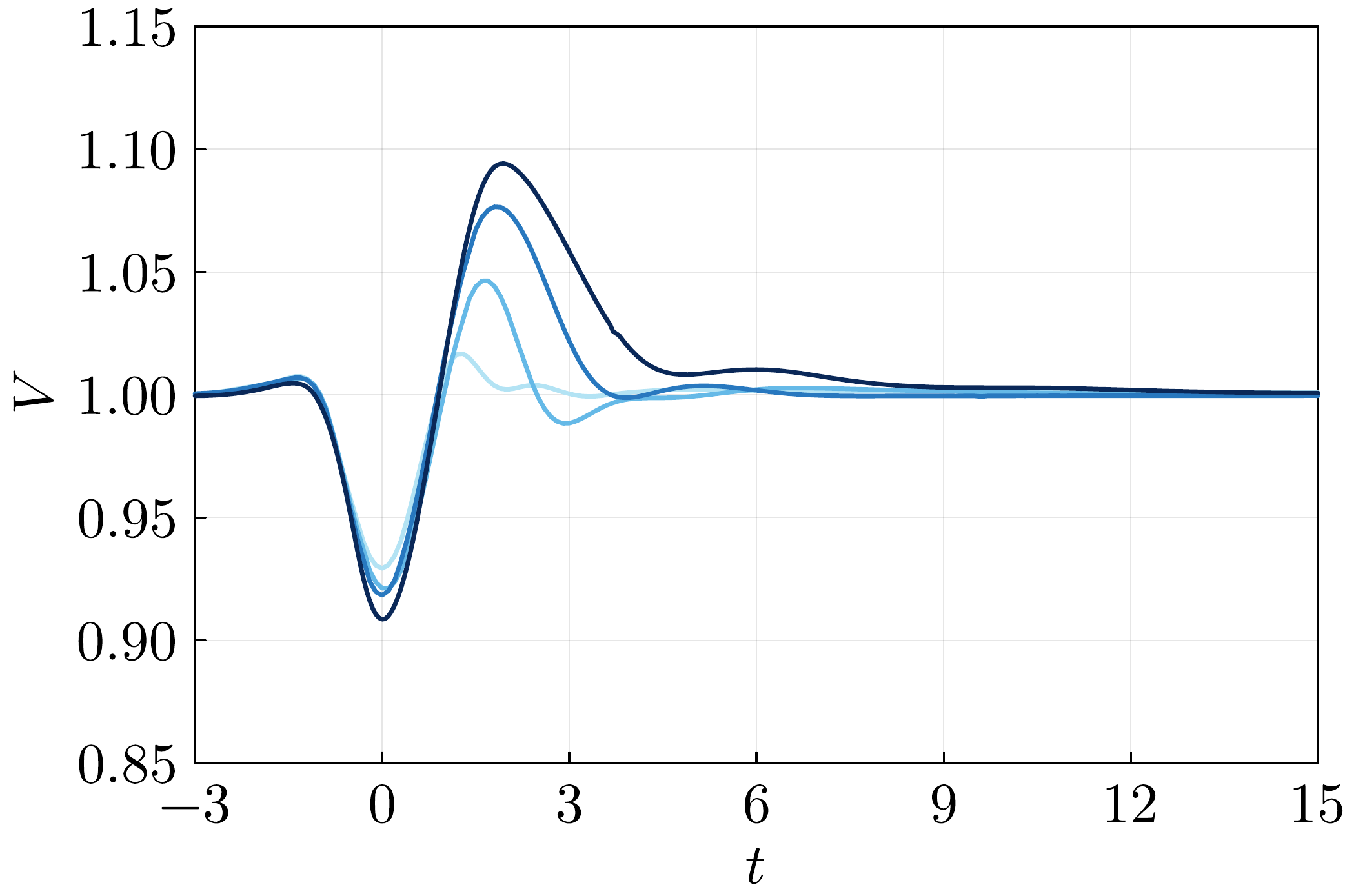}
        \caption{$Re=25$}
    \end{subfigure}
    \begin{subfigure}{.49\textwidth}
        \centering
        \includegraphics[width=.8\columnwidth, trim=0 0 0 0, clip]{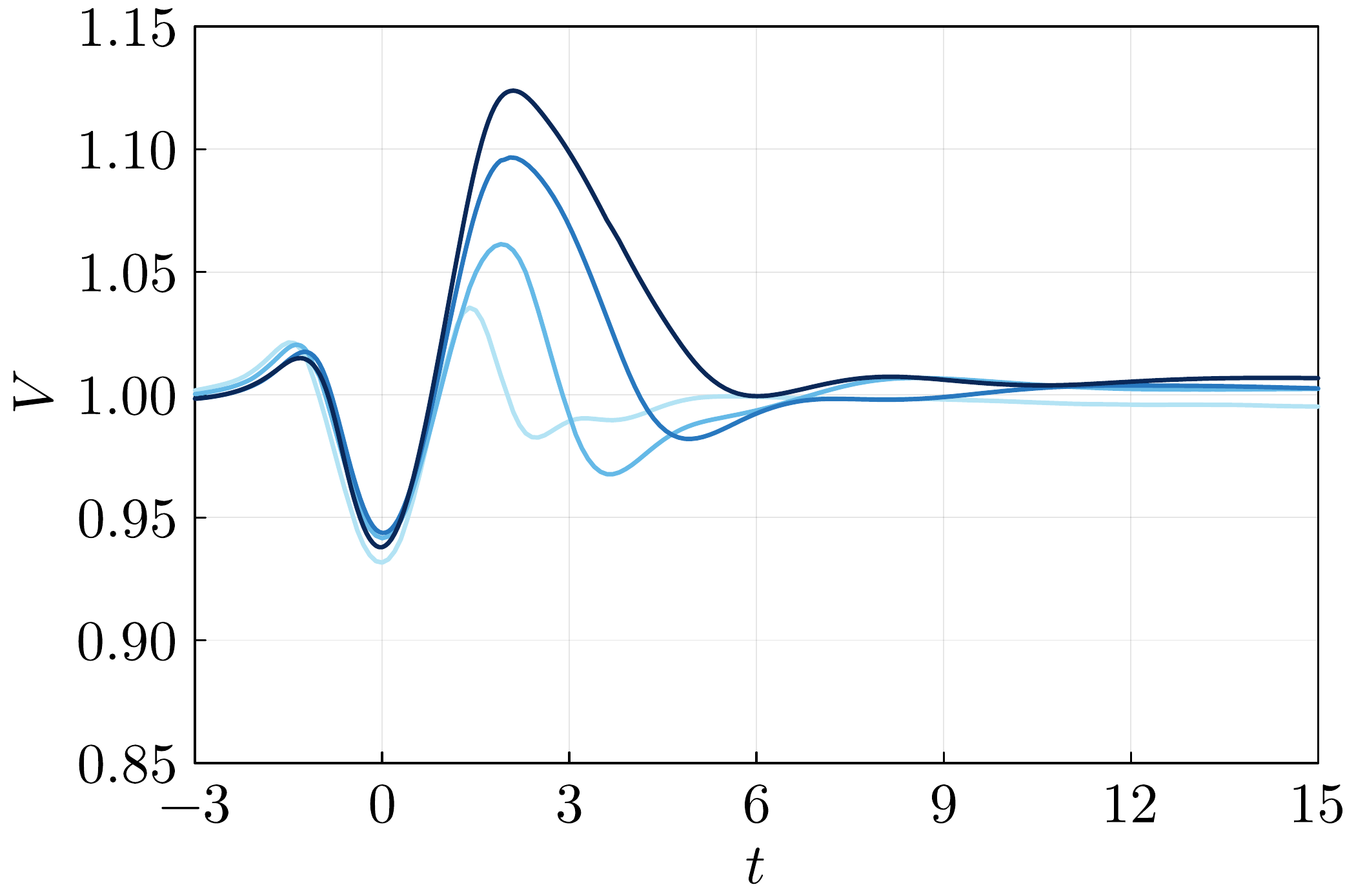}
        \caption{$Re=50$}
    \end{subfigure}
    \caption{Temporal evolution of the capsule centroid velocity $V$ at fixed Reynolds numbers.}
    \label{fig:allvelRe}
\end{figure}

\begin{figure}
    \centering
 \begin{subfigure}{.49\textwidth}
        \centering
        \includegraphics[width=.8\columnwidth, trim=0 0 0 0, clip]{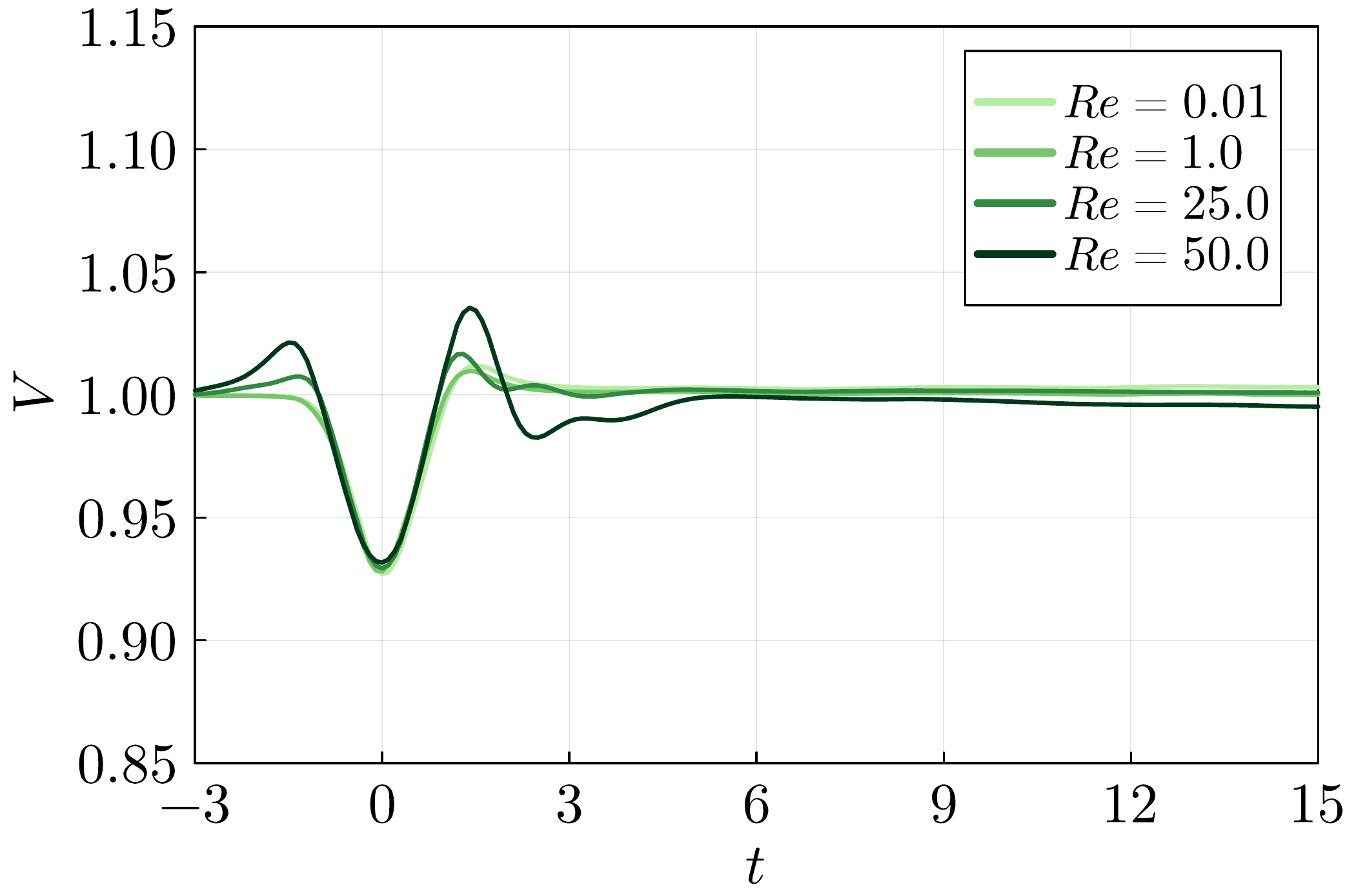}
        \caption{$Ca=0.075$}
        \label{fig:Vc_allReCa=0.075}
    \end{subfigure}
    \begin{subfigure}{.49\textwidth}
        \centering
        \includegraphics[width=.8\columnwidth, trim=0 0 0 0, clip]{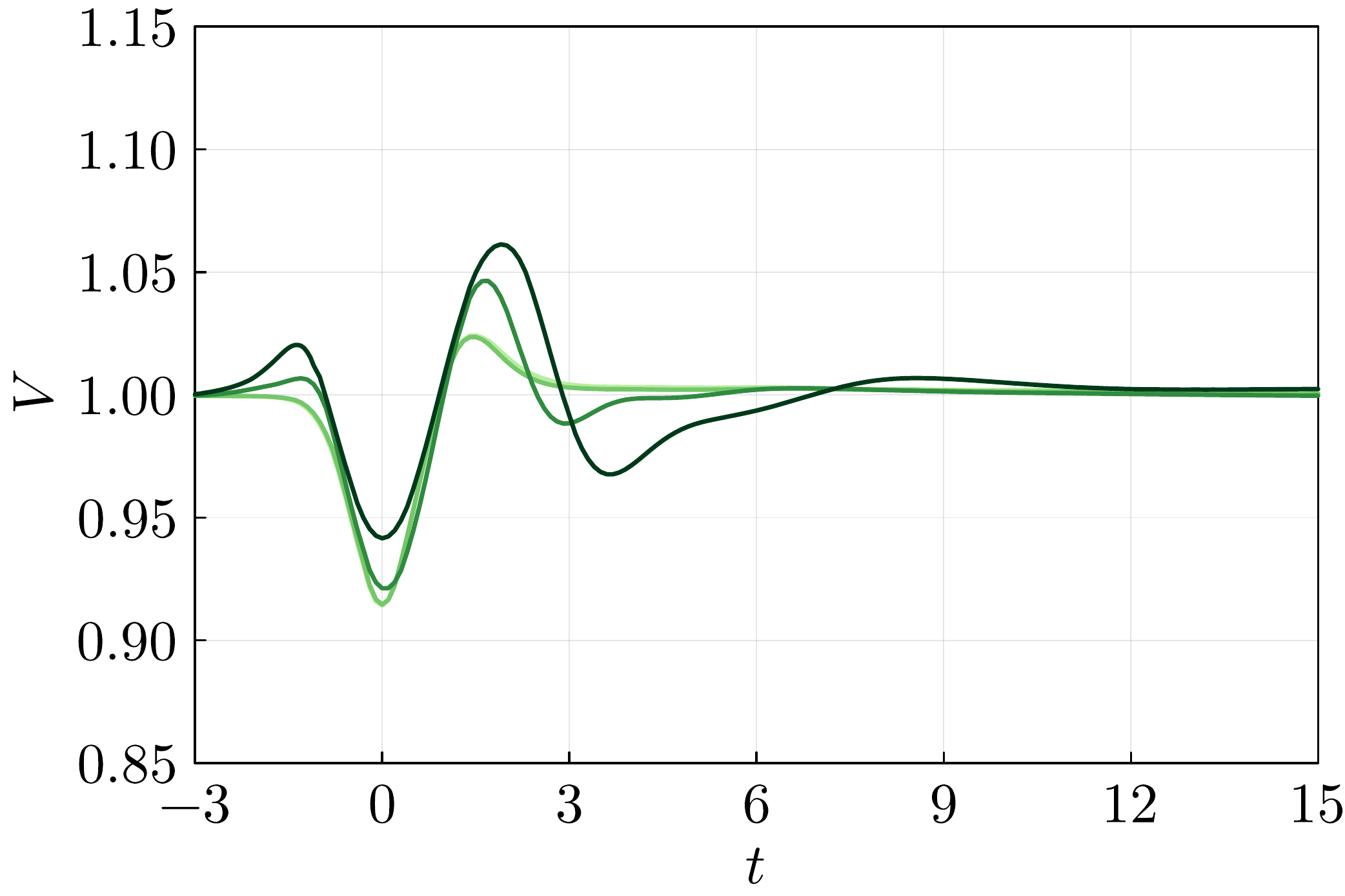}
        \caption{$Ca=0.15$}
    \end{subfigure}\\

    \begin{subfigure}{.49\textwidth}
        \centering
        \includegraphics[width=.8\columnwidth, trim=0 0 0 0, clip]{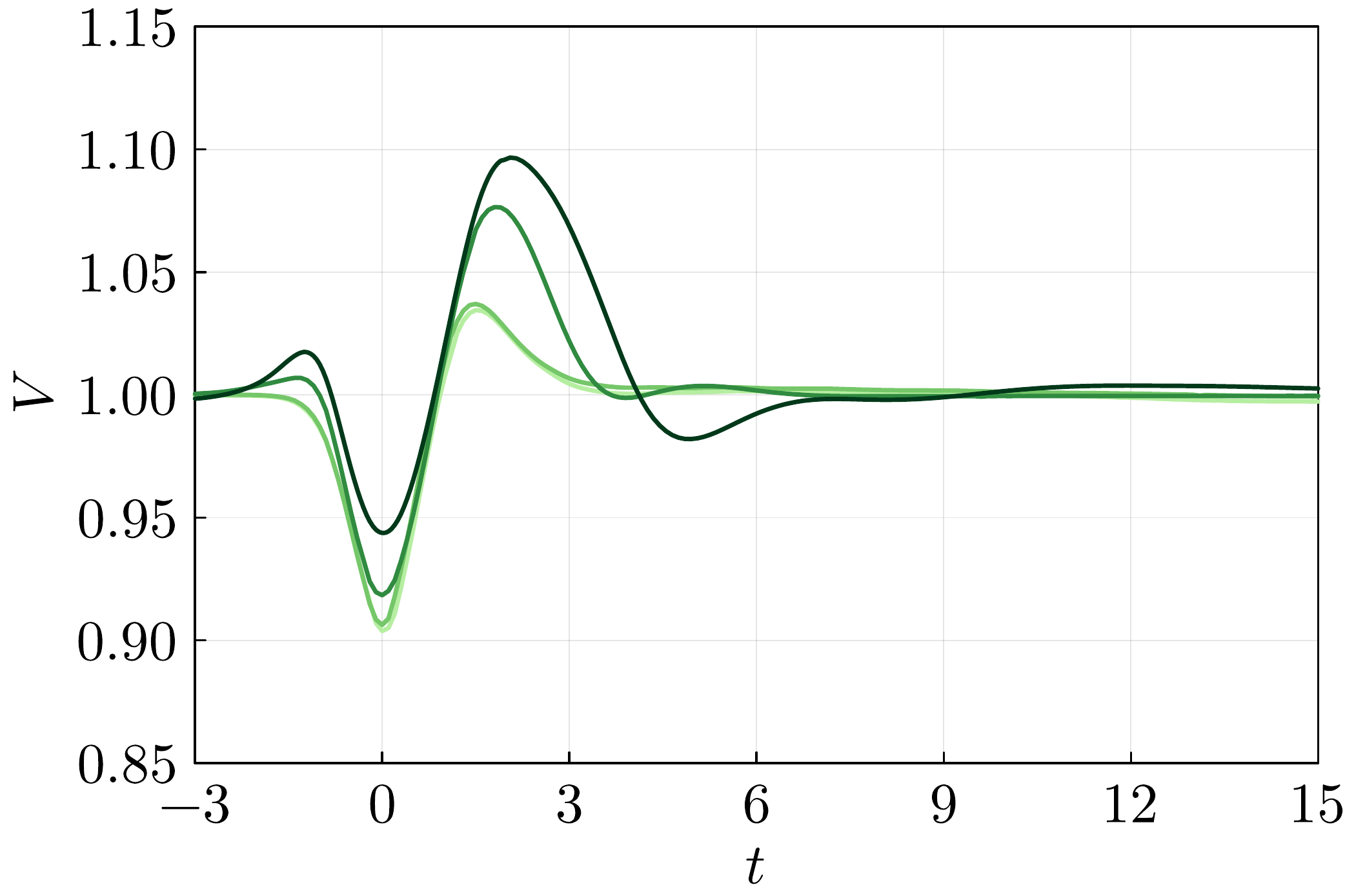}
        \caption{$Ca=0.25$}
    \end{subfigure}
    \begin{subfigure}{.49\textwidth}
        \centering
        \includegraphics[width=.8\columnwidth, trim=0 0 0 0, clip]{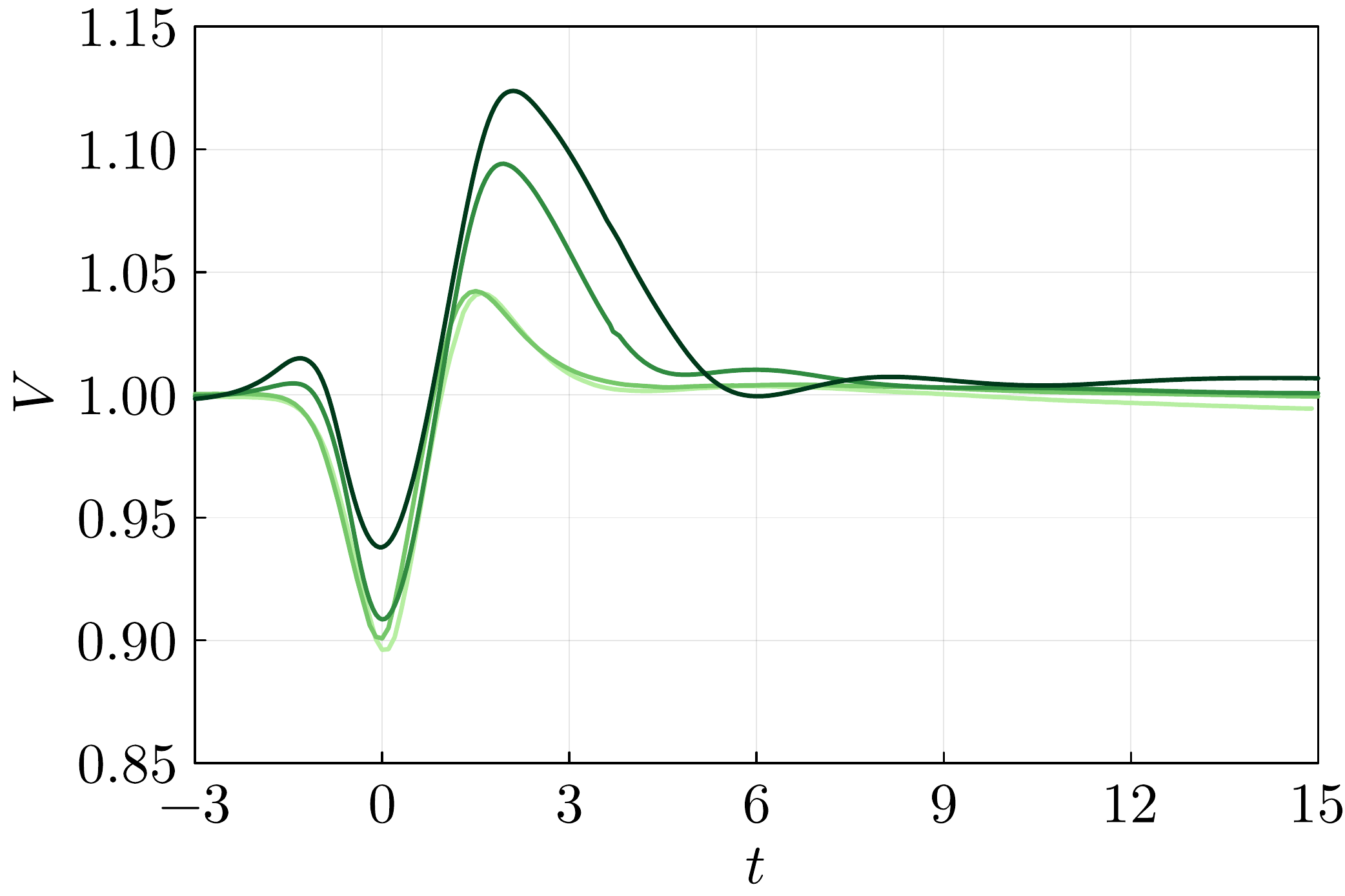}
        \caption{$Ca=0.35$}
    \end{subfigure}
    \caption{Temporal evolution of the capsule centroid velocity $V$ at fixed Capillary numbers.}
    \label{fig:allvelCa}
\end{figure}

To characterize the dynamics of the capsule as it flows through the corner, we analyze the time evolution of the centroid velocity $V$ and the surface area $\mathcal{A}$. \Reffig{fig:allvelRe} shows the velocity of the capsule centroid for $Ca$ ranging from 0.075 to 0.35. $Re$ is constant for each subfigure of \reffig{fig:allvelRe}. Conversely, \reffig{fig:allvelCa} shows the same data as \reffig{fig:allvelRe}, but with each subfigure corresponding to a constant $Ca$.
From both figures, we observe a general trend for all cases: the capsule approaches the corner with a steady velocity $V_{eq}$, then reaches a global minimum $V_{min}$ and a global maximum $V_{max}$ as it flows through the corner, and relaxes back to $V_{eq}$ downstream of the corner. Moreover, we observe in \reffig{fig:allvelRe} that the velocity extrema increase with increasing $Ca$. In the more inertial regimes especially, the maximum velocity deviation of the capsule at $Ca = 0.35$ is close to three times that of the capsule at $Ca = 0.075$.

We note from \reffig{fig:allvelCa} that the curves corresponding to $Re = 0.01$ and $Re = 1$ practically overlap, indicating that the capsule motion in low inertial regimes is very similar to that in the non-inertial regime. As the Reynolds number is increased to 25 and 50, major deviations from the non-inertial regime appear. First, as the capsule enters the corner zone, a local maximum appears in the capsule velocity, which is independent of the Capillary number, and is about 1\% greater than $V_{eq}$ at $Re = 25$ and 2\% greater than $V_{eq}$ at $Re = 50$. This local maximum is due to the migration of the capsule across the centerline of the secondary channel: in this process the capsule is located far away from the channel walls and is therefore less subject to their confinement effect. Then, the minimum velocity $V_{min}$ is reached in the heart of the corner.
Interestingly, at small $Ca$, $V_{min}$ is observed to be independent of the $Re$, as can be seen in \reffig{fig:Vc_allReCa=0.075} at $Ca = 0.075$. In contrast, in the case of larger $Ca$ the minimum velocity of the capsule increases slightly with $Re$. A difference of about 4\% is observed for $V_{min}$ as $Re$ increases from 0.01 to 50 for both $Ca = 0.25$ and $Ca = 0.35$.

As the capsule exits the corner zone and migrates to the channel centerline, its velocity reaches its maximum value $V_{max}$ which increases with increasing $Re$ and $Ca$: at $Ca = 0.075$, $V_{max}$ increases by 3\% between $Re = 0.01$ and $Re = 50$ while at $Ca = 0.35$, $V_{max}$ increases by about 8\% between $Re = 0.01$ and $Re = 50$. Then, the capsule velocity relaxes back to its equilibrium value and its relaxation time increases with increasing $Re$. Interestingly, velocity undershoots are observed during the relaxation stage in the inertial regime, which magnitude increases with $Re$. The relaxation time does not depend on $Ca$.

\begin{figure}
    \centering
 \begin{subfigure}{.49\textwidth}
        \centering
        \includegraphics[width=.8\columnwidth, trim=0 0 0 0, clip]{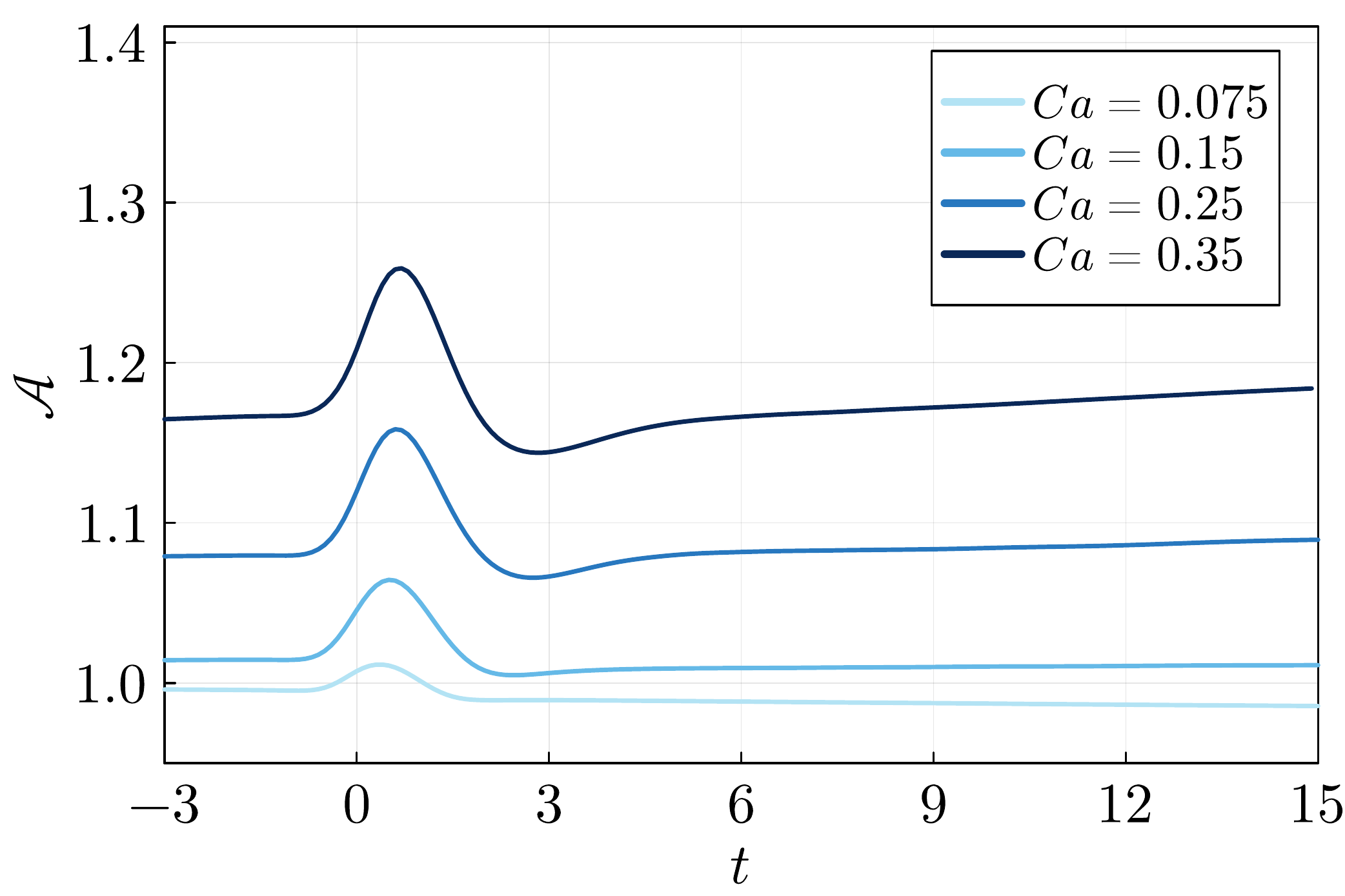}
        \caption{$Re=0.01$}
    \end{subfigure}
    \begin{subfigure}{.49\textwidth}
        \centering
        \includegraphics[width=.8\columnwidth, trim=0 0 0 0, clip]{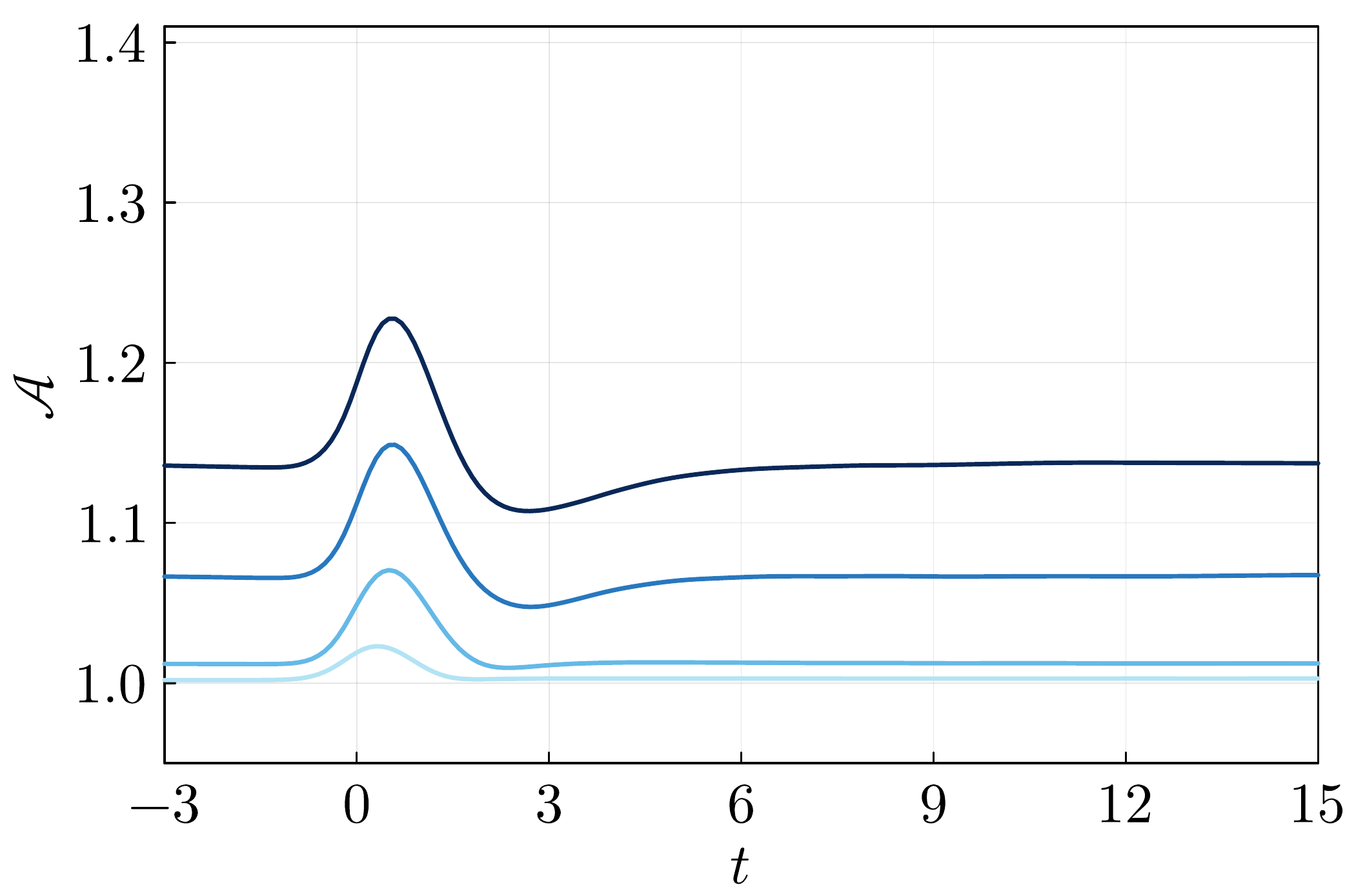}
        \caption{$Re=1$}
    \end{subfigure}\\

    \begin{subfigure}{.49\textwidth}
        \centering
        \includegraphics[width=.8\columnwidth, trim=0 0 0 0, clip]{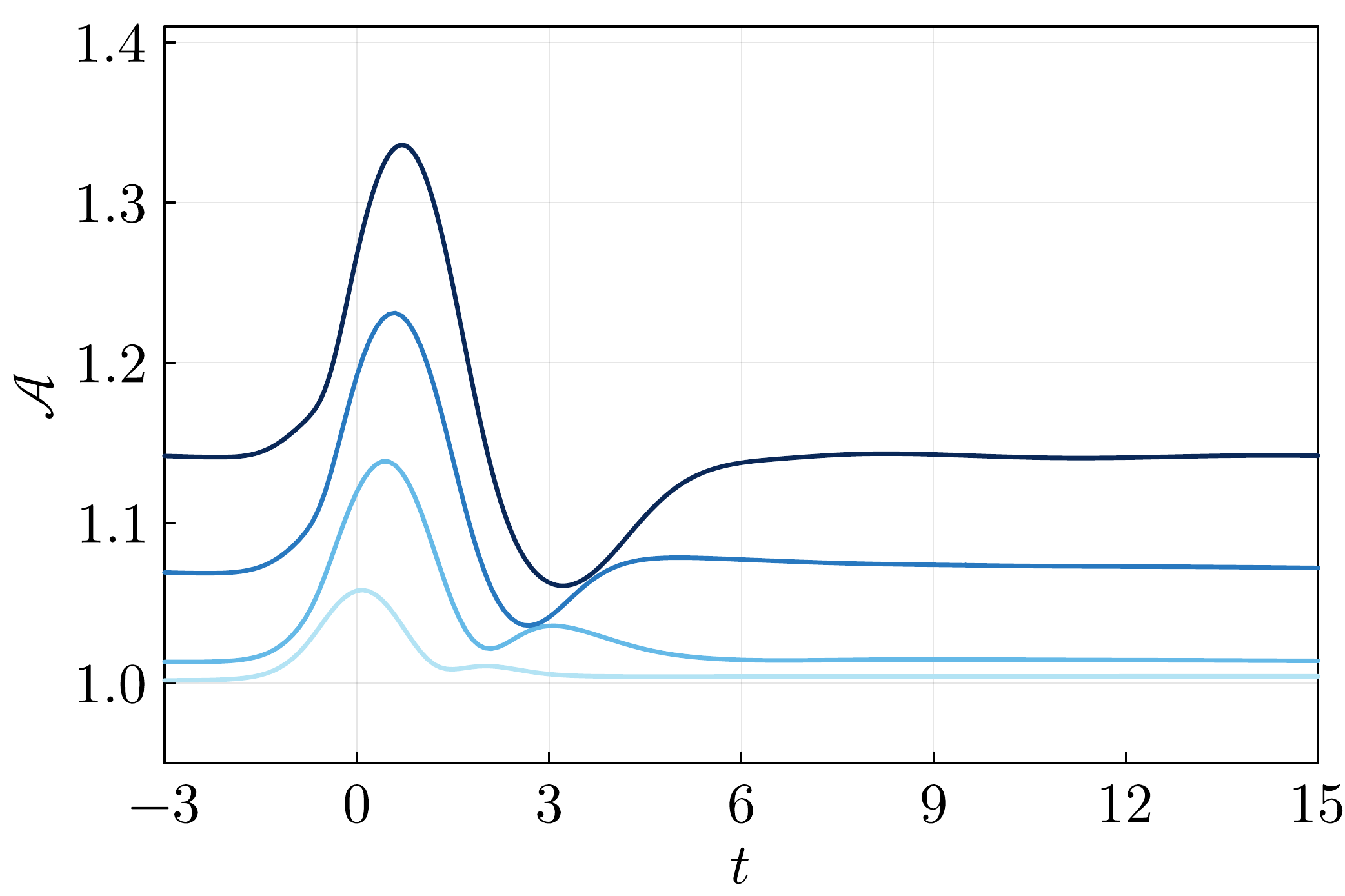}
        \caption{$Re=25$}
    \end{subfigure}
    \begin{subfigure}{.49\textwidth}
        \centering
        \includegraphics[width=.8\columnwidth, trim=0 0 0 0, clip]{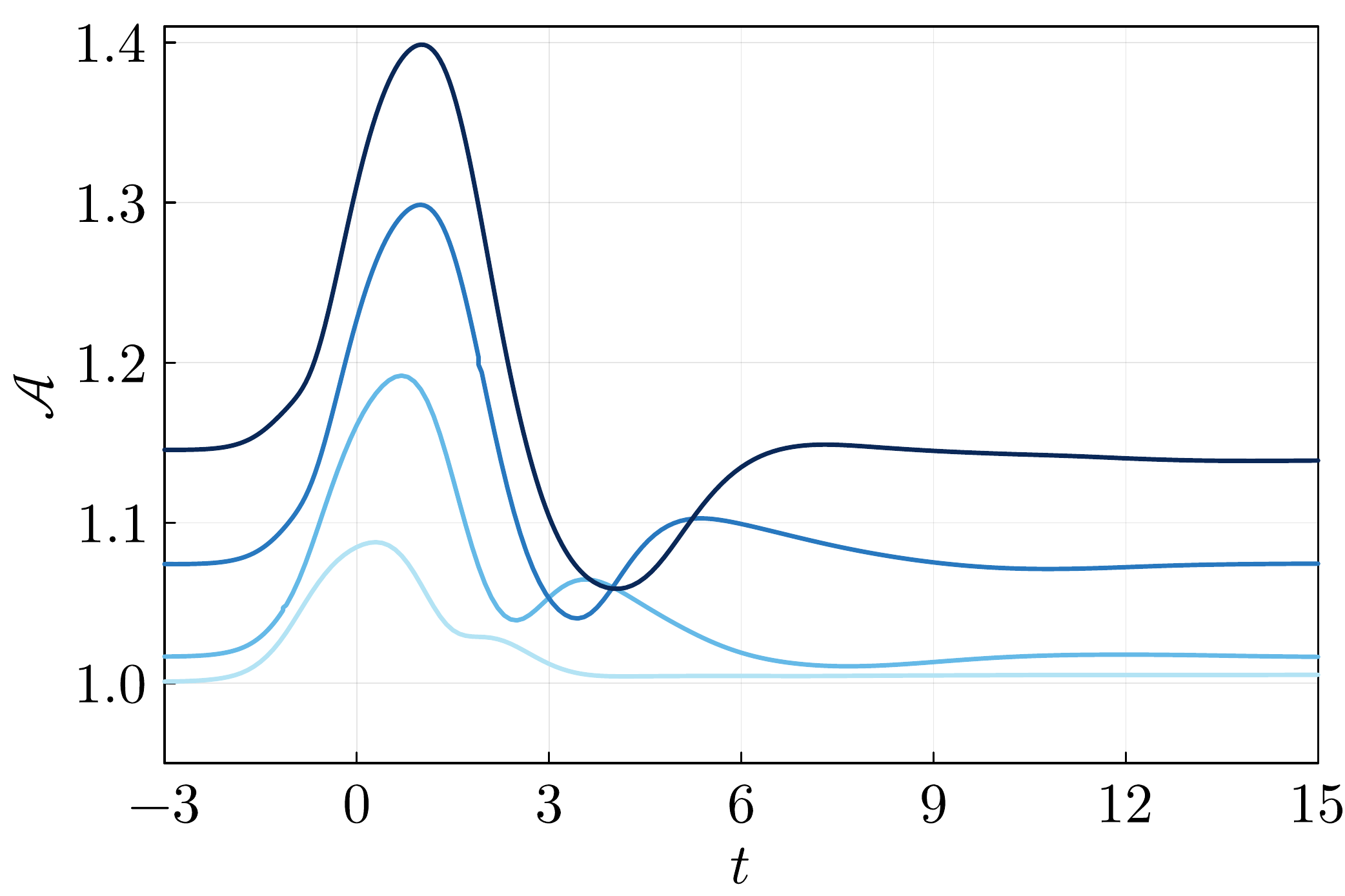}
        \caption{$Re=50$}
    \end{subfigure}
    \caption{Temporal evolution of the capsule surface area $\mathcal{A}$ at fixed Reynolds numbers.}
    \label{fig:allareaRe}
\end{figure}

\begin{figure}
    \centering
 \begin{subfigure}{.49\textwidth}
        \centering
        \includegraphics[width=.8\columnwidth, trim=0 0 0 0, clip]{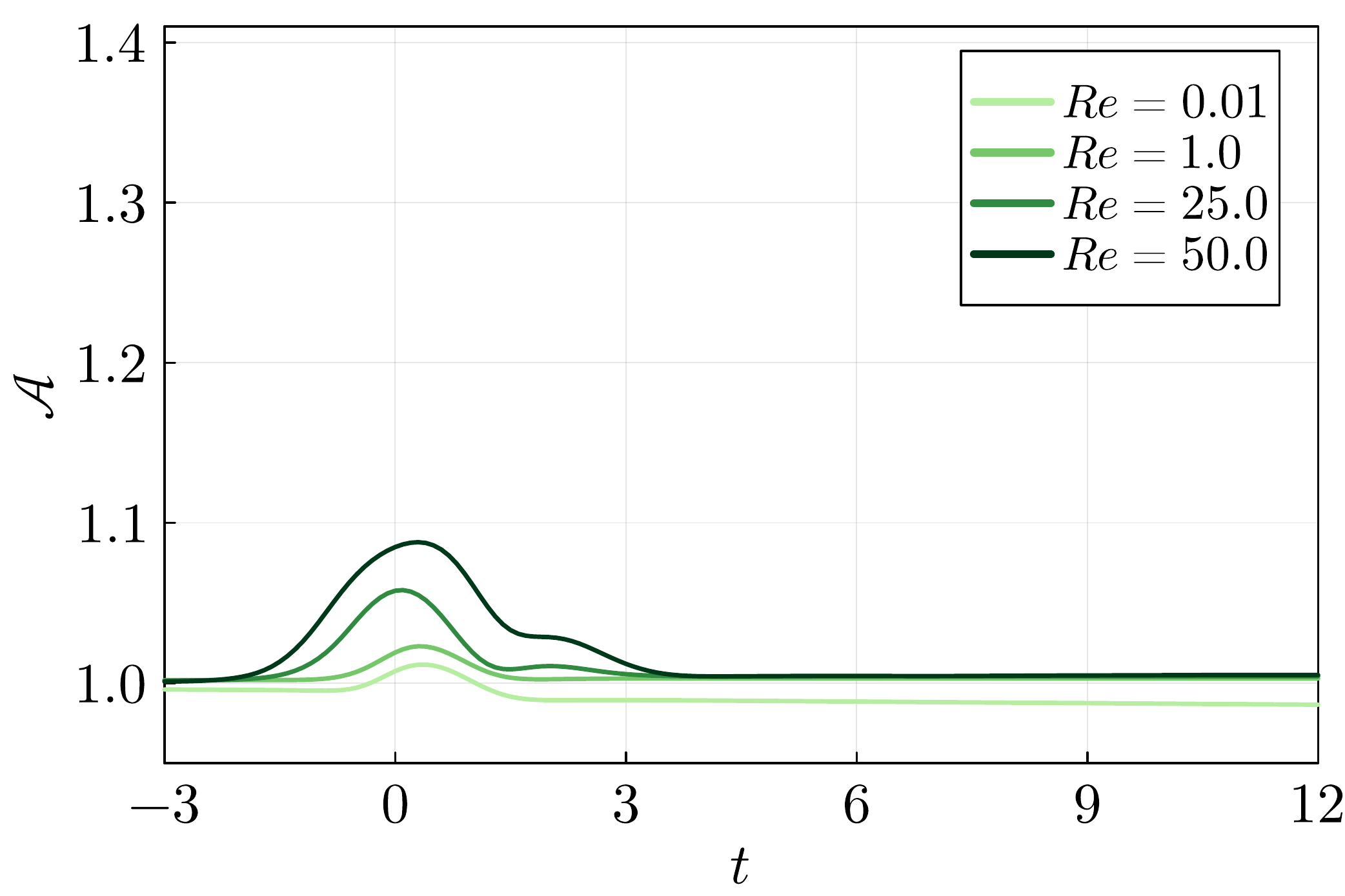}
        \caption{$Ca=0.075$}
        \label{fig:allareaCa1}
    \end{subfigure}
    \begin{subfigure}{.49\textwidth}
        \centering
        \includegraphics[width=.8\columnwidth, trim=0 0 0 0, clip]{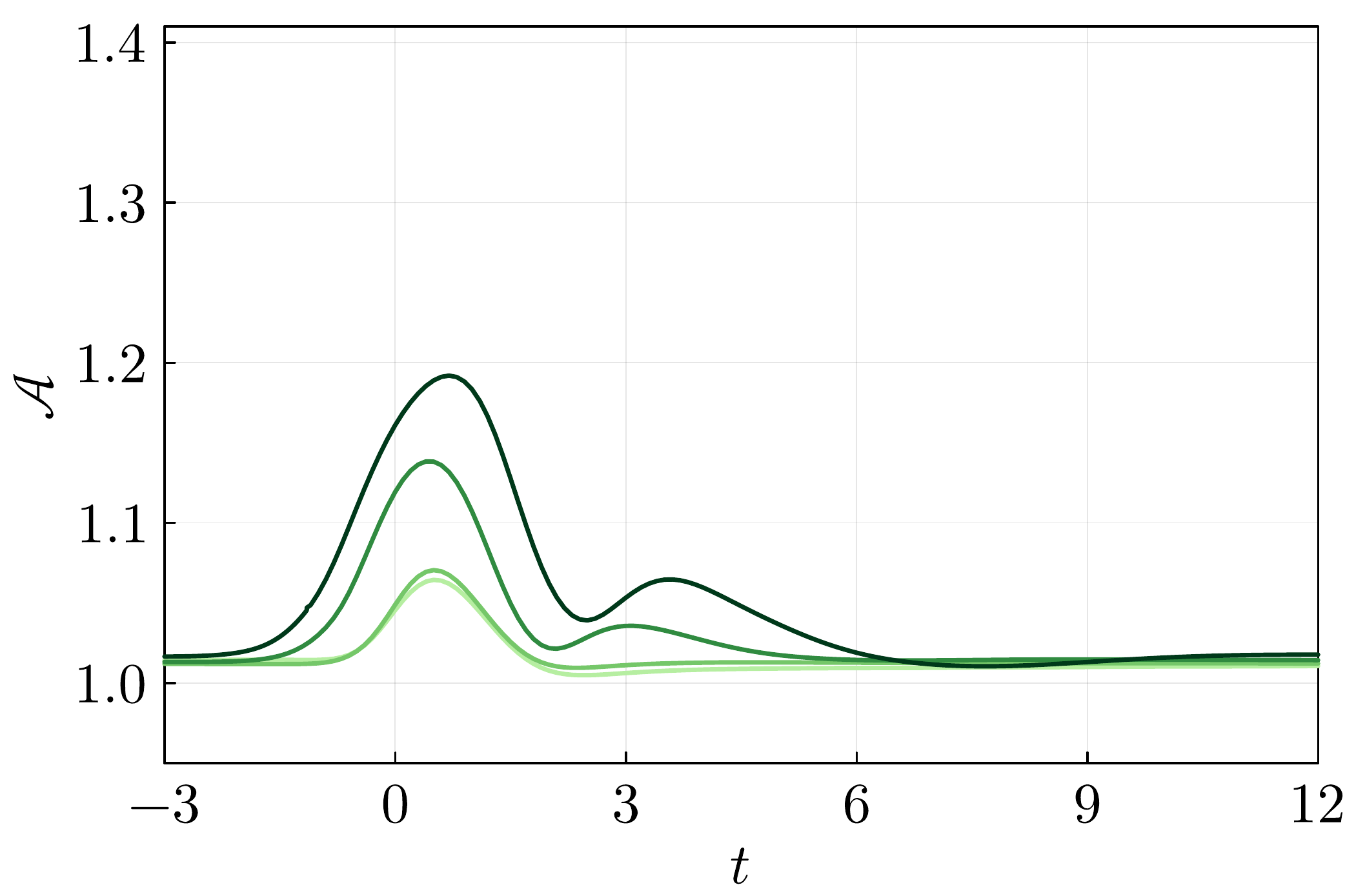}
       \caption{$Ca=0.15$}
       \label{fig:allareaCa2}
    \end{subfigure}\\

    \begin{subfigure}{.49\textwidth}
        \centering
        \includegraphics[width=.8\columnwidth, trim=0 0 0 0, clip]{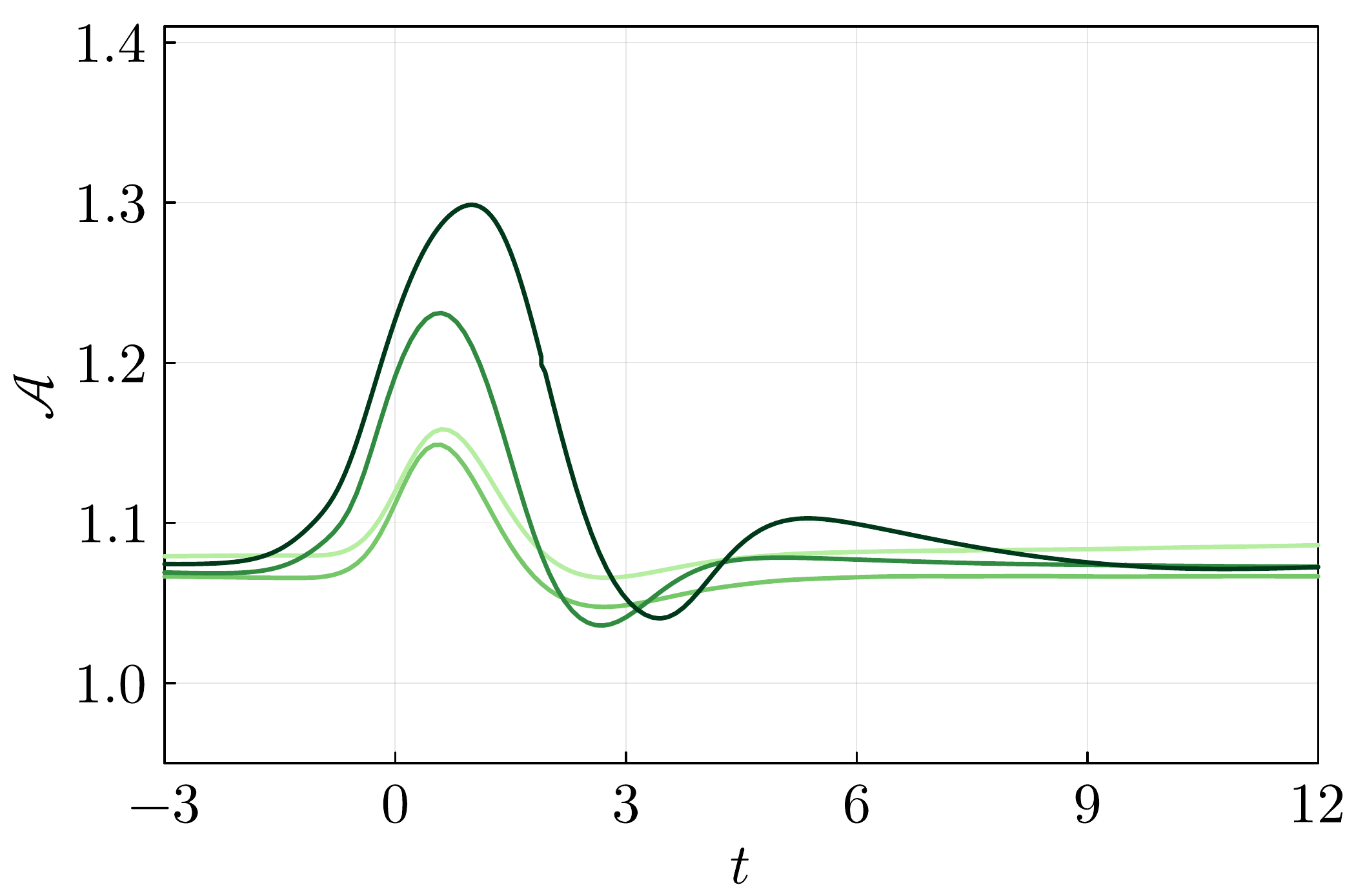}
        \caption{$Ca=0.25$}
        \label{fig:allareaCa3}
    \end{subfigure}
    \begin{subfigure}{.49\textwidth}
        \centering
        \includegraphics[width=.8\columnwidth, trim=0 0 0 0, clip]{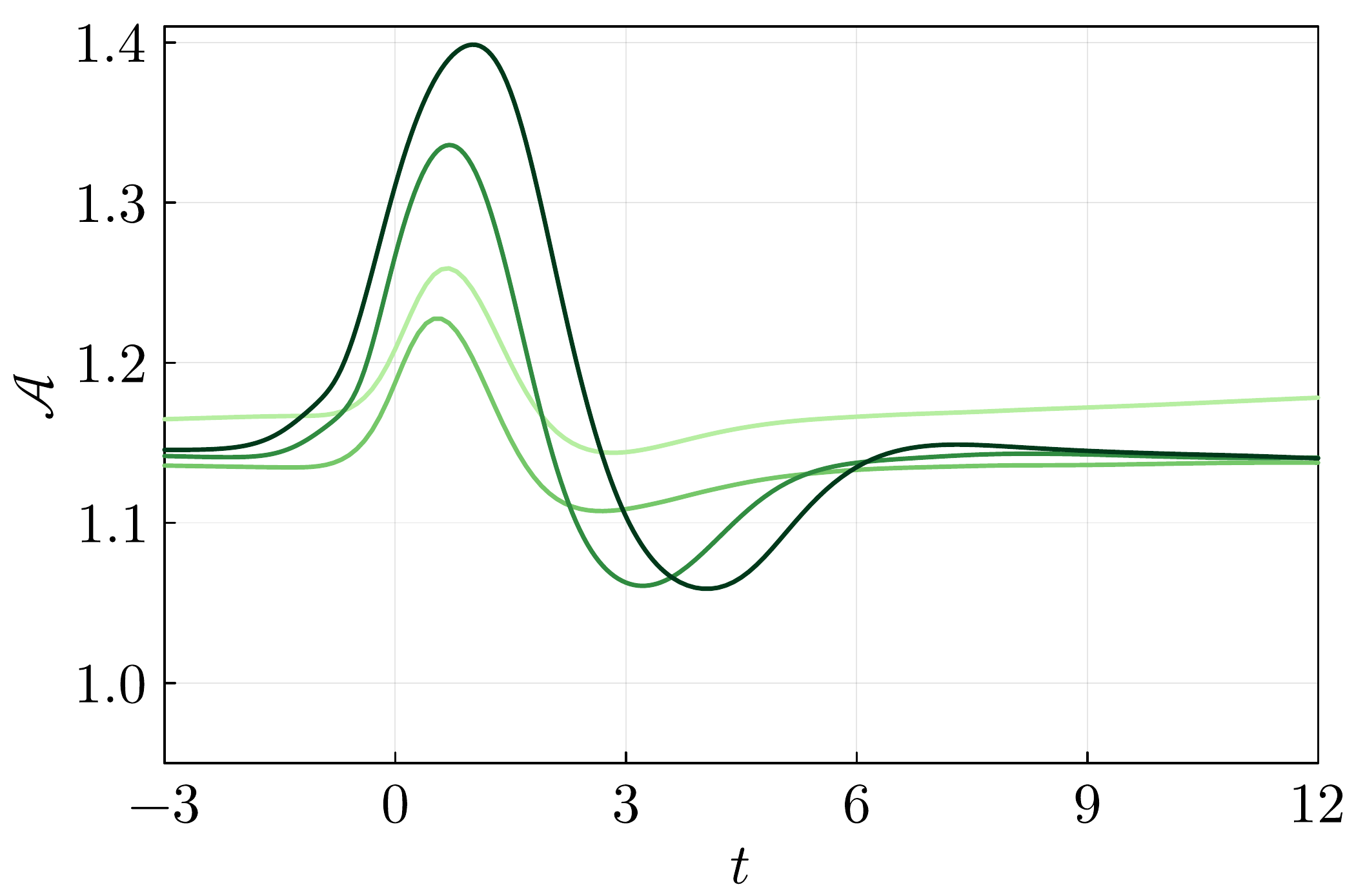}
        \caption{$Ca=0.35$}
        \label{fig:allareaCa4}
    \end{subfigure}
    \caption{Temporal evolution of the capsule surface area $\mathcal{A}$ at fixed Capillary numbers.}
    \label{fig:allareaCa}
\end{figure}

The time evolution of the normalized capsule surface area $\mathcal{A}$ is shown in \reffig{fig:allareaRe} for fixed $Re$ and in \reffig{fig:allareaCa} for fixed $Ca$. We observe that the surface area presents a maximum $\mathcal{A}_{max}$ at around $t = 1$ before relaxing to its equilibrium value $\mathcal{A}_{eq}$.
Unsurprisingly, \reffig{fig:allareaRe} confirms that a large $Ca$, i.e. a highly deformable capsule, results in a greater surface area than for lower $Ca$. \Reffig{fig:allareaRe} also shows that the magnitude of the maximum surface area increases with $Ca$. Moreover, when large $Ca$ are considered the time evolution of the capsule surface area presents some undershoots that are more pronounced as $Re$ is increased. Additionally, \reffig{fig:allareaCa} reveals that $Re$ has a very strong influence on the deformation of the capsule, especially at large $Ca$: at $Ca = 0.075$, $\mathcal{A}_{max}/\mathcal{A}_{eq}$ increases from 2\% to 8\% between $Re = 0.01$ and $Re = 50$, and at $Ca = 0.35$ it increases from 8\% to a staggering 22\% between $Re = 0.01$ to $Re = 0.35$. In particular, at $Ca = 0.35$ the maximum capsule surface area increases from 9\% to 40\% of the surface area of a sphere between the non-inertial and the highly inertial regimes. These surface area deviations are very large and are discussed further in the next section.


\subsection{Maximum deformation of the capsule\label{sec:corner_single_area}}



 \begin{figure}
\centering
    \begin{subfigure}{.49\textwidth}
        \centering
        \includegraphics[width=.8\textwidth]{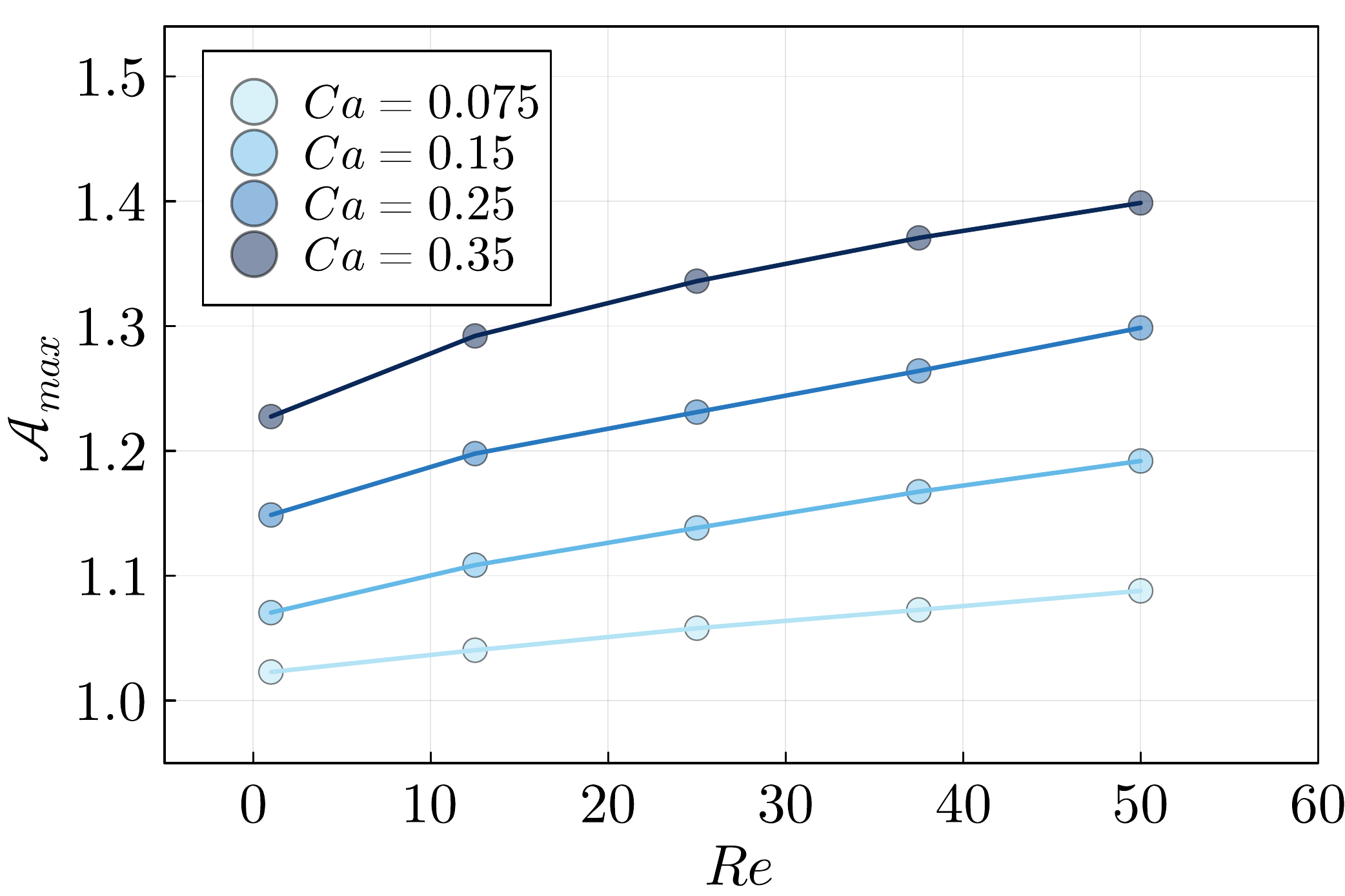}
        \caption{$\mathcal{A}_{max}$ as a function of $Re$}
    \end{subfigure}
    \begin{subfigure}{.49\textwidth}
        \centering
        \includegraphics[width=.8\textwidth]{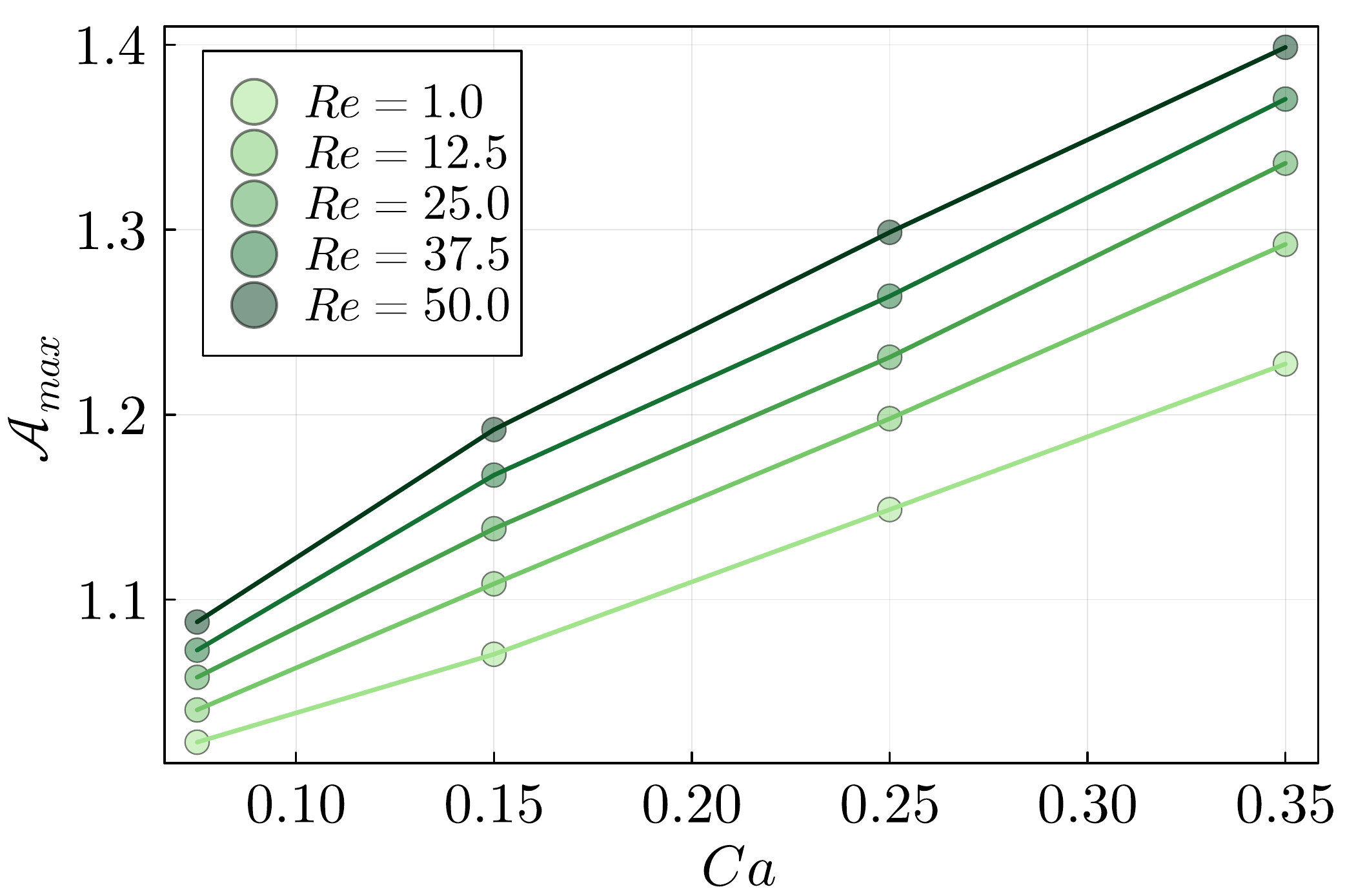}
        \caption{$\mathcal{A}_{max}$ as a function of $Ca$}
    \end{subfigure}
  \caption{Maximum surface area $\mathcal{A}_{max}$ as a function of $Re$ and $Ca$ for a single capsule passing through the corner.}
\label{Cca_Amax}
\end{figure}

The maximum surface area $\mathcal{A}_{max}$ of the capsule is presented in \reffig{Cca_Amax}, as a function of both the Reynolds number and the Capillary number. To better analyze the trends in this figure, we also report the maximum area at intermediate Reynolds numbers, namely at $Re = 12.5$ and $37.5$.
The data reported in \reffig{Cca_Amax} clearly exhibits a double linear scaling of $\mathcal{A}_{max}$ 
with both $Ca$ and $Re$ as long as $Ca$ is below 0.35 $-$ at $Ca = 0.35$, the shape of the curve $\mathcal{A}_{max}(Re)$ is slightly concave. The slope of the scaling is about $0.003$ for $\mathcal{A}_{max}(Ca)$ and $1.12$ for $\mathcal{A}_{max}(Re)$. This means that the capsule maximum deformation responds proportionally to the Capillary number, but also to the Reynolds number. To our knowledge, this is the first time such a trend has been reported and established for low ($Re=1$) to moderate ($Re=12.5, 25, 37.5, 50$) inertial regimes. We believe that this result can be used as a predictive tool for many studies involving single capsules travelling through duct corners, as the maximum deformation observed for a capsule is a measure of its mechanical integrity, which is of major interest in many microfluidic applications.

 \begin{figure}
\centering
\begin{subfigure}{.49\textwidth}
        \centering
        \includegraphics[width=.8\columnwidth, trim=0 0 0 0, clip]{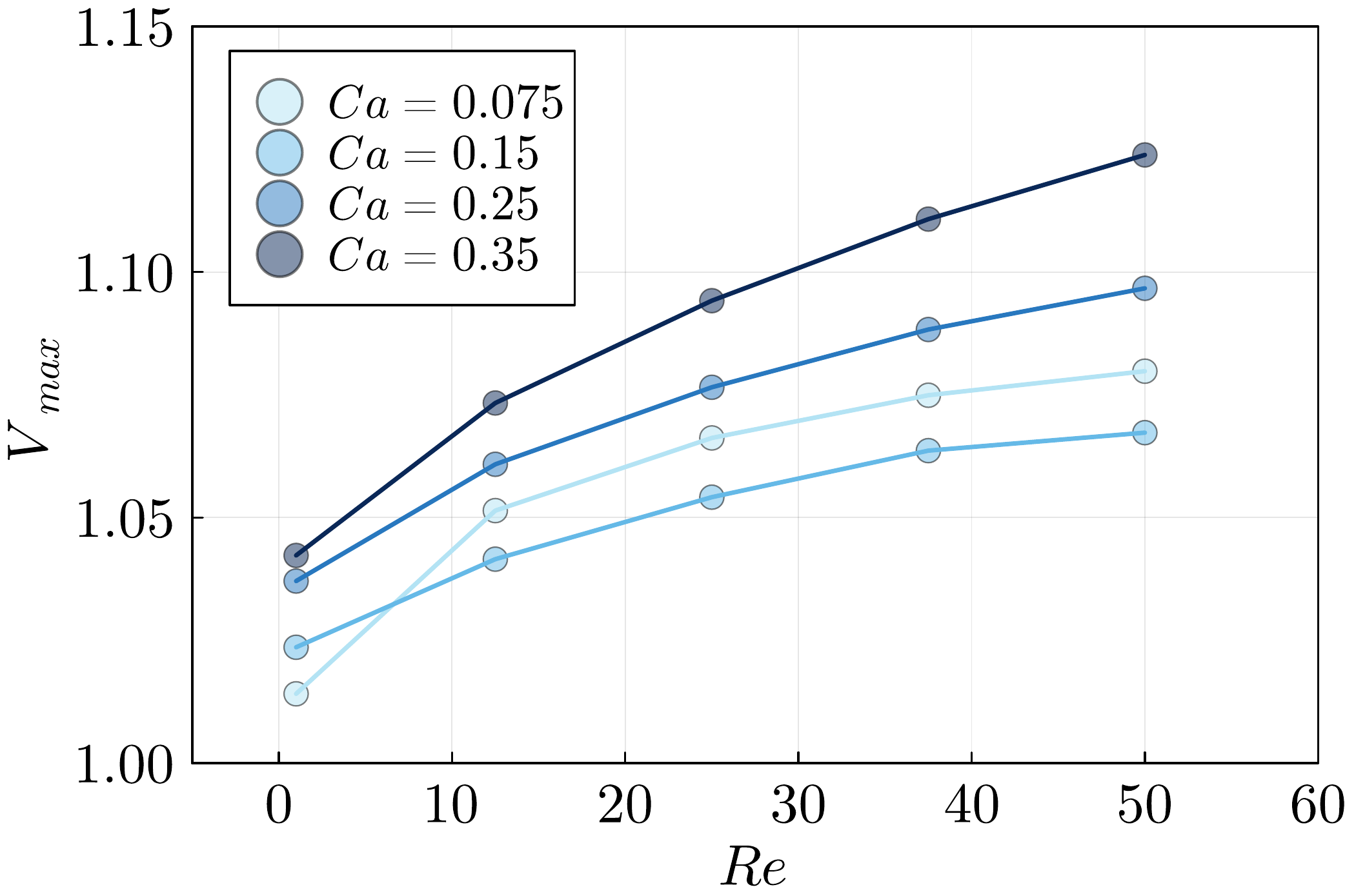}
        \caption{$V_{max}$}
        \label{Cca_Vmax}
\end{subfigure}
\begin{subfigure}{.49\textwidth}
        \centering
        \includegraphics[width=.8\columnwidth, trim=0 0 0 0, clip]{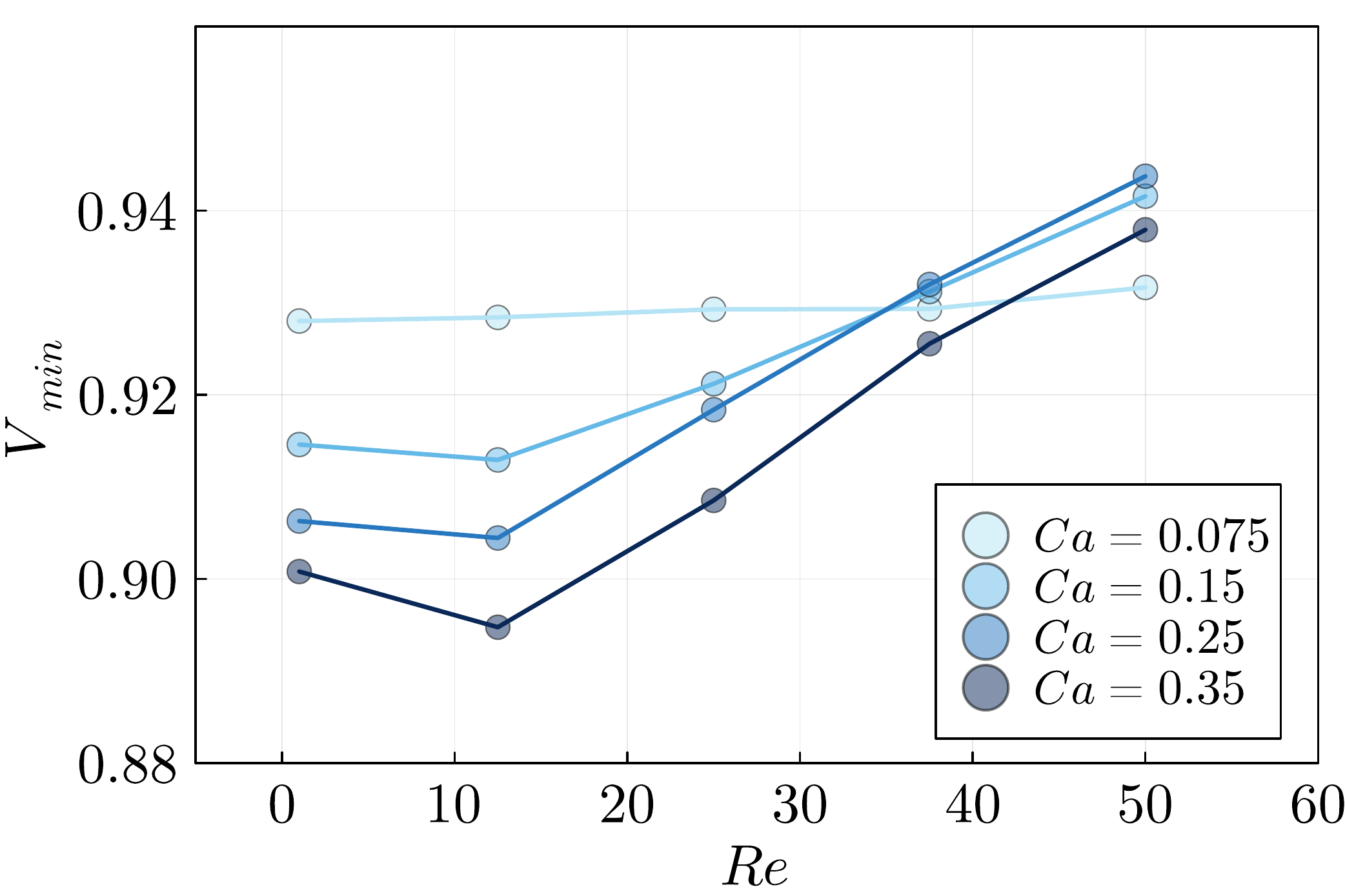}
        \caption{$V_{min}$}
        \label{Cca_Vmin}
\end{subfigure}
  \caption{Maximum (minimum) velocity $V_{max}$ ($V_{min}$) as a function of $Re$ and $Ca$ for a single capsule passing through the corner.}
\label{Cca_VmaxVmin}
\end{figure}

Additionally, we present in \reffig{Cca_VmaxVmin} the maximum and minimum velocity of the single capsule flowing through the corner. In the non-inertial regime, the maximum velocity of the capsule increases with $Ca$, as shown in \reffig{Cca_Vmax}. In inertial conditions we observe that $V_{max}$ increases for $Re$ ranging from 1 to 50. The increase in $V_{max}$ between $Re = 1$ and $Re = 50$ is significant in \reffig{Cca_Vmax}, especially for large $Ca$. For instance, at $Ca = 0.35$, $V_{max}$ increases by about 8\% between the non-inertial and the highly inertial regimes.
We then consider the evolution of the minimum velocity $V_{min}$ for a single capsule at various $Ca$ and $Re$ in \reffig{Cca_Vmin}. In general, we observe that the minimum velocity decreases with $Ca$ in both the non-inertial and the inertial regimes for $Re\leq 25$. 
In \reffig{Cca_Vmin}, we also observe a non-monotonous behavior of $V_{min}$ at low inertia and at sufficiently high $Ca$: for $Ca \geq 0.15$, $V_{min}$ first decreases with increasing $Re$, reaching a minimum for $Re = 12.5$, before increasing sharply at $Re > 12.5$.
Overall, we observe from \reffig{Cca_VmaxVmin} that the presence of inertia tends to increase both velocity extrema of the capsule, especially at large $Ca$.

\begin{figure}
    \centering
    \begin{minipage}{.55\columnwidth}
        \includegraphics[width=.8\columnwidth]{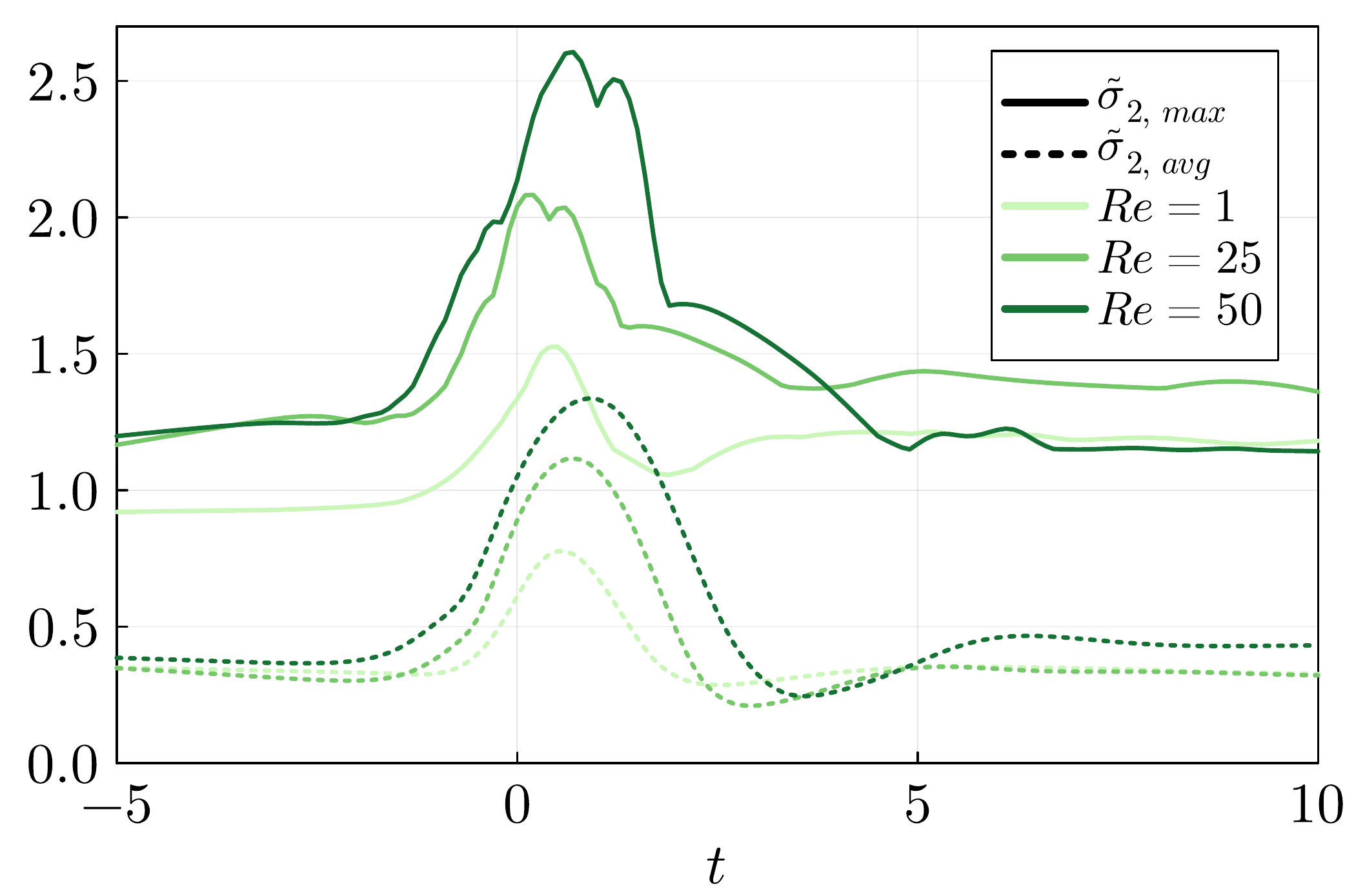}
    \end{minipage}
    \hspace{4em}
    \begin{minipage}{.23\columnwidth}
        \vspace{-2em}
        \centering
        \includegraphics[width=.7\columnwidth]{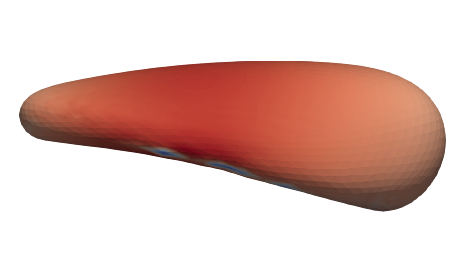}\\
        \vspace{-.75em}\small{$Re = 50$}\\
        \includegraphics[width=.7\columnwidth]{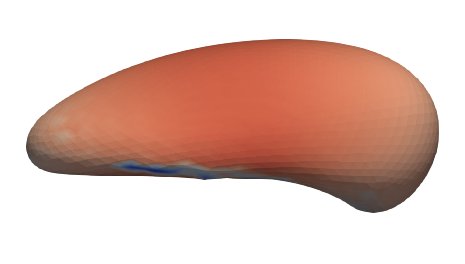}\\
        \vspace{-.75em}\small{$Re = 25$}\\
        \includegraphics[width=.7\columnwidth]{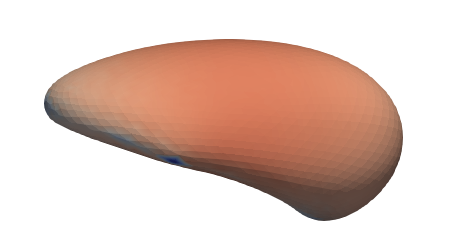}\\
        \vspace{-.5em}\small{$Re = 1$}\\

    \end{minipage}
    \caption{Left: Maximum and average tensions in the capsule at $Ca = 0.35$ and $Re = 1$, 25 and 50. Right: Capsule shape colored by $\tilde{\sigma}_2$ when $\tilde{\sigma}_{2, \, max}$ reaches its maximum.}
    \label{fig:tensions}
\end{figure}

A quantity of practical interest to experimentalists is the maximum stress experienced by the capsule, as it can be used to predict \textit{a priori} if a given geometry can induce plastic deformation or even breakup of the capsule membrane \cite{haner2021sorting}. More specifically, it is the largest eigenvalue $\tilde{\sigma}_2$ of the stress tensor $\bm{\tilde{\sigma}}$ that can bring insight into the mechanical integrity of the membrane.
In \reffig{fig:tensions}, we show the maximum and average values of $\tilde{\sigma}_2$ over the membrane surface as the capsule approaches and flows through the corner at $Ca = 0.35$ and $Re = 1$, 25 and 50. We observe that $\tilde{\sigma}_{2, \, avg}$ follows a trend very similar to that of the capsule surface area observed in \reffig{fig:allareaCa4}: $\tilde{\sigma}_{2, \, avg}$ varies smoothly with time, presents a maximum near $t = 1$ and a local minimum near $t = 2.5$, and the value of the maximum deviation from steady state nearly doubles between the low and moderate inertial cases $Re = 1$ and $Re = 50$.
We also note that the steady state value of $\tilde{\sigma}_{2, \, avg}$ prior to entering the corner is independent of $Re$, as was observed in the case of the capsule surface area in \reffig{fig:allareaCa4}. In particular, we find by comparing \reffigs{fig:allareaCa4} and \ref{fig:tensions} that at $Ca = 0.35$, a non-dimensional surface area $\mathcal{A}$ of about 1.14 leads to an average non-dimensional membrane stress of about 0.4.
The steady state of the maximum stress $\tilde{\sigma}_{2, \, max}$, however, increases by about 40\% between the low inertial case ($Re = 1$) and the moderate inertial cases ($Re = 25$, 50). Inside the corner, $\tilde{\sigma}_{2, \, max}$ increases by nearly 75\% between $Re = 1$ and $Re = 50$, confirming that a capsule in a moderate inertial regime has a higher risk of breakup than in a low inertial regime.

It is worth noting that for all $Re$, the value of the maximum stress $\tilde{\sigma}_{2, \, max}$ is about double that of the average stress $\tilde{\sigma}_{2, \, avg}$: since we showed previously that $\tilde{\sigma}_{2, \, avg}$ is closely related to the capsule surface area $-$ a quantity that is relatively easy to measure experimentally $-$, this observation can be used by experimentalists as a rule of thumb to estimate the maximum stress in the capsule  membrane and assess the mechanical integrity of the membrane.

\subsection{Evolution of the capsule shape}
\begin{figure}
    \centering
 \begin{subfigure}{.49\textwidth}
        \centering
        \includegraphics[width=.8\columnwidth, trim=0 0 0 0, clip]{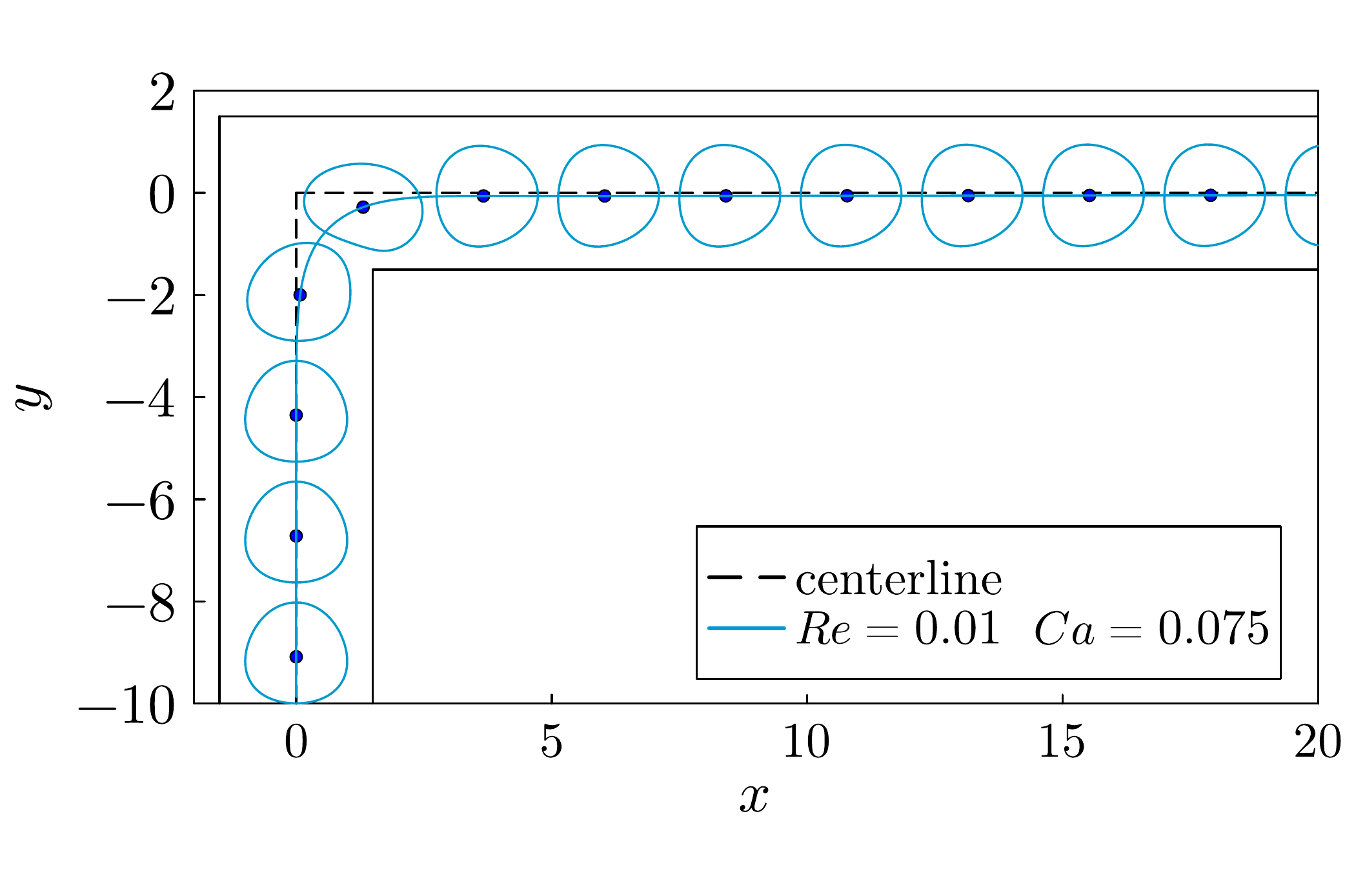}
        \caption{}
        \label{Cca_Seq_Ca0.075Re0.01}
    \end{subfigure}
    \begin{subfigure}{.49\textwidth}
        \centering
        \includegraphics[width=.8\columnwidth, trim=0 0 0 0, clip]{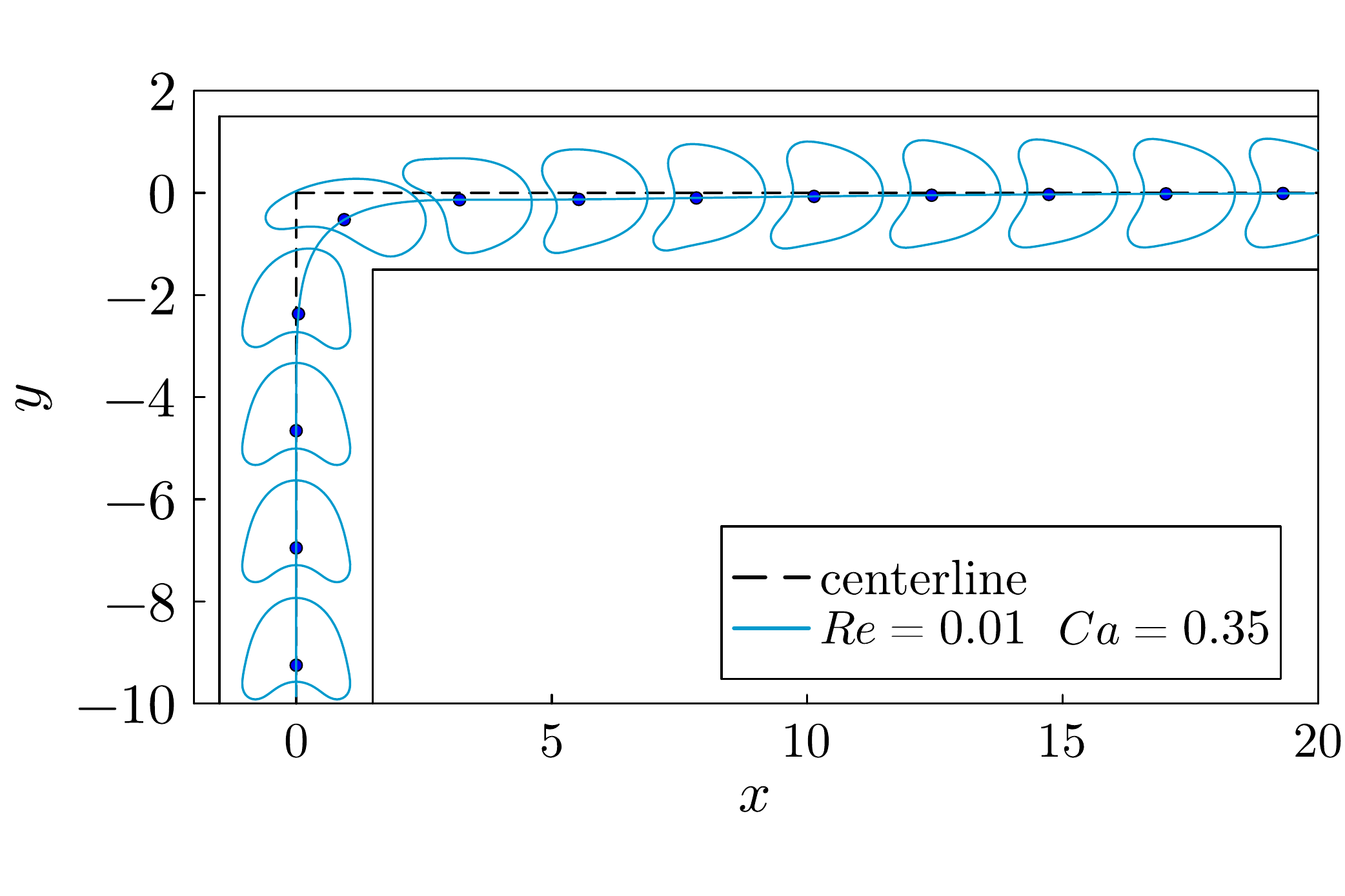}
        \caption{}
       \label{Cca_Seq_Ca0.35Re0.01}
    \end{subfigure}\\
    \begin{subfigure}{.49\textwidth}
        \centering
        \includegraphics[width=.8\columnwidth, trim=0 0 0 0, clip]{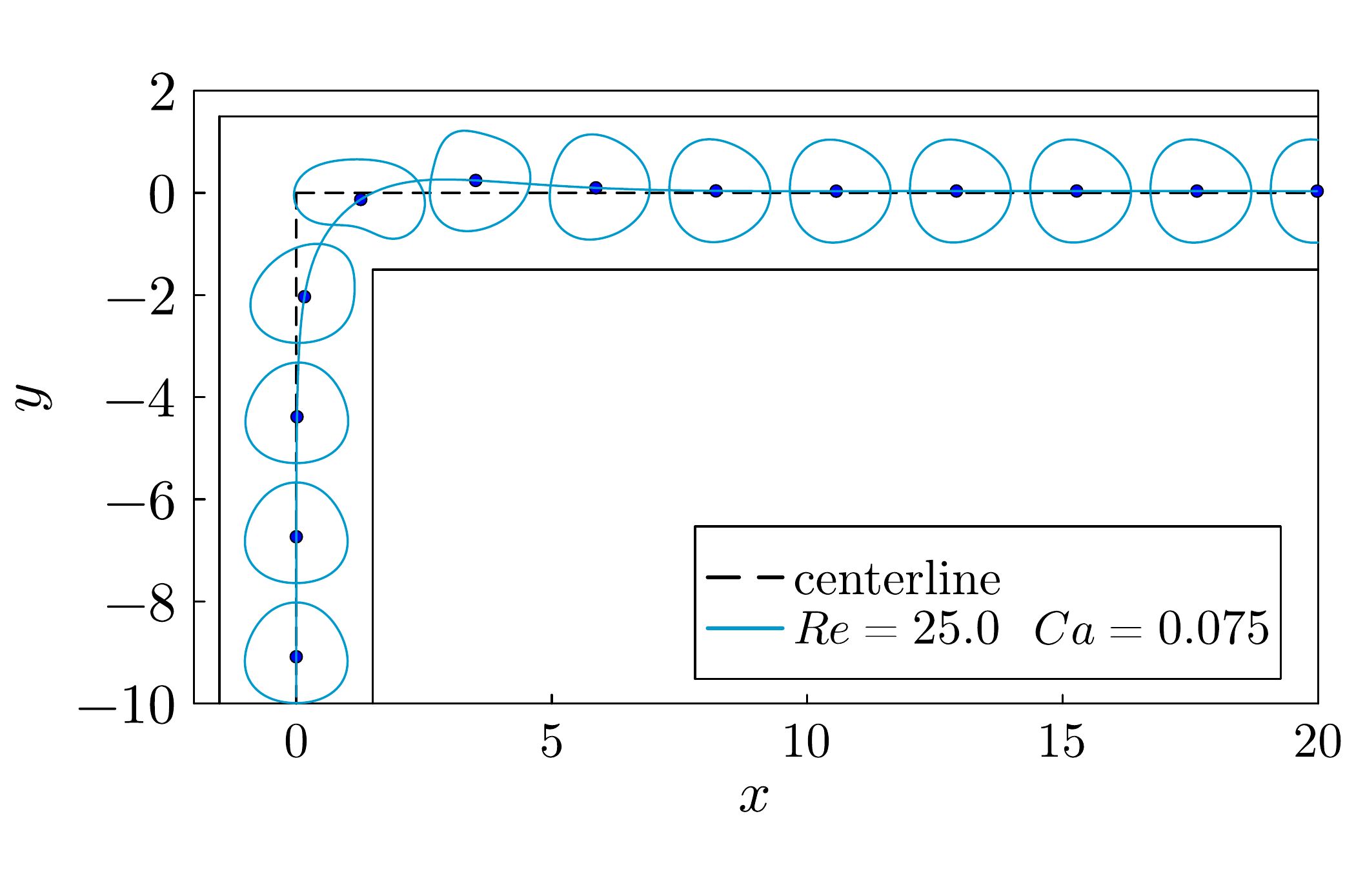}
        \caption{}
        \label{Cca_Seq_Ca0.075Re25.0}
    \end{subfigure}
    \begin{subfigure}{.49\textwidth}
        \centering
        \includegraphics[width=.8\columnwidth, trim=0 0 0 0, clip]{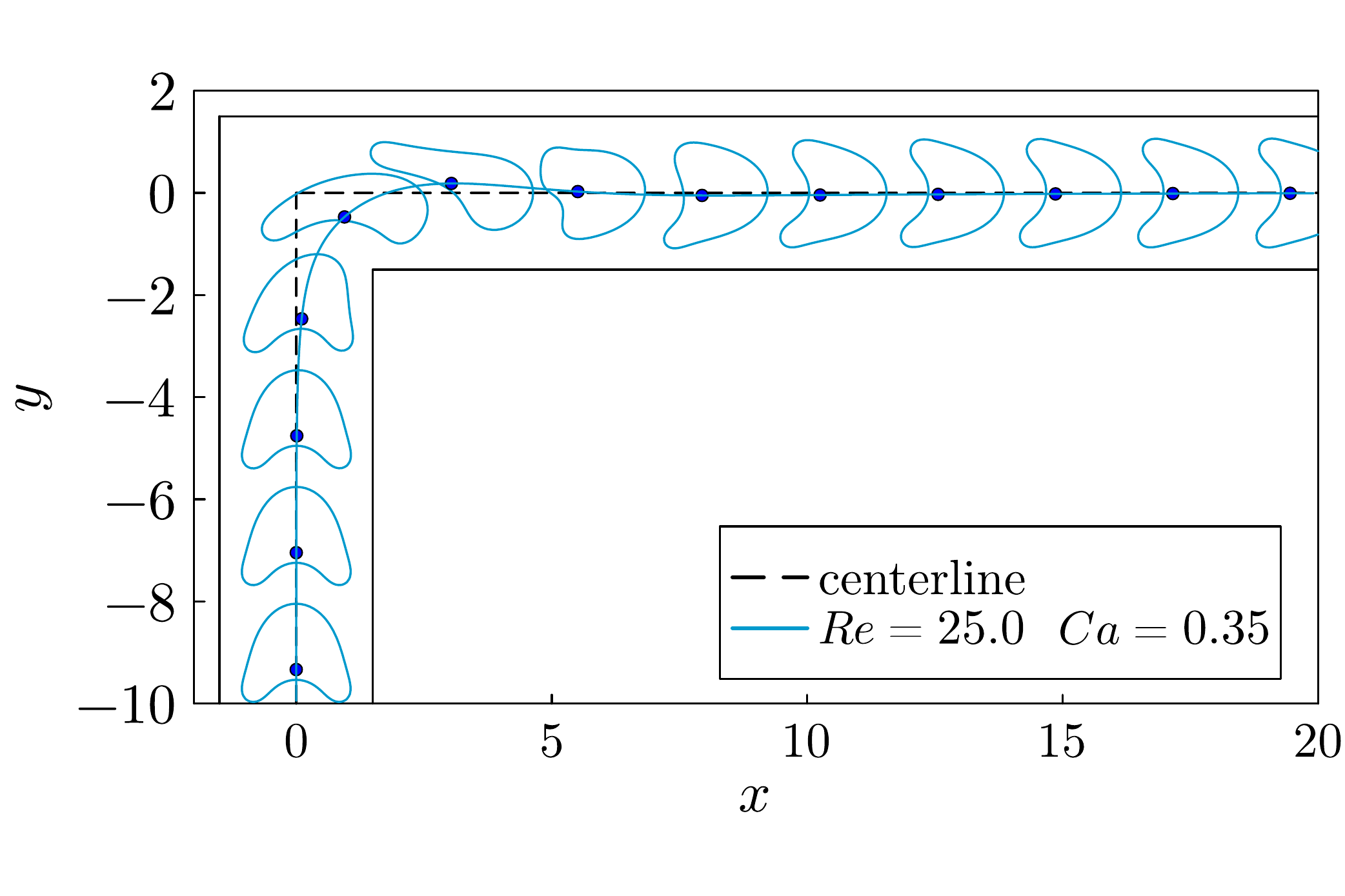}
        \caption{}
        \label{Cca_Seq_Ca0.35Re25.0}
    \end{subfigure}  \\
    \begin{subfigure}{.49\textwidth}
        \centering
        \includegraphics[width=.8\columnwidth, trim=0 0 0 0, clip]{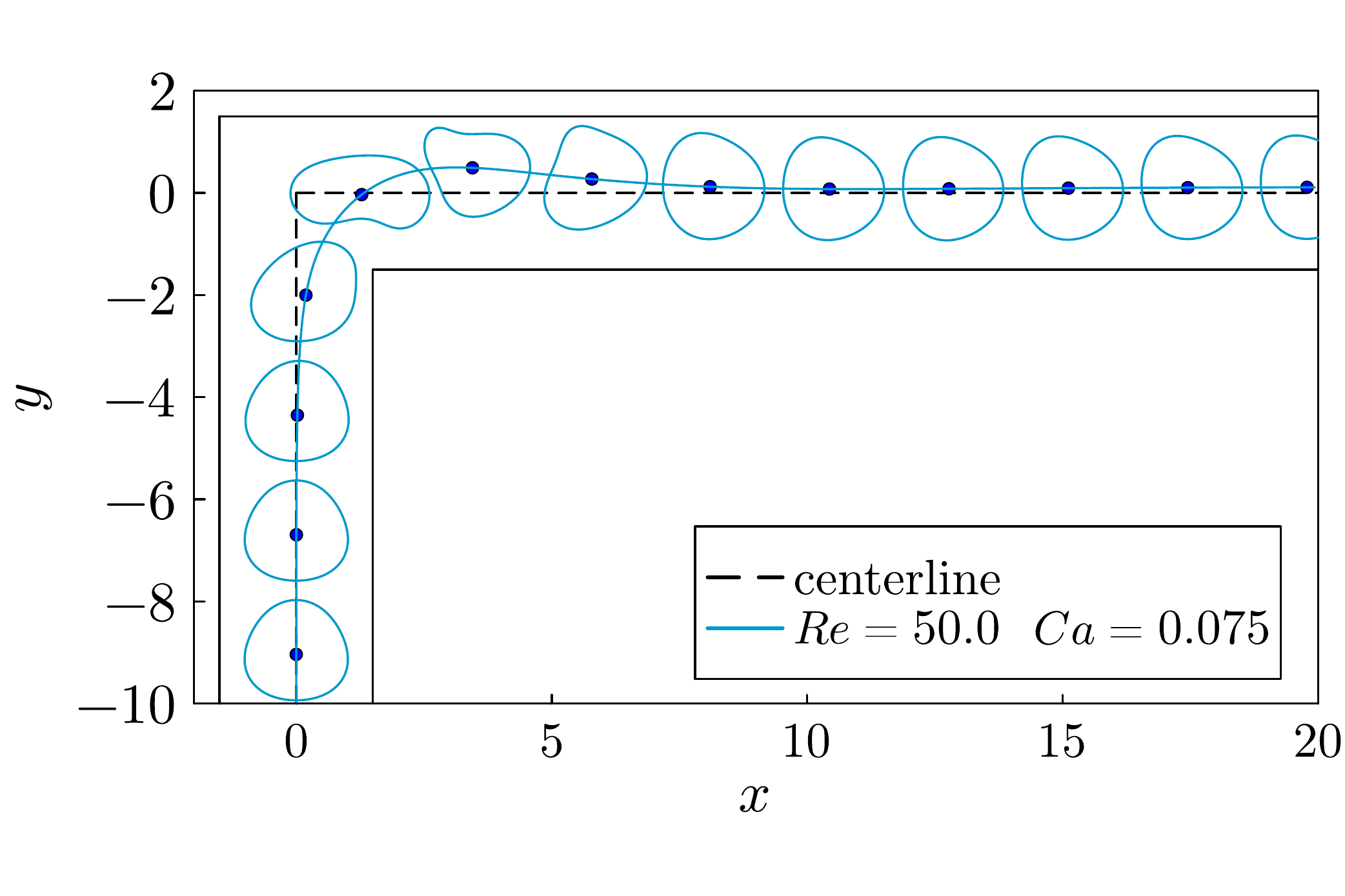}
        \caption{}
        \label{Cca_Seq_Ca0.075Re50.0}
    \end{subfigure}
    \begin{subfigure}{.49\textwidth}
        \centering
        \includegraphics[width=.8\columnwidth, trim=0 0 0 0, clip]{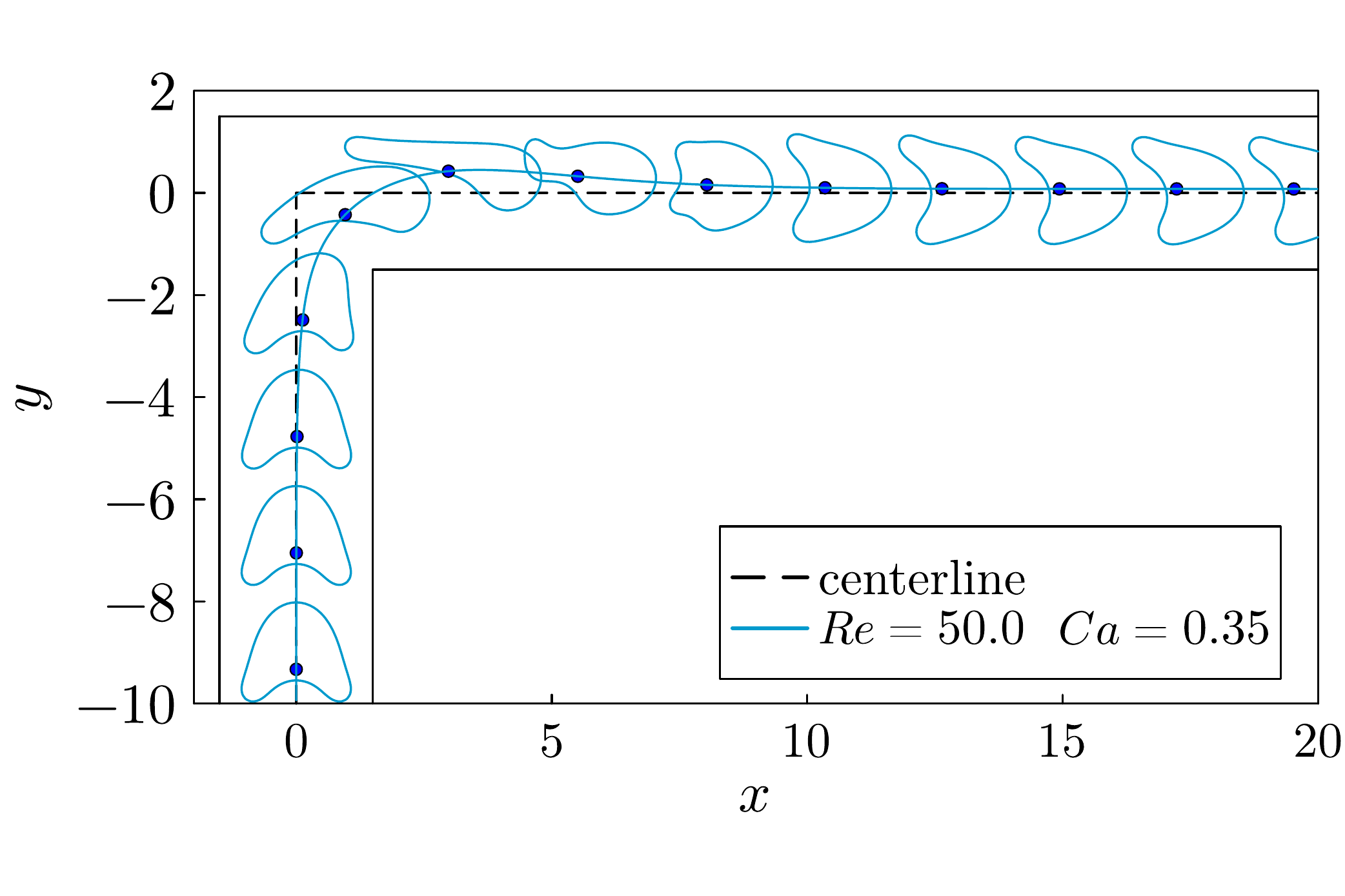}
        \caption{}
        \label{Cca_Seq_Ca0.35Re50.0}
    \end{subfigure}
    \caption{Sequence of Capsule outlines for different $Ca$ and $Re$. The time between each frame is $t = 1.5$.}
    \label{Cca_Seq}
\end{figure}

We now illustrate the temporal evolution of the capsule travelling through the corner. \Reffig{Cca_Seq} shows the outline of the capsule in the symmetry plane $z=0$ for successive discrete times. The capsule outlines are given for $Ca = 0.075$ and $Ca = 0.35$ and $Re = 0.01$, 25 and 50. Prior to entering the corner, the capsule adopts a steady shape that is determined by the confinement of the walls. In the case of $Ca = 0.35$, we observe the well-known ``parachute" shape. Upstream of the corner, the trajectory of the capsule coincides with the centerline of the primary (vertical) channel. As the capsule flows through the corner, the capsule deviates from the channel centerline: in the non-inertial regime, Zhu \& Brandt \cite{zhu2015motion} showed that the capsule trajectory closely matches the flow streamlines. We obtain the same conclusion in the inertial regime. When inertia is considered, the capsule trajectory crosses the horizontal centerline of the secondary channel and comes increasingly close to the upper wall as $Re$ increases, before relaxing to the channel centerline.

Figures \ref{Cca_Seq_Ca0.075Re0.01} and \ref{Cca_Seq_Ca0.35Re0.01} show clear differences in the effects of $Ca$ in the Stokes regime. Increasing $Ca$ from $0.075$ to $0.35$ causes the equilibrium shape of the capsule to change from an slightly deformed spheroid to a concave ``parachute" shape.
For a small $Ca=0.075$, the equilibrium shapes of the capsule remain similar as $Re$ increases from $Re=0.01$ to $Re=50$ (see \reffigs{Cca_Seq_Ca0.075Re0.01}, \ref{Cca_Seq_Ca0.075Re25.0}, and \ref{Cca_Seq_Ca0.075Re50.0}).
However, the deformation of the capsule becomes more evident inside the corner at higher $Re$, particularly in \reffig{Cca_Seq_Ca0.075Re50.0}. After passing the corner, the capsule shape returns to its steady spheroid shape observed in the Stokes regime for all values of $Re$.
In the case of a high $Ca=0.35$, we observe that the equilibrium shape of the capsule is more and more concave as $Re$ increases. Inside the corner, the capsule is highly elongated and presents an increasingly long tail for increasing $Re$ $-$ e.g. \reffig{Cca_Seq_Ca0.35Re50.0} in the case of $Re = 50$. In the highly inertial regime, strong lubrication interactions occur between the capsule and the top wall, resulting in a flat top surface.

\begin{figure}
\centering
\begin{subfigure}{.49\textwidth}
        \centering
        \includegraphics[width=.9\columnwidth, trim=0 0 0 0, clip]{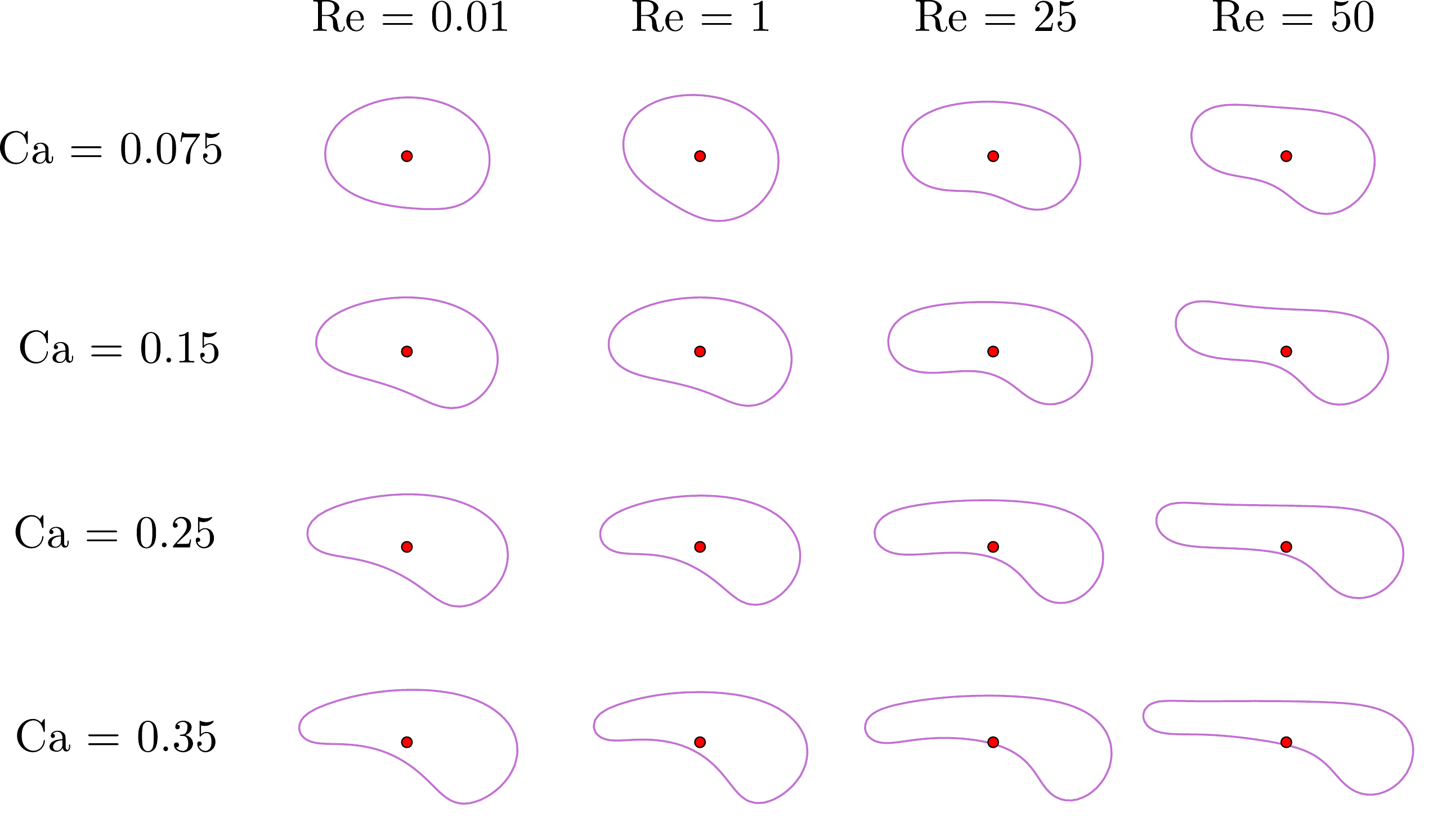}
        \caption{}
        \label{Cca_area_labels}
\end{subfigure}
\hfill
\begin{subfigure}{.49\textwidth}
        \centering
        \includegraphics[width=.9\columnwidth, trim=0 0 0 0, clip]{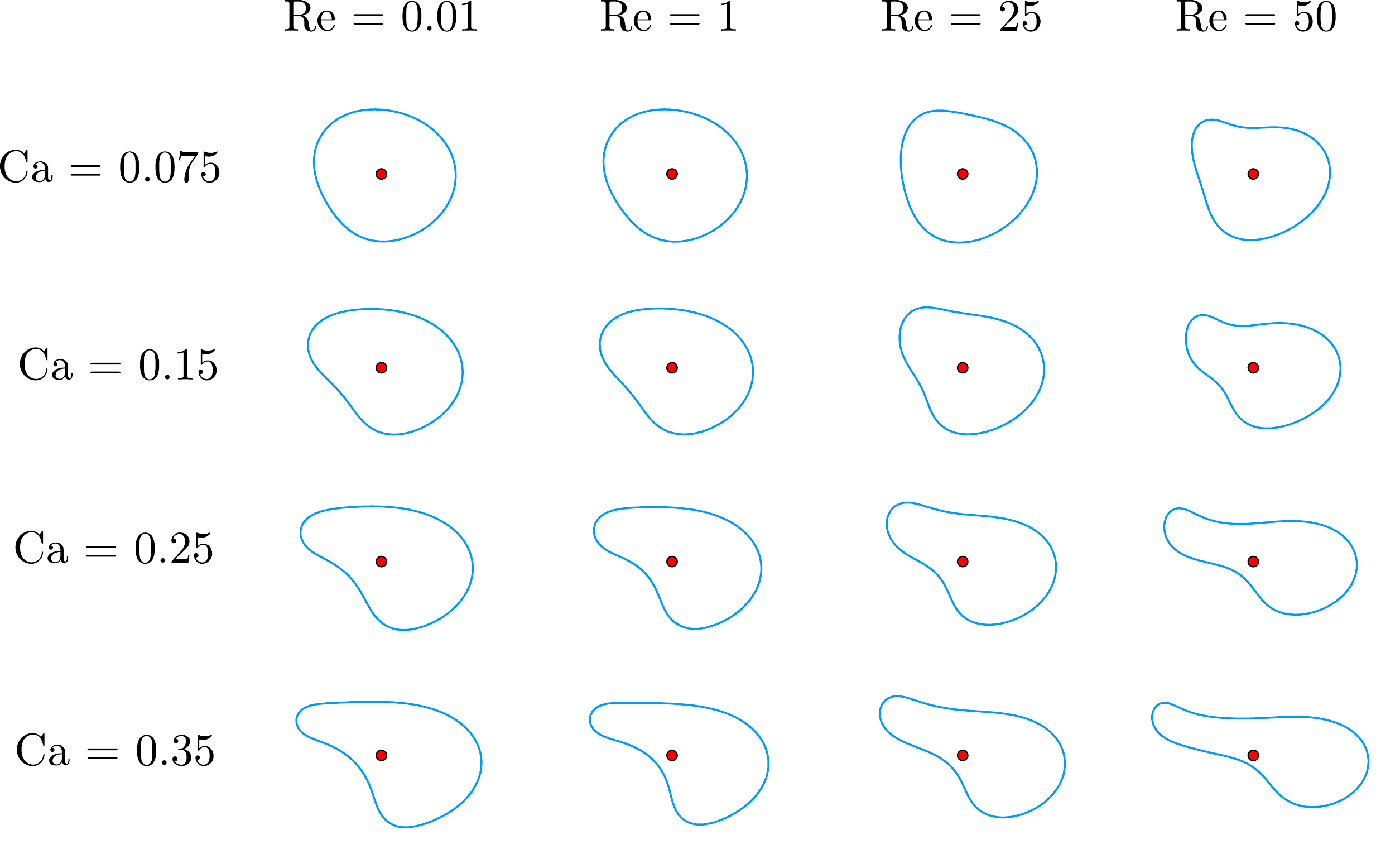}
        \caption{}
        \label{Cca_vel_labels}
\end{subfigure}
\caption{Outlines of a single capsule passing a corner with (a) maximal surface area $\mathcal{A}$ and (b) maximal velocity $v_{max}$ at various $Re$ and $Ca$.}
\end{figure}


In figures \ref{Cca_area_labels} and \ref{Cca_vel_labels}, we present the single capsule outline with the maximum surface area $\mathcal{A}_{max}$ and the maximum velocity $V_{max}$ inside the corner for all the cases investigated in this section. Inside the corner, the maximum surface area of the single capsule is reached when it approaches the upper wall and it is quickly followed by the maximum velocity.
From figures \ref{Cca_area_labels} and \ref{Cca_vel_labels}, we observe in particular that a high $Re$ leads to an elongation of the capsule in the streamwise direction, while a high $Ca$ increases the concavity of the capsule.
Moreover, we note that the centroid of the capsule moves closer to the rim of the outline at high values of $Ca$: note that the centroid drawn in figures \reffig{Cca_Seq}-\ref{Cca_vel_labels} corresponds to the centroid of the three-dimensional capsule, not to that of the two-dimensional outline.
The results shown in figures \ref{Cca_Seq}-\ref{Cca_vel_labels} indicate that $Ca$ has a significant effect on capsule deformation, while $Re$ has a more pronounced effect on the trajectory of the capsule as well as its deformation resulting from the lubrication layer against the top wall of the corner. In particular, at high $Re$, the capsule undergoes significant stretching, which may cause damage or even rupture in microfluidic devices. Understanding the effects of $Re$ on capsule deformation and the resulting damage is crucial in designing efficient and reliable microfluidic devices.

\subsection{Discussion on the Stokes regime}
We observe in figures \ref{fig:allareaCa} a surprising, non-monotonous behavior of the capsule surface area around $Re = 1$: at large $Ca$, the surface area of the capsule is smaller at $Re = 1$ than at $Re = 0.01$ and $Re = 25$. Additionally, in \reffig{fig:allareaCa1} the steady surface area of the capsule at $Re = 0.01$ and $Ca = 0.075$ downstream of the corner is about 1\% lower than the initial spherical surface area of the capsule, indicating a small loss of the internal capsule volume.
The cause of these observations may be related to the limitations of the FTM coupled with a sub-optimal choice of numerical parameters in the case of $Re = 0.01$ only.
Indeed, the immersed boundary method is known to conserve volume asymptotically rather than to machine precision. In earlier IBM studies involving capsules, the volume loss is always small, typically below 1\% \cite{balogh2017computational, lu2021path, wang2016motion, wang2018path}. Moreover, Stokes conditions are known to be challenging for PDE-based incompressible Navier-Stokes solvers, as the matrix inverted in the velocity viscous Poisson problem is less well conditioned at low $Re$. 
While it is worth noting that the capsule surface area in the Stokes regime should be interpreted with caution, these limitations only affect the capsule surface area and not the centroid velocity. Moreover, our solver was extensively validated in Stokes conditions in \cite{huet2022cartesian} and showed excellent agreement with the BIM as well as other FTM solvers. As such, while further investigation should be conducted in the Stokes regime, it cannot be excluded that at high $Ca$ the capsule surface area at $Re = 0.01$ is physically slightly greater than that at $Re = 1$. Finally, the main focus of the present work is to investigate the inertial motion and deformation of capsules through a sharp corner, i.e. in conditions where our FTM solver does not suffer from the limitations outlined above.


\section{System of two capsules\label{sec:corner_double}}
In this section, we consider two identical capsules flowing through the corner as we vary the normalized interspacing distance $d = \tilde{d}/2\tilde{a} -1$ between the capsules as well as the Reynolds and Capillary numbers. Lu et al. \cite{lu2021path} previously considered the binary interaction of capsules flowing through a T-junction: they showed that when $d_0 \geq 1.3$ the trailing capsule has minimal impact on the motion of the leading capsule. In contrast, in their T-junction geometry Lu et al. observed that the motion of the trailing capsule is significantly affected by the presence of the leading capsule. To gain insight into the physical features relevant to capsule interactions through a corner in the inertial and non-inertial regimes, we select small values for the normalized interspacing distance $d_0=1, 1/2$, and $1/4$ and we examine phenomena such as migration, dynamics and deformation of the leading and the trailing capsules.

\subsection{Qualitative analysis: trajectory and capsule shape}
\begin{figure}
    \centering
    \begin{subfigure}{.49\textwidth}
        \centering
        \includegraphics[width=.9\columnwidth, trim=0 0 0 0, clip]{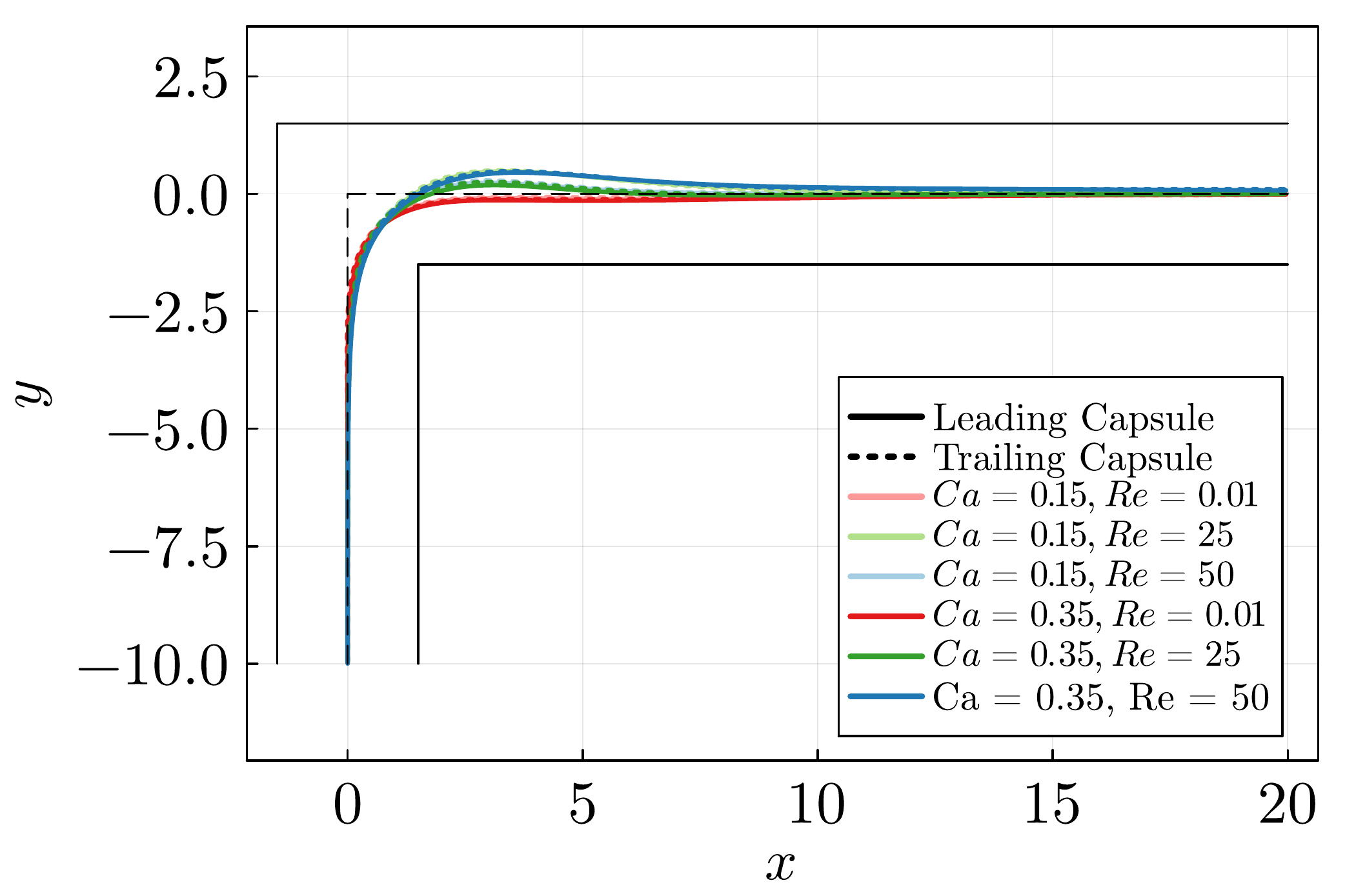}
        \caption{}
        \label{Dca_Traj_ReCa_d1}
    \end{subfigure}
    \begin{subfigure}{.49\textwidth}
        \centering
        \includegraphics[width=.9\columnwidth, trim=0 0 0 0, clip]{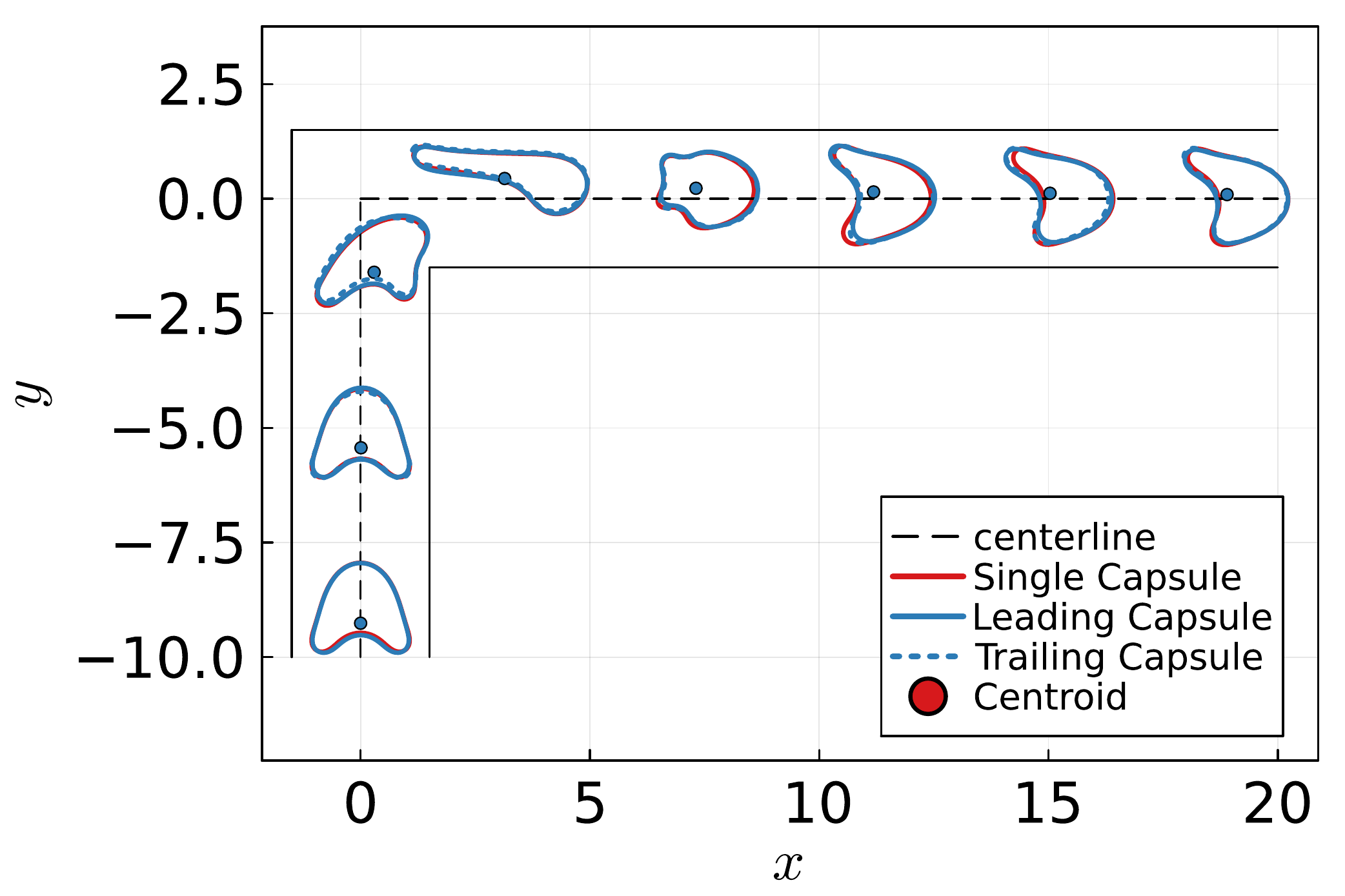}
        \caption{}
        \label{Dca_Traj_SLF_Ca0.36Re50_d0.25}
    \end{subfigure}\\
    \caption{(a) Trajectory of the two capsules at different $Ca$ and $Re$. (b) Outlines of the leading and trailing capsules at $Ca=0.35,Re=50,d_0=0.25$, with comparison to a single capsule.}
    \label{binary_setup}
\end{figure}

We first analyze the trajectory and the qualitative shapes of the pair of capsules as they flow through the corner. \Reffig{Dca_Traj_ReCa_d1} shows the trajectory of the capsules at $Re = 0.01,$ 25 and 50 and $Ca = 0.15$ and 0.35. We note that all curves corresponding to the same $Ca$ overlap: $Ca$ has no impact on the path of either the leading or the trailing capsule. Likewise, we observe no significant difference in the trajectories of the leading and the trailing capsules, unlike the strikingly different paths reported in the case of a T-junction \cite{lu2021path}. In fact, the key parameter that controls the capsule trajectory is the Reynolds number. As $Re$ increases, the inertia drives the capsule closer to the upper channel wall, as observed in \refsec{sec:corner_single} in the case of a single capsule.
We then illustrate the capsule shape on the symmetry plane $z=0$ in \reffig{Dca_Traj_SLF_Ca0.36Re50_d0.25} for the most deformed capsule configuration corresponding to $Ca=0.35$ and $Re=50$ with an initial interspacing distance $d_0=0.25$. We compare the outlines of the leading and the trailing capsules to that of a single capsule in the same conditions. Qualitatively, the deformation of interacting capsules is not significantly different than that observed in the case of a single capsule. Perhaps more surprisingly, the qualitative outlines of the leading and the trailing capsules are also very similar, almost overlapped, even in the strongly interacting configuration corresponding to $d_0 = 0.25$. Note that this qualitative shape analysis relies on the outline of the capsule in the plane of symmetry $z=0$, while the actual three-dimensional shape of the leading and trailing capsules may differ more strongly.

\subsection{Quantitative analysis: velocity and membrane surface area}
\begin{figure}
    \centering
    \begin{subfigure}{.49\textwidth}
        \centering
        \includegraphics[width=.8\columnwidth, trim=0 0 0 0, clip]{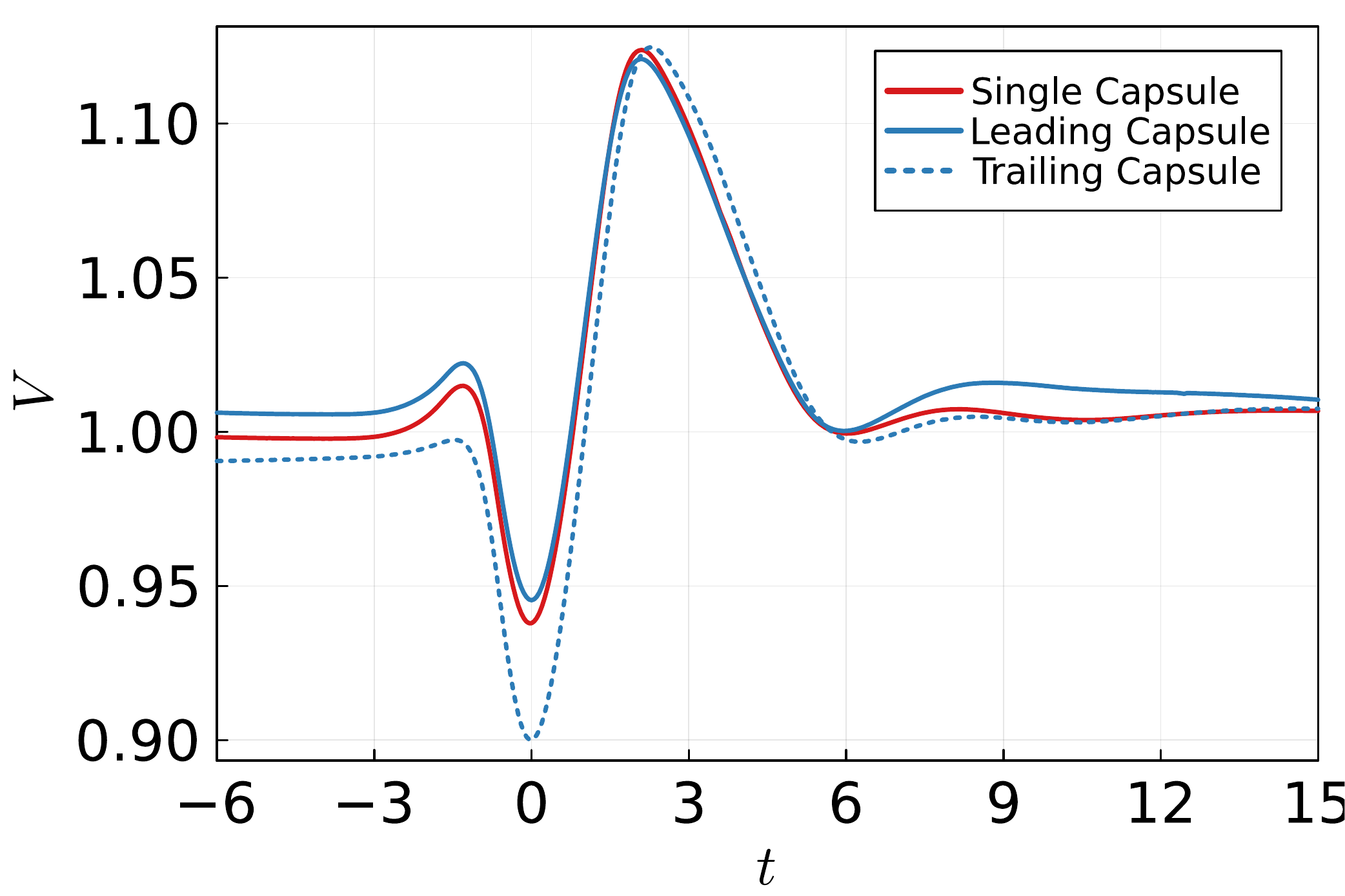}
        \caption{}
        \label{Dca_Velo_SLF}
    \end{subfigure}
    \begin{subfigure}{.49\textwidth}
        \centering
        \includegraphics[width=.8\columnwidth, trim=0 0 0 0, clip]{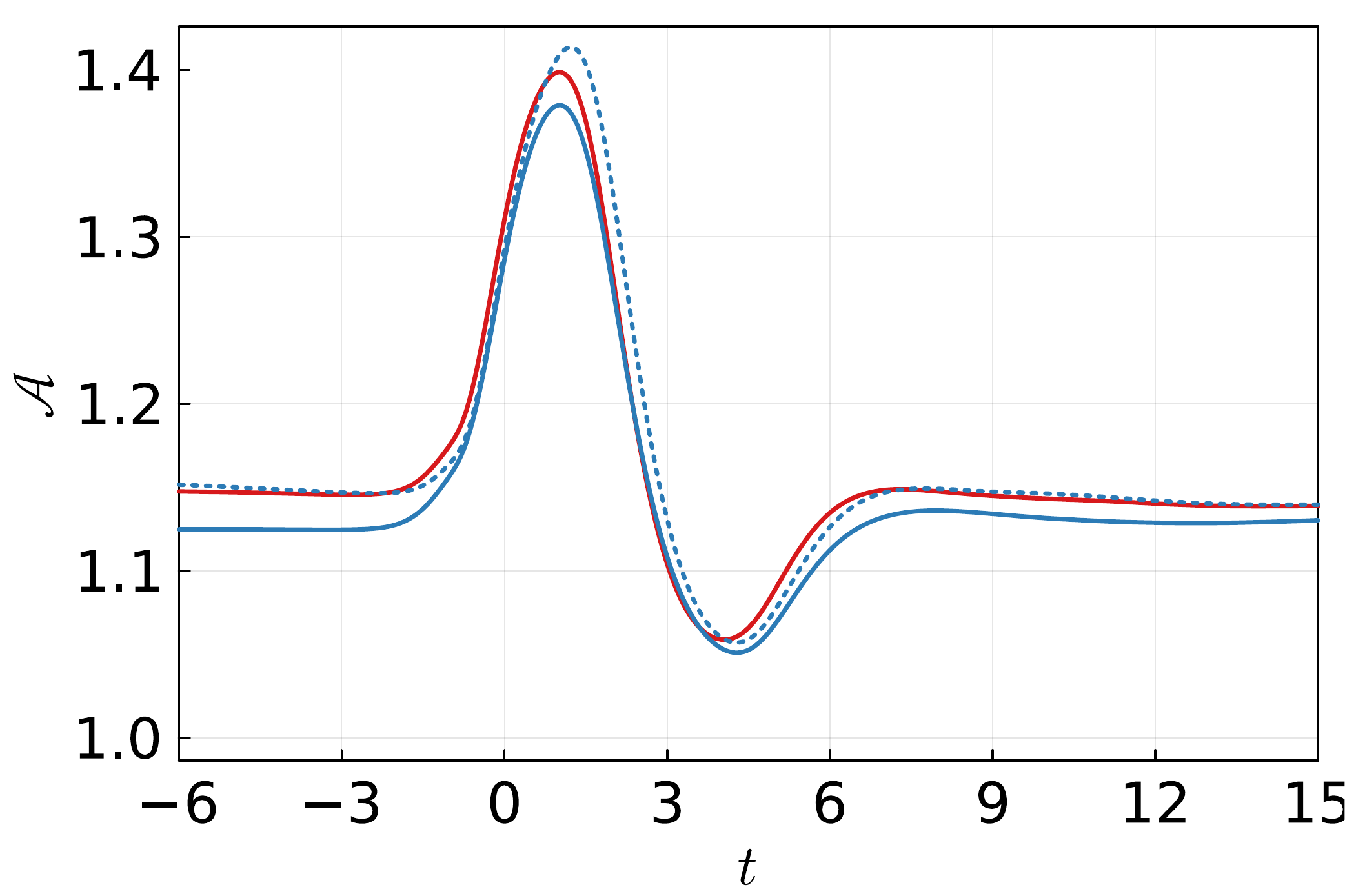}
        \caption{}
        \label{Dca_Area_SLF}
    \end{subfigure}  \\
    \caption{Temporal evolution of the velocity $V$ and the surface area $\mathcal{A}$ of capsules at $Ca=0.35, Re=50$ with $d_0=0.25$: a comparison of the leading, trailing and a single capsule.}
    \label{Dca_LSF}
\end{figure}

We now compare the temporal evolution of the velocity of the centroids of the capsules as well as the time evolution of their surface areas, as plotted in \reffig{Dca_LSF}. To simplify the identification of interaction features, we first focus on the most deformed configuration corresponding to $Ca=0.35$, $Re=50$ and $d_0=0.25$. For reference, we also plot the evolution of a single capsule under the same conditions in red. Throughout the remainder of this study, and unless otherwise stated, the velocity of interacting capsules is normalized by the equilibrium velocity $V_{eq}$ of a single capsule for the same Capillary and Reynolds numbers. This normalization choice allows for an unbiased comparison between the velocities of the leading and the trailing capsules. In this section we also denote the reduced velocity of the single capsule by $V_s$, that of the leading capsule by $V_l$ and that of the trailing capsule by $V_{t}$. Similarly, we denote by $\mathcal{A}_s$, $\mathcal{A}_l$, $\mathcal{A}_t$ the normalized surface areas of respectively the single, leading and trailing capsules.

In \reffig{Dca_Velo_SLF}, we observe that the velocity of the leading capsule is affected by the presence of the trailing capsule before it reaches the corner, as it is about 1\% higher than that of a single capsule. However, the extrema of $V_l$ as it flows through the corner closely match those of $V_s$. After the corner, $V_l$ is about 2\% larger than $V_s$ but slowly relaxes back to $V_s$ further downstream. With regards to the trailing capsule, we note that its velocity is more markedly affected by the presence of the leading capsule. Prior to reaching the corner, $V_{t}$ is about 1\% lower than $V_s$, but inside the corner its minimum value is 4\% lower than $V_s$. However, the maximum of $V_{t}$ is identical to that of both $V_l$ and $V_s$. Downstream of the corner, $V_{t}$ quickly relaxes back to $V_s$ and maintains a similar value thereafter, eventually converging to $V_{eq}$.
The time evolution of the surface areas of the pair of capsules is shown in \reffig{Dca_Area_SLF}. The normalized surface area of the leading capsule $\mathcal{A}_l$ is clearly influenced by the presence of the trailing capsule, as was observed above in the case of its velocity. The steady and maximum surface areas of the leading capsule are about 2\% lower than that of the single capsule. In contrast, the steady surface area of the trailing capsule closely matches that of the single capsule upstream and downstream of the corner, while its maximum value is about 1\% higher than that of the single capsule. We postulate that the small interspacing distance between the two capsules disturbs the wake behind the leading capsule, which tends to mitigate its deformation and therefore decreases its surface area. Conversely, as the wake of the trailing capsule is unaffected, the discrepancies between its surface area and that of the single capsule are less pronounced.

\begin{figure}
    \centering
 \begin{subfigure}{.48\textwidth}
        \centering
        \includegraphics[width=.9\columnwidth, trim=0 0 0 0, clip]{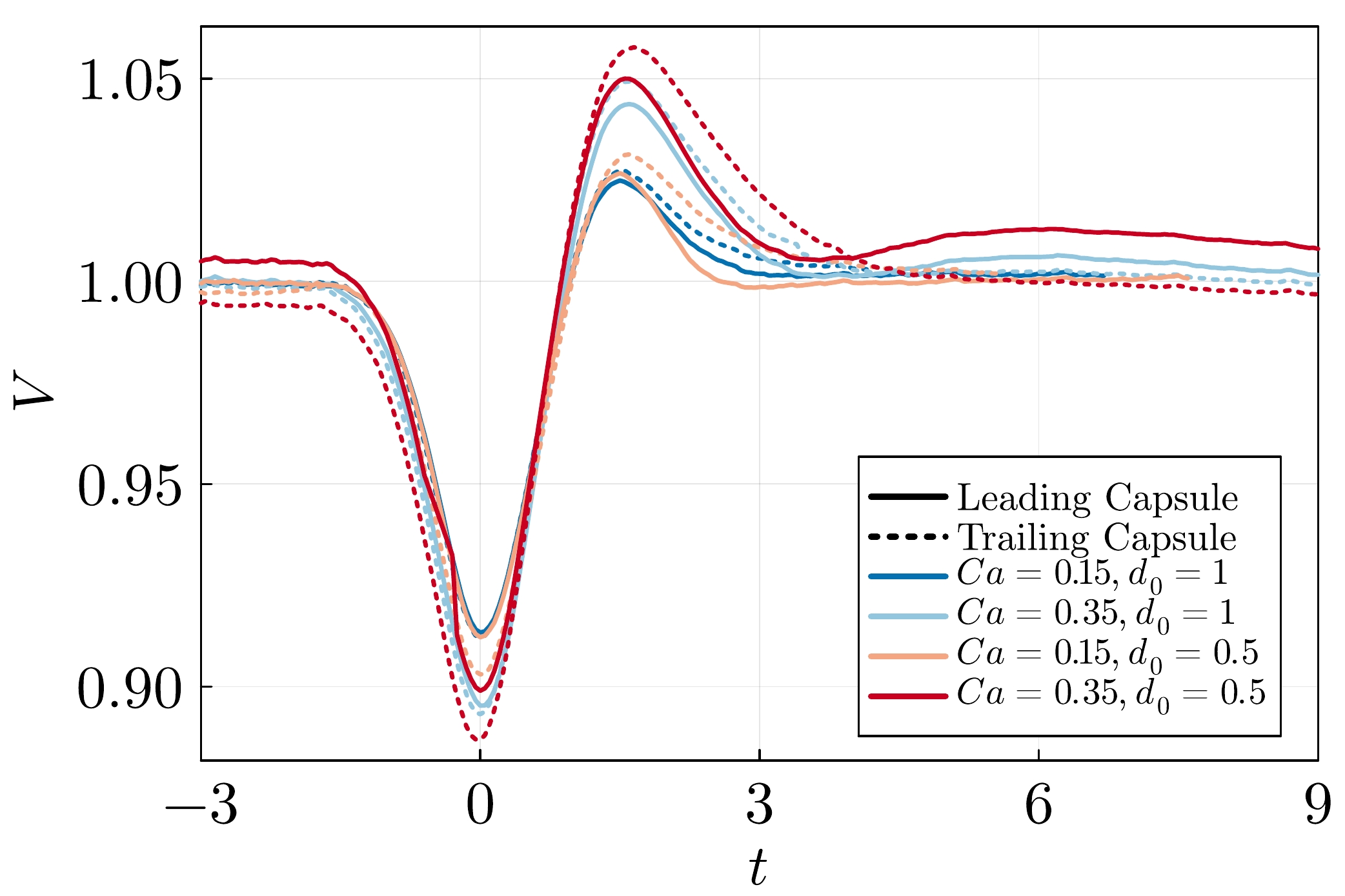}
        \caption{$Re=0.01$}
        \label{Dca_Velo_NIn}
    \end{subfigure}\\
    \begin{subfigure}{.48\textwidth}
        \centering
        \includegraphics[width=.9\columnwidth, trim=0 0 0 0, clip]{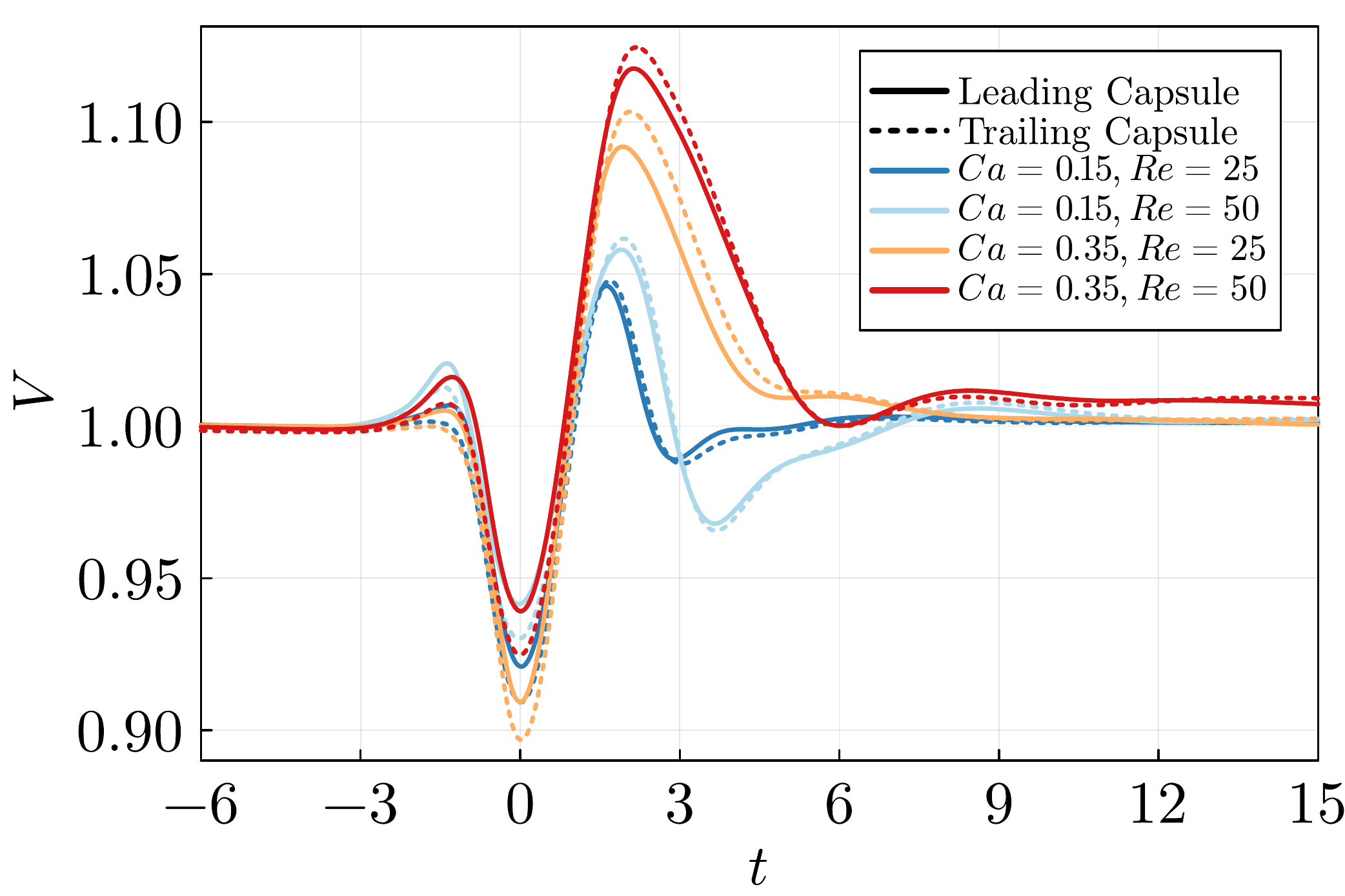}
        \caption{$d_0=1$}
        \label{Dca_Velo_In_d1}
    \end{subfigure}\hfill
    \begin{subfigure}{.48\textwidth}
        \centering
        \includegraphics[width=.9\columnwidth, trim=0 0 0 0, clip]{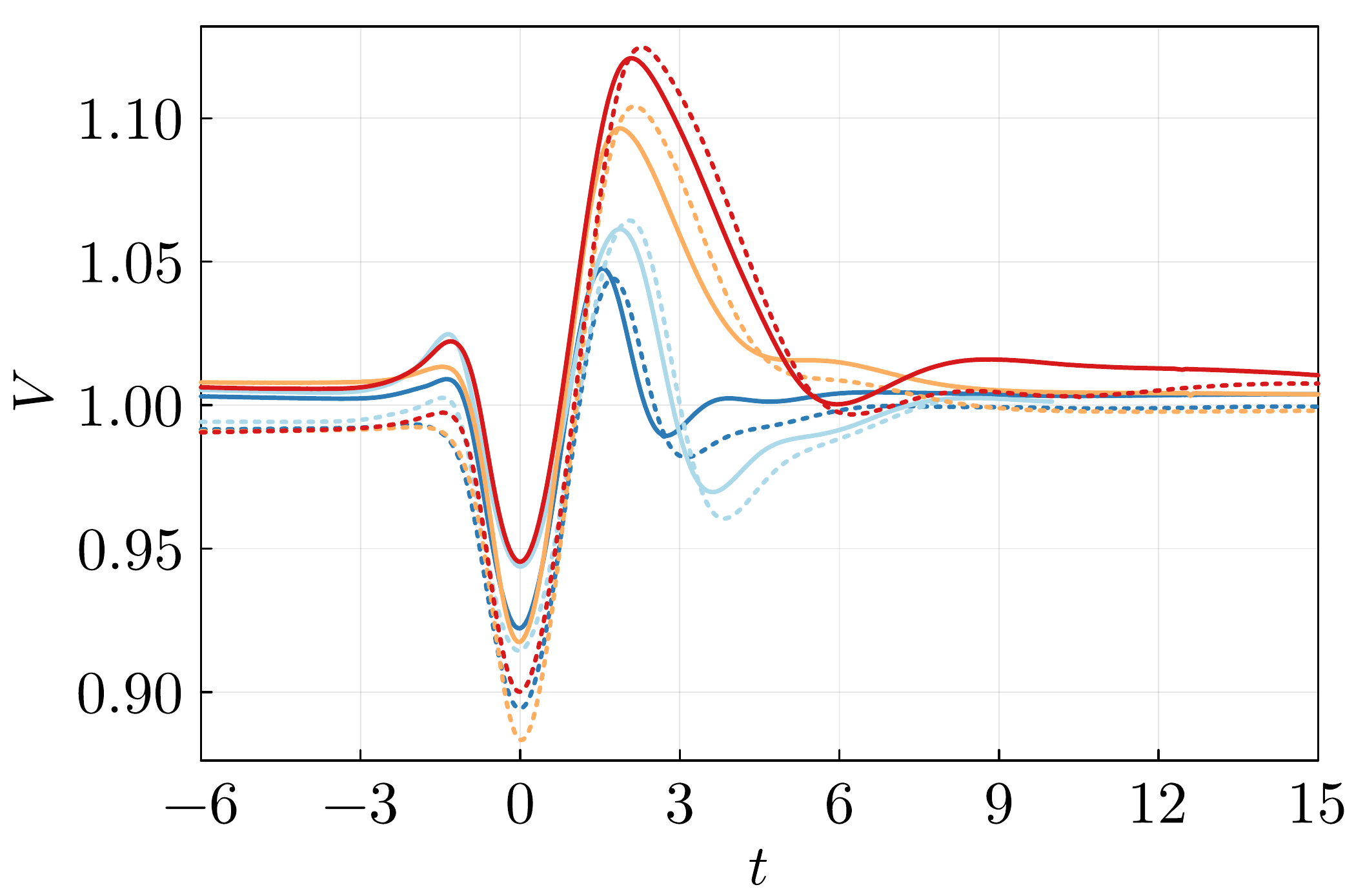}
        \caption{$d_0=0.25$}
        \label{Dca_Velo_In_d0.25}
    \end{subfigure}
    \caption{Temporal evolution of $V$ of the leading and trailing capsules at different $Ca$, $Re$ and $d_0$.}
    \label{Dca_Velo}
\end{figure}

We then present the time evolution of the velocity and surface area of the leading and the trailing capsules at various $Ca$, $Re$ and $d_0$.
We first focus on the velocity of the capsules, displayed in \reffig{Dca_Velo} for $Ca = 0.15$ and 0.35 and for $d_0 = 0.5$ and 1. The velocity of both capsules displays a minimum at $t=0$ and a maximum at $t\approx2$ at $Ca=0.15$ and $Ca=0.35$. The extrema of the velocity are more pronounced as $Ca$ increases. The effects of the initial interspacing distance $d_0$ on these extrema are less evident but still present: the velocity maxima of both the leading and the trailing capsules are increased by about 1\% as $d_0$ is halved from 1 to 0.5. Interestingly, the relaxation time of $V_{t}$ to $V_{eq}$ is significantly reduced when compared to that of $V_l$: about 3 time units in the case of $V_{t}$ with respect to more than 10 time units in the case of $V_l$. Capsule velocities in the inertial regimes at $Re = 25$ and 50 and at $Ca=0.15$ and $Ca=0.35$ are plotted in \reffig{Dca_Velo_In_d1} and \reffig{Dca_Velo_In_d0.25} for $d_0 = 1$ and 0.25, respectively. The results are similar to that of the non-inertial regime: $Ca$ enhances the velocity deviations and the extrema are more pronounced in the case of the trailing capsule. Surprisingly, we note that \reffig{Dca_Velo_In_d1} and \reffig{Dca_Velo_In_d0.25} display very similar behaviors: therefore, the interspacing distance does not seem to impact the capsule velocities inside the corner: its effects are bounded to the capsule velocities upstream and downstream from the corner. We will come back to this observation in \refsec{sec:dcaps_interspacing distance}.

When analyzing the capsule surface areas for varying $Re$, $Ca$ and $d_0$, a similar behavior is found: the surface area of the trailing capsule is consistently greater than that of the leading capsule, and increasing Capillary and Reynolds numbers and decreasing the initial interspacing distance enhance this phenomenon. In particular we report in Table \ref{Dca_Area_Tab_leading} the maximum surface areas of the leading capsule and in Table \ref{Dca_Area_Tab_trailing} that of the trailing capsule. As can be seen from Table \ref{Dca_Area_Tab_leading} and Table \ref{Dca_Area_Tab_trailing}, the maximum surface area of the leading capsule exceeds that of the trailing capsule by up to 5\%. The full time-dependant data is provided in Appendix \ref{app:Dca_area}.

\begin{table}
\centering
\begin{tabular}{c|cccc}
\hline
\toprule
$d_0$ & & $Re=0.01$ & $Re=25$ & $Re=50$   \\
\midrule
\multirow{ 2}{*}{$1$} & $Ca=0.15$ & 1.065 & 1.138 &1.193 \\
& $Ca=0.35$ & 1.263 & 1.334 & 1.399 \\
\midrule
\multirow{ 2}{*}{$0.5$} & $Ca=0.15$ & 1.065 & 1.135 &1.186 \\
& $Ca=0.35$ & 1.247 & 1.323 & 1.383 \\
\midrule
\multirow{ 2}{*}{$0.25$} & $Ca=0.15$ & 1.068 & 1.129 & 1.180 \\
& $Ca=0.35$ & 1.236 & 1.308 & 1.379 \\
\bottomrule
\hline
\end{tabular}
\caption{Maximum surface area $\mathcal{A}_{max}$ of the leading capsule at different $Ca$, $Re$ and $d_0$.}
\label{Dca_Area_Tab_leading}
\end{table}
\begin{table}
\centering
\begin{tabular}{c|cccc}
\hline
\toprule
$d_0$ & & $Re=0.01$ & $Re=25$ & $Re=50$   \\
\midrule
\multirow{2}{*}{$1$} & $Ca=0.15$ & 1.068 & 1.143 &1.201 \\
& $Ca=0.35$ & 1.271 & 1.342 & 1.417 \\
\midrule
\multirow{2}{*}{$0.5$} & $Ca=0.15$ & 1.070 & 1.144 &1.204 \\
& $Ca=0.35$ & 1.277 & 1.345 & 1.414 \\
\midrule
\multirow{2}{*}{$0.25$} & $Ca=0.15$ & 1.069 & 1.148 & 1.204 \\
& $Ca=0.35$ & 1.277 & 1.344 & 1.41 \\
\bottomrule
\hline
\end{tabular}
\caption{Maximum surface area $\mathcal{A}_{max}$ of the trailing capsule at different $Ca$, $Re$ and $d_0$.}
\label{Dca_Area_Tab_trailing}
\end{table}

\subsection{Time evolution of the interspacing distance\label{sec:dcaps_interspacing distance}}
\begin{figure}
    \centering
    \begin{subfigure}{.49\textwidth}
        \centering
        \includegraphics[width=.8\columnwidth, trim=0 0 0 0, clip]{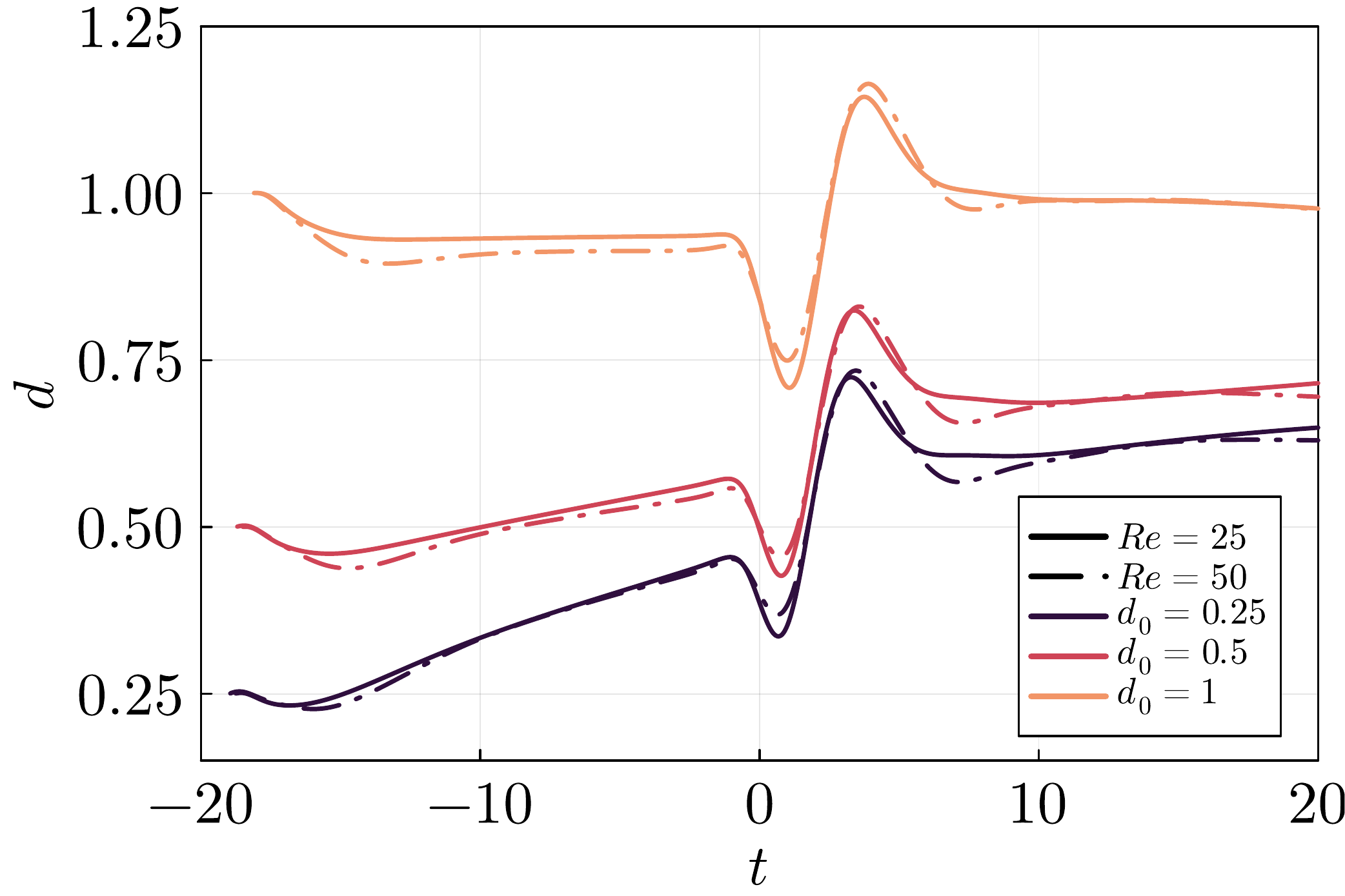}
        \caption{$Ca=0.15$}
        \label{Dca_Isp_Ca0.15}
    \end{subfigure}
    \begin{subfigure}{.49\textwidth}
        \centering
        \includegraphics[width=.8\columnwidth, trim=0 0 0 0, clip]{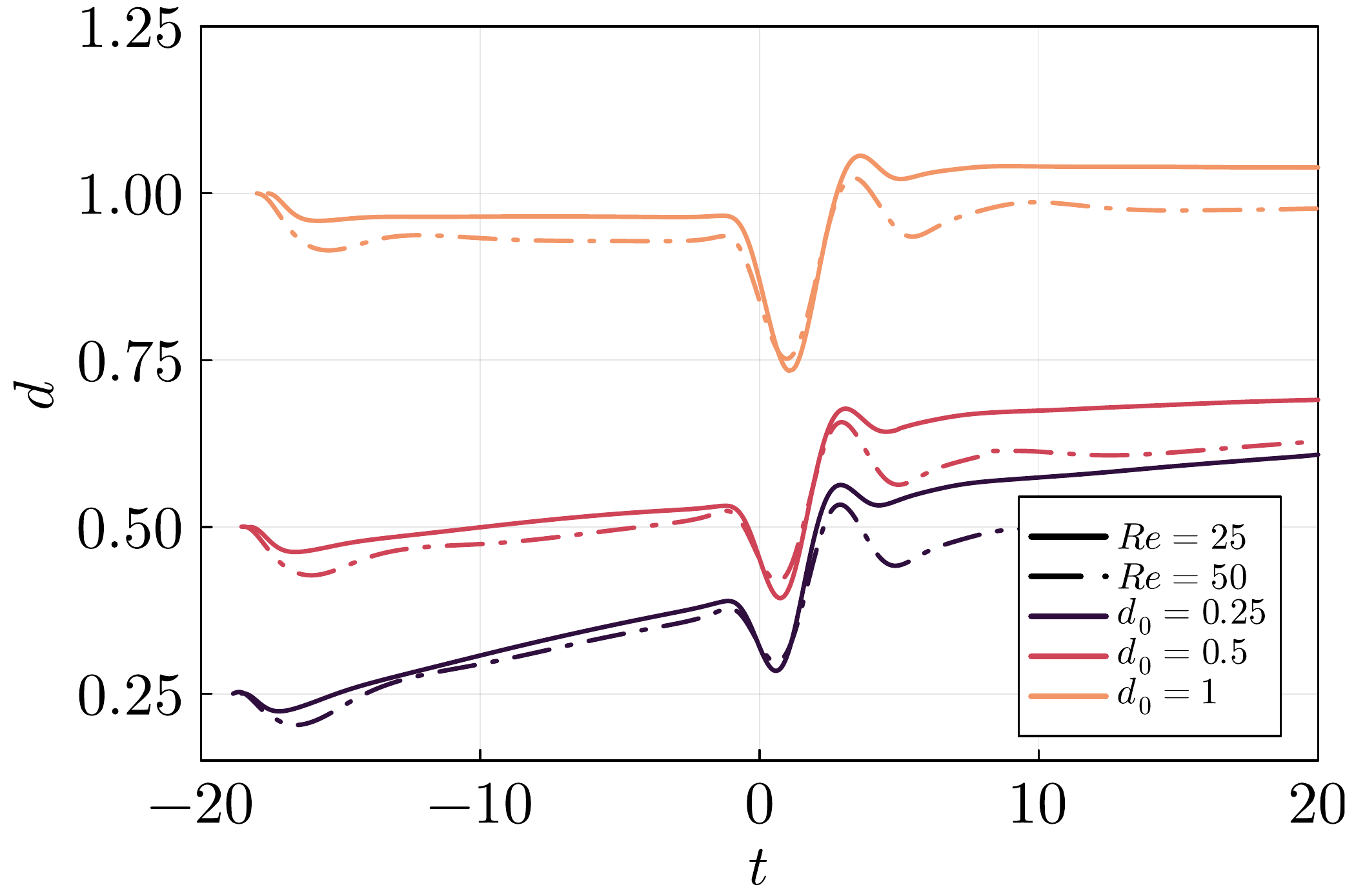}
        \caption{$Ca=0.35$}
        \label{Dca_Isp_Ca0.35}
    \end{subfigure}\\
    \caption{Temporal evolution of $d$ for different initial interspacing distance $d_0$ and Reynolds number $Re$.}
    \label{Dca_Isp}
\end{figure}

\begin{figure}
    \centering
    \includegraphics[width=.49\textwidth, trim=0 0 0 0, clip]{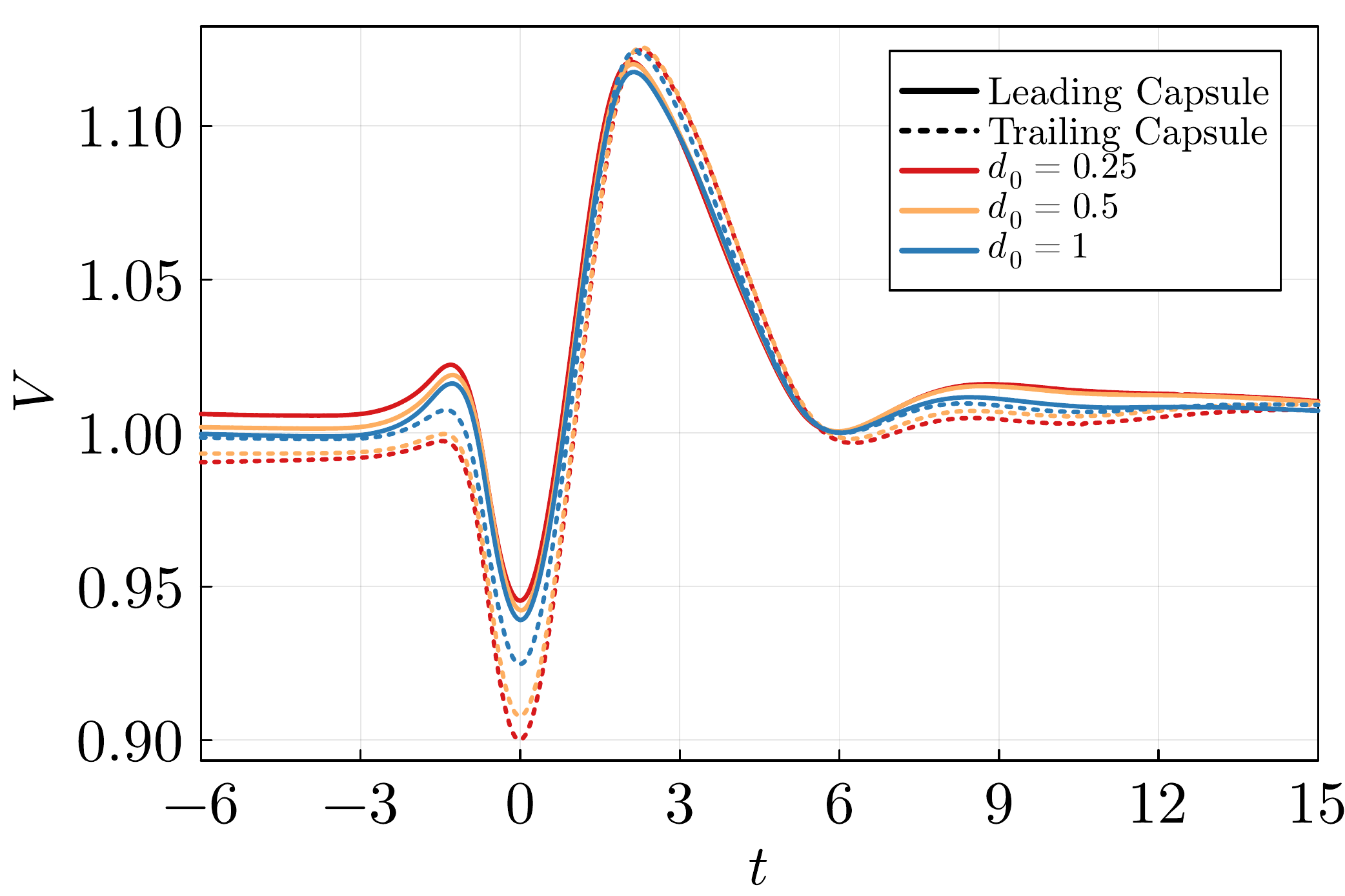}
    \caption{Effects of the initial interspacing distance $d_0$ on the evolution of the capsules velocities $V$ at $Ca=0.35, Re=50$.}
    \label{Dca_LSF_Gap}
\end{figure}

We now analyze the time evolution of the interspacing distance between the two confined capsules considered in this section. \Reffig{Dca_Isp} shows the time-dependent interspacing distance for $Ca = 0.15$ and 0.35, $Re = 25$ and 50 and $d_0 = 1$, 0.5 and 0.25. In this figure, we note that in all cases, the interspacing distance decrease immediately after the trailing capsule is released. This is due to the fact that upon release, the trailing capsule is spherical and therefore located farther away from the channel walls than is the leading capsule, resulting in its initial acceleration before a steady shape is found $-$ typically within less than five time units. In the case where $d_0 = 1$, the interspacing distance $d$ is steady until the leading capsule approaches the corner, reaches a minimum then a maximum value inside the corner and becomes steady again as the trailing capsule leaves the corner region. Interestingly, the steady interspacing distance after the corner is up to 10\% greater than its steady value prior to the corner, suggesting that the corner separates the two capsules. Moreover, the initial interspacing distance is greater in the case $Re = 25$ than in the case $Re = 50$: this is only an artifact of our release mechanism. Indeed, the steady ``parachute" shape of the capsule is deployed faster at $Re = 50$ than at $Re = 25$, leading to a shorter initial acceleration phase of the trailing capsule towards the leading capsule at $Re = 50$ than at $Re = 25$.
When $d_0 = 0.5$ and $d_0 = 0.25$, we observe that the interspacing distance steadily increases until the capsules reach the corner region where it displays the same behavior as in the case of $d_0 = 1$, and continues to increase downstream of the corner. While a steady value of $d$ is not clearly reached within the considered time range, we can extrapolate the trend and conclude that the interspacing distance seems to saturate to values ranging from 0.6 to 0.8 depending on $Re$, $Ca$ and $d_0$. Therefore, the pair of confined capsules we consider exhibit a minimum stable interspacing distance $d_{min}$. Moreover, we note that the slope of $d$ is greater in the case of lower initial interspacing distances, suggesting that the relative velocity of the capsules is a function of their interspacing distance.
To investigate further this behavior, we show in \reffig{Dca_LSF_Gap} the velocity of the two capsules at $Ca = 0.35$, $Re = 50$ and $d_0$ ranging from 0.25 to 1. We observe that the velocity of the trailing capsule is lower than that of the leading capsule prior to entering and downstream of the corner, and that the velocity difference increases with decreasing interspacing distance. This velocity difference confirms the above observations in terms of interspacing distance, in particular that a lower interspacing distance results in a greater relative velocity between the two capsules, i.e. an enhanced repulsive behavior. Moreover, we note in \reffig{Dca_LSF_Gap} that the difference in velocity minima between the leading and the trailing capsules is always greater than the difference between their velocity maxima. As a result, the residence time of the trailing capsule inside the corner region is always greater than that of the leading capsule, and the corner tends to separate the pair of capsules. 
The present analysis of the binary interaction of capsules through a corner reveals that the two considered capsules do interact in this geometry, affecting their motion and deformation. In particular, the trailing capsule tends to be more deformed than the leading capsule, and the corner tends to separate the pair of capsules. A natural question that arises is that of the accumulation of such effects if more than two capsules are considered.

\section{Train of ten capsules\label{sec:corner_train}}

\begin{figure}
    \centering
 \begin{subfigure}{.49\textwidth}
        \centering
        \includegraphics[width=.8\columnwidth, trim=0 0 0 0, clip]{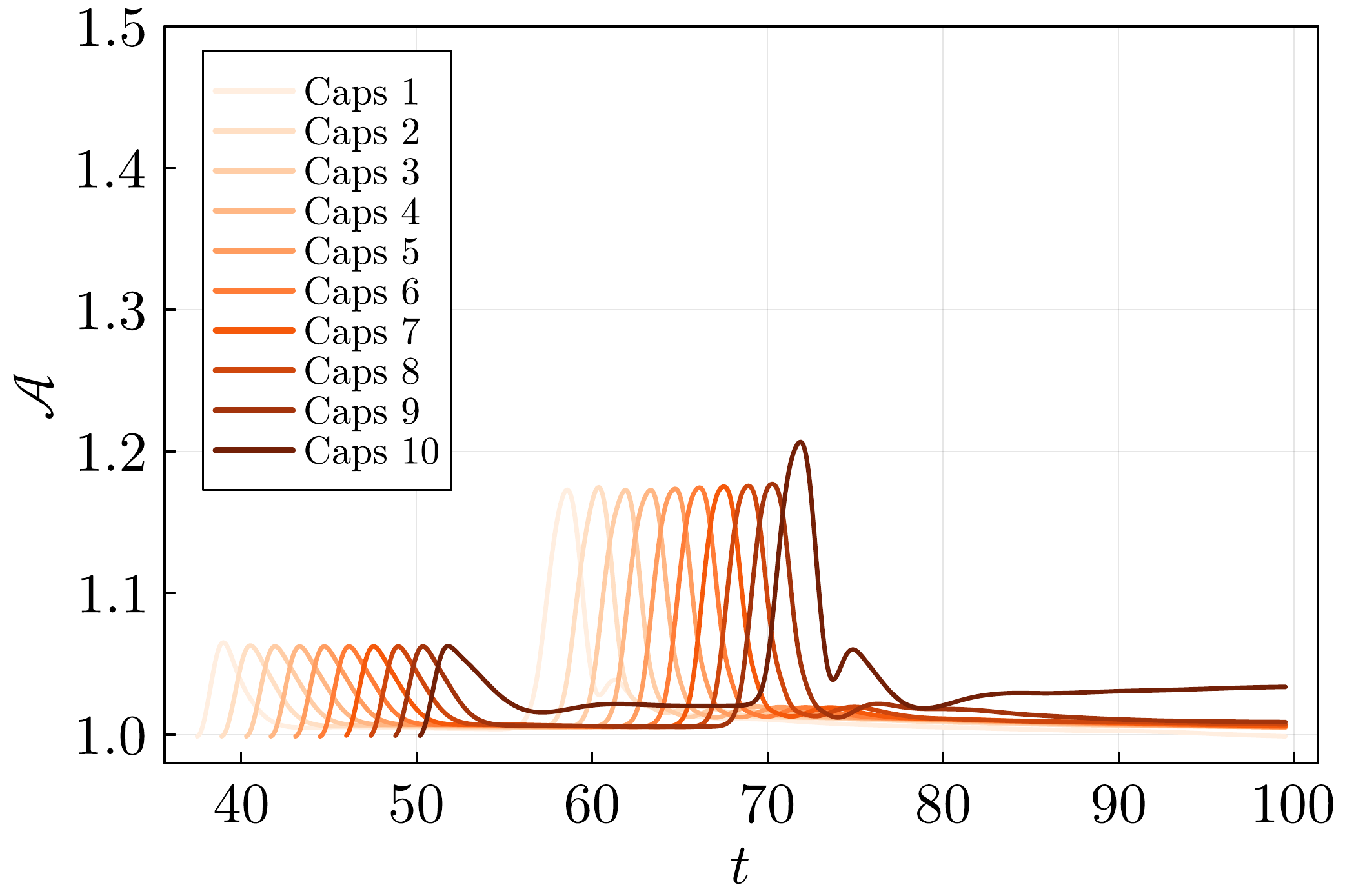}
        \caption{$Ca = 0.15$}
    \end{subfigure}
    \begin{subfigure}{.49\textwidth}
        \centering
        \includegraphics[width=.8\columnwidth, trim=0 0 0 0, clip]{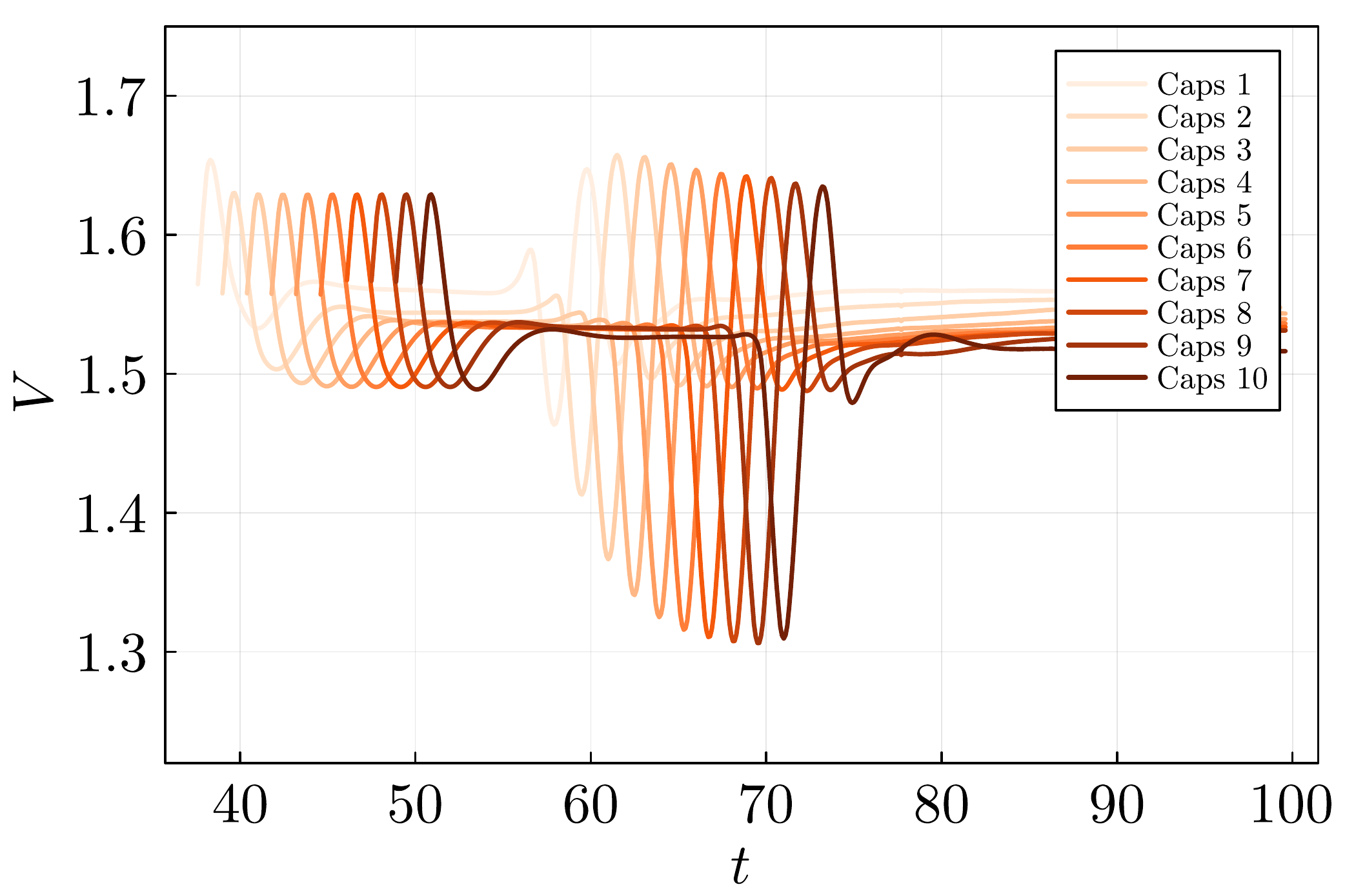}
        \caption{$Ca = 0.15$}
    \end{subfigure}\\

    \begin{subfigure}{.49\textwidth}
        \centering
        \includegraphics[width=.8\columnwidth, trim=0 0 0 0, clip]{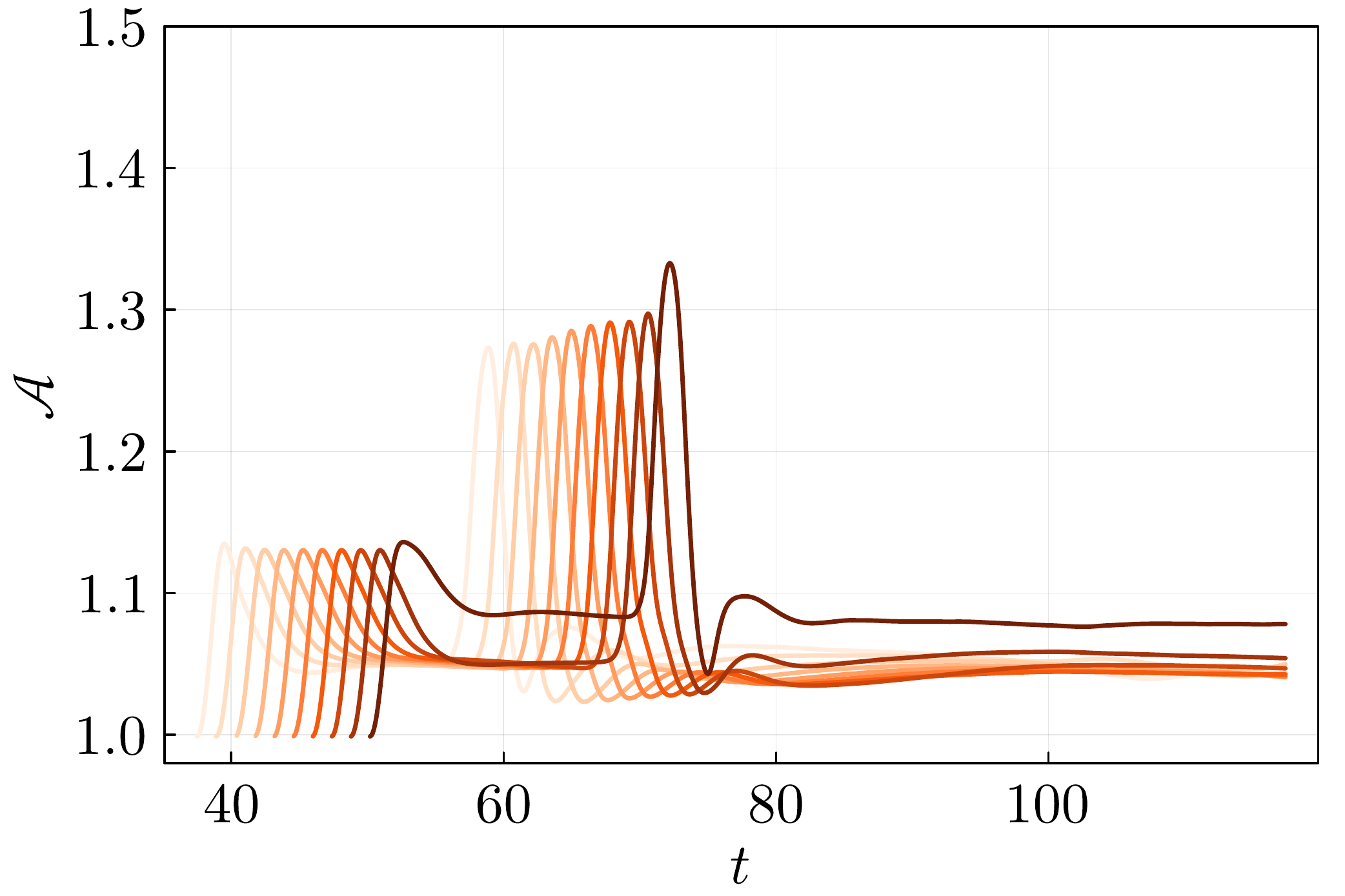}
        \caption{$Ca = 0.25$}
    \end{subfigure}
    \begin{subfigure}{.49\textwidth}
        \centering
        \includegraphics[width=.8\columnwidth, trim=0 0 0 0, clip]{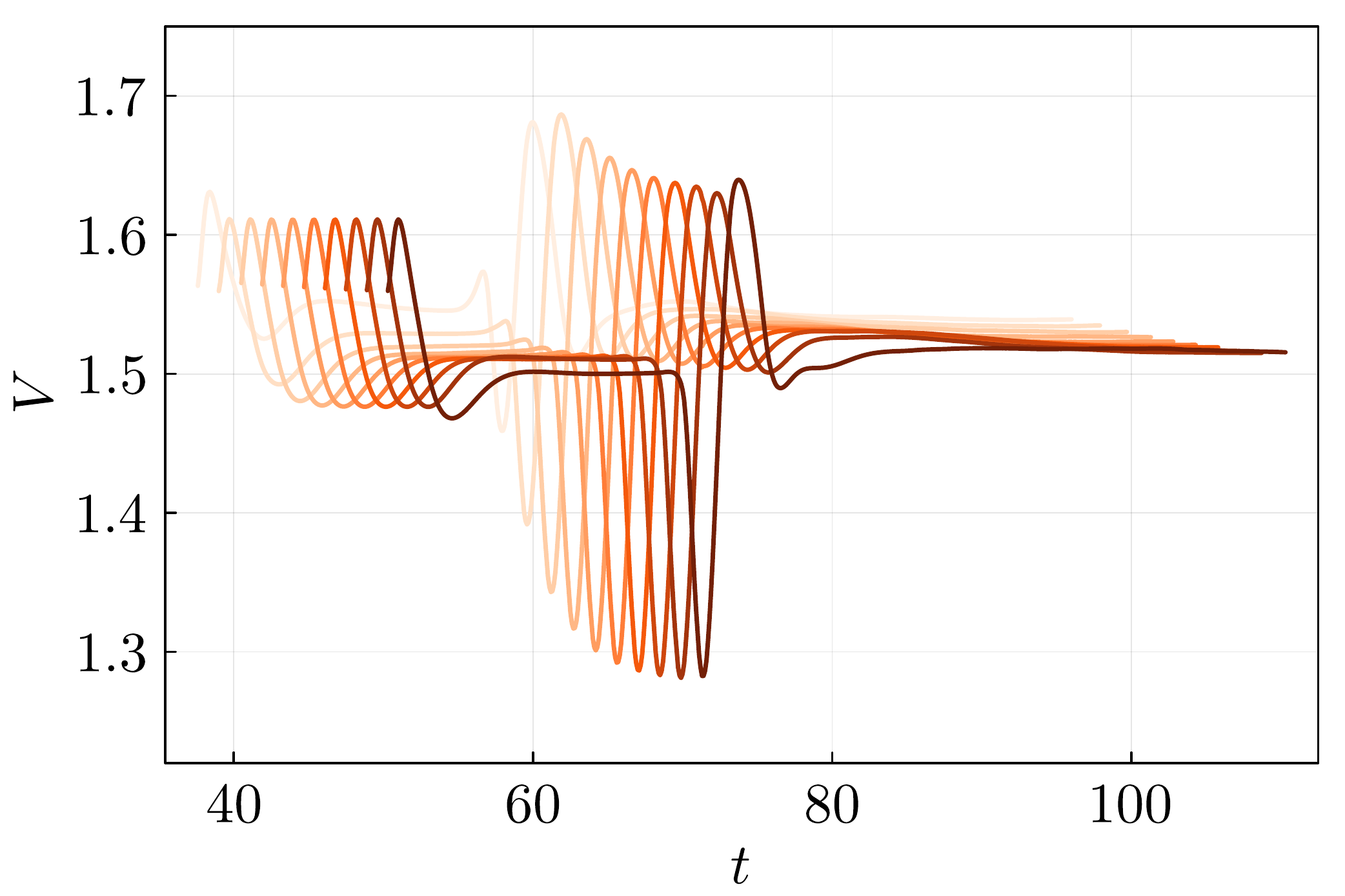}
        \caption{$Ca = 0.25$}
    \end{subfigure}  \\

    \begin{subfigure}{.49\textwidth}
        \centering
        \includegraphics[width=.8\columnwidth, trim=0 0 0 0, clip]{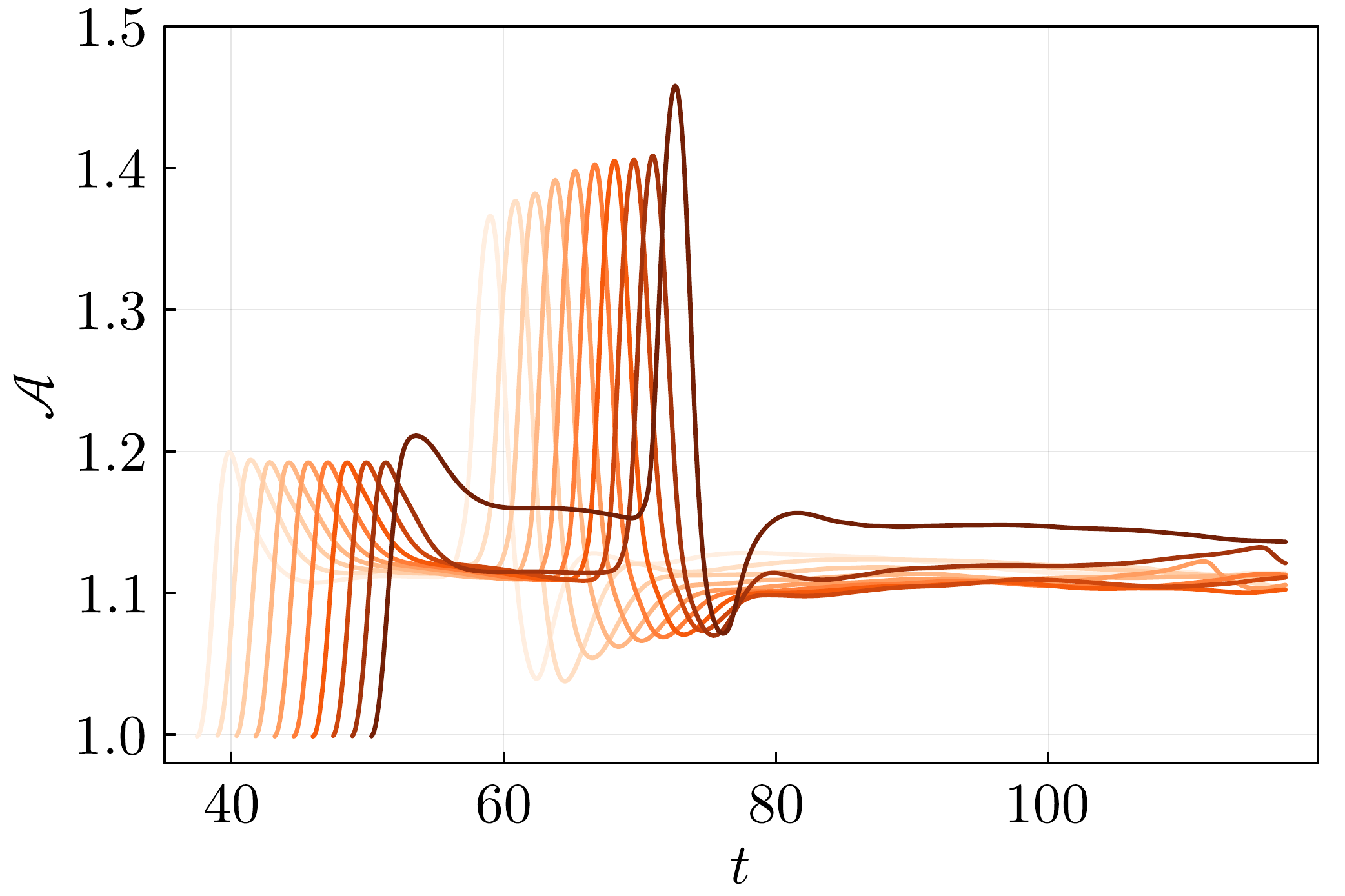}
        \caption{$Ca = 0.35$}
    \end{subfigure}
    \begin{subfigure}{.49\textwidth}
        \centering
        \includegraphics[width=.8\columnwidth, trim=0 0 0 0, clip]{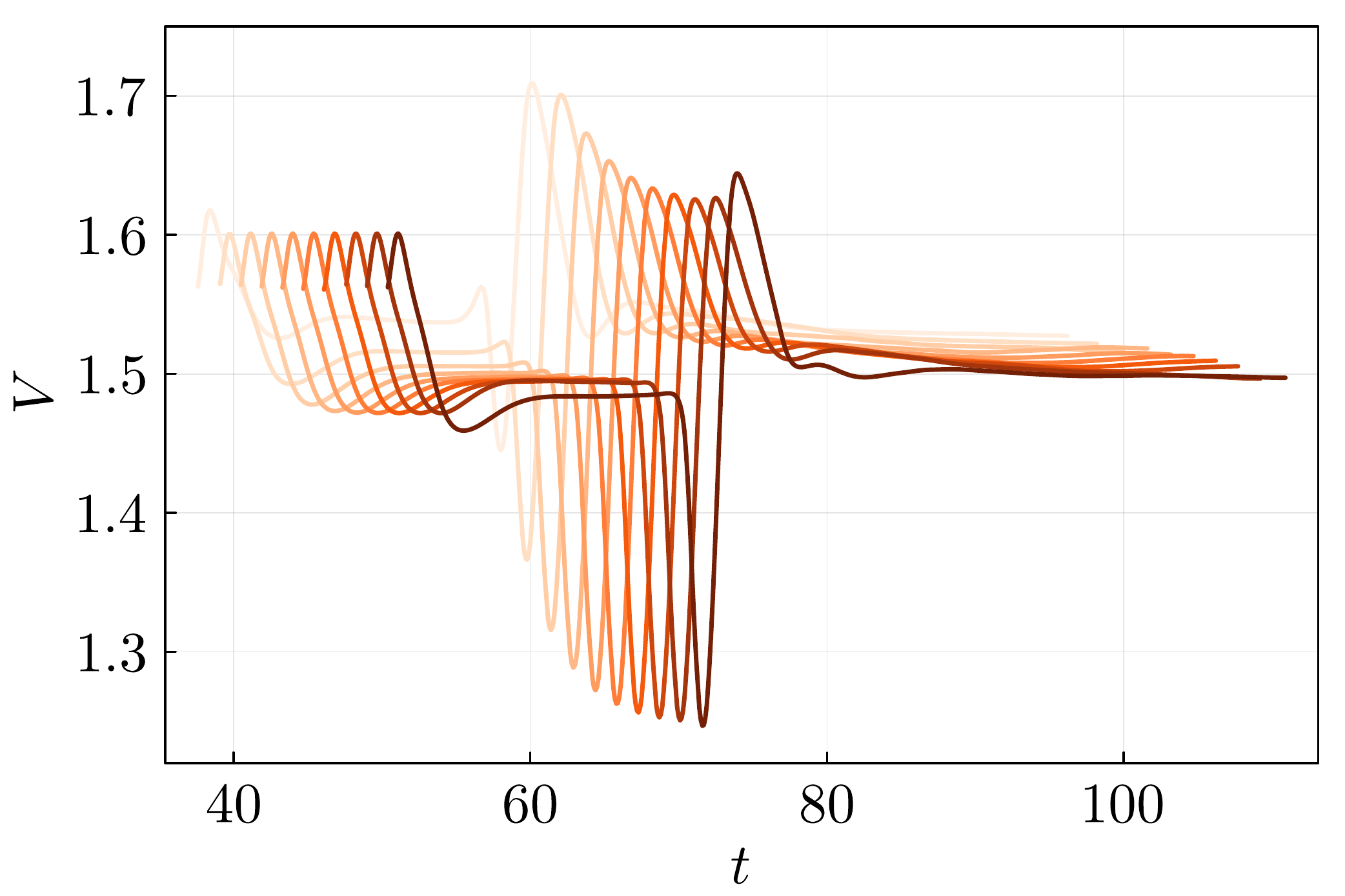}
        \caption{$Ca = 0.35$}
    \end{subfigure}
    \caption{Time evolution of the reduced surface areas and velocities of ten capsules at $Re = 50$ and $d_0 = 1/8$ for $Ca = 0.15$, 0.25 and 0.35.}
    \label{fig:train_time}
\end{figure}

\begin{figure}
    \centering
    \begin{subfigure}{.49\textwidth}
        \centering
        \includegraphics[width=.8\textwidth]{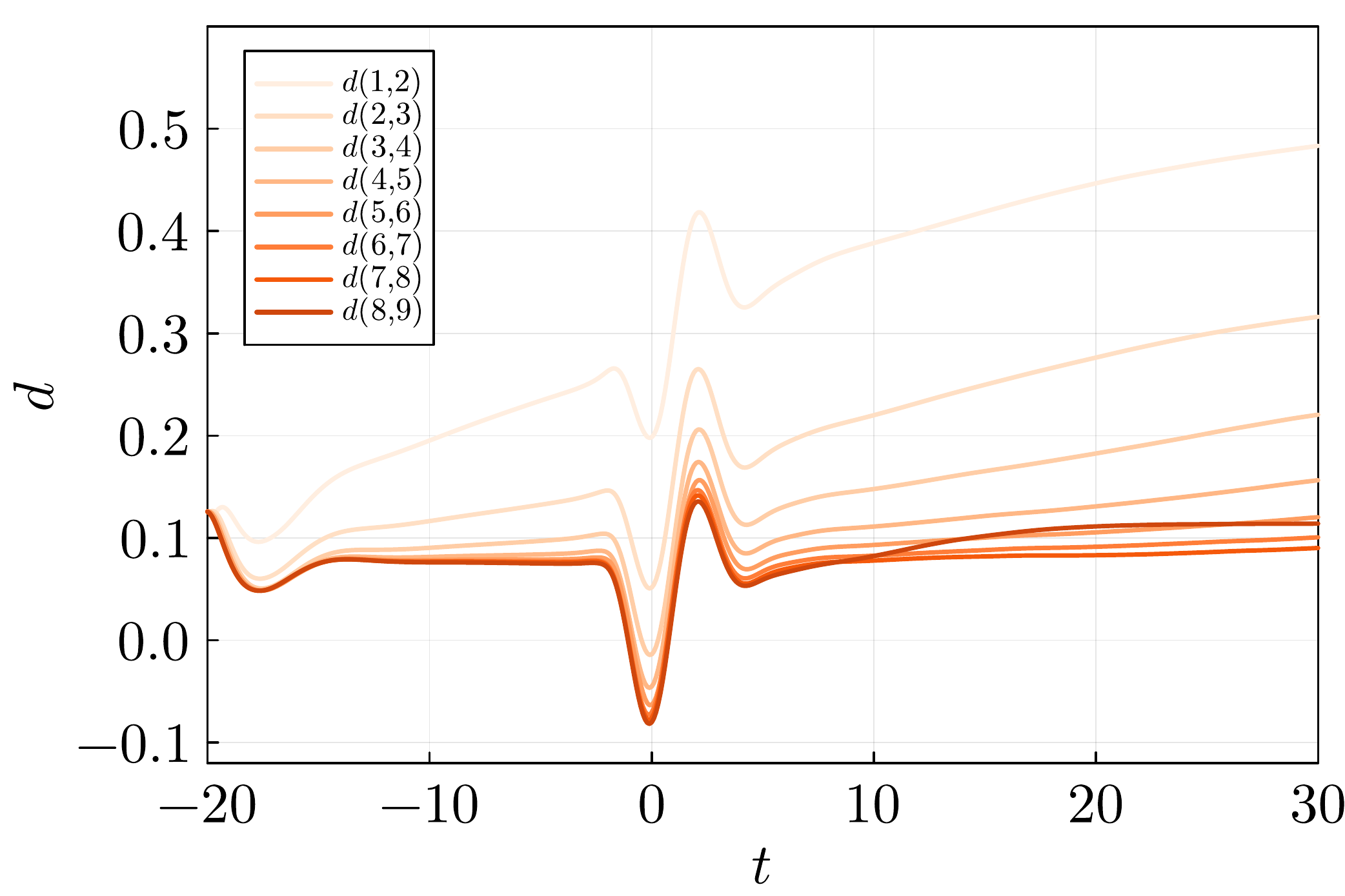}
        \caption{$Ca = 0.15$}
    \end{subfigure}
    \hfill
    \begin{subfigure}{.49\textwidth}
        \centering
        \includegraphics[width=.8\textwidth]{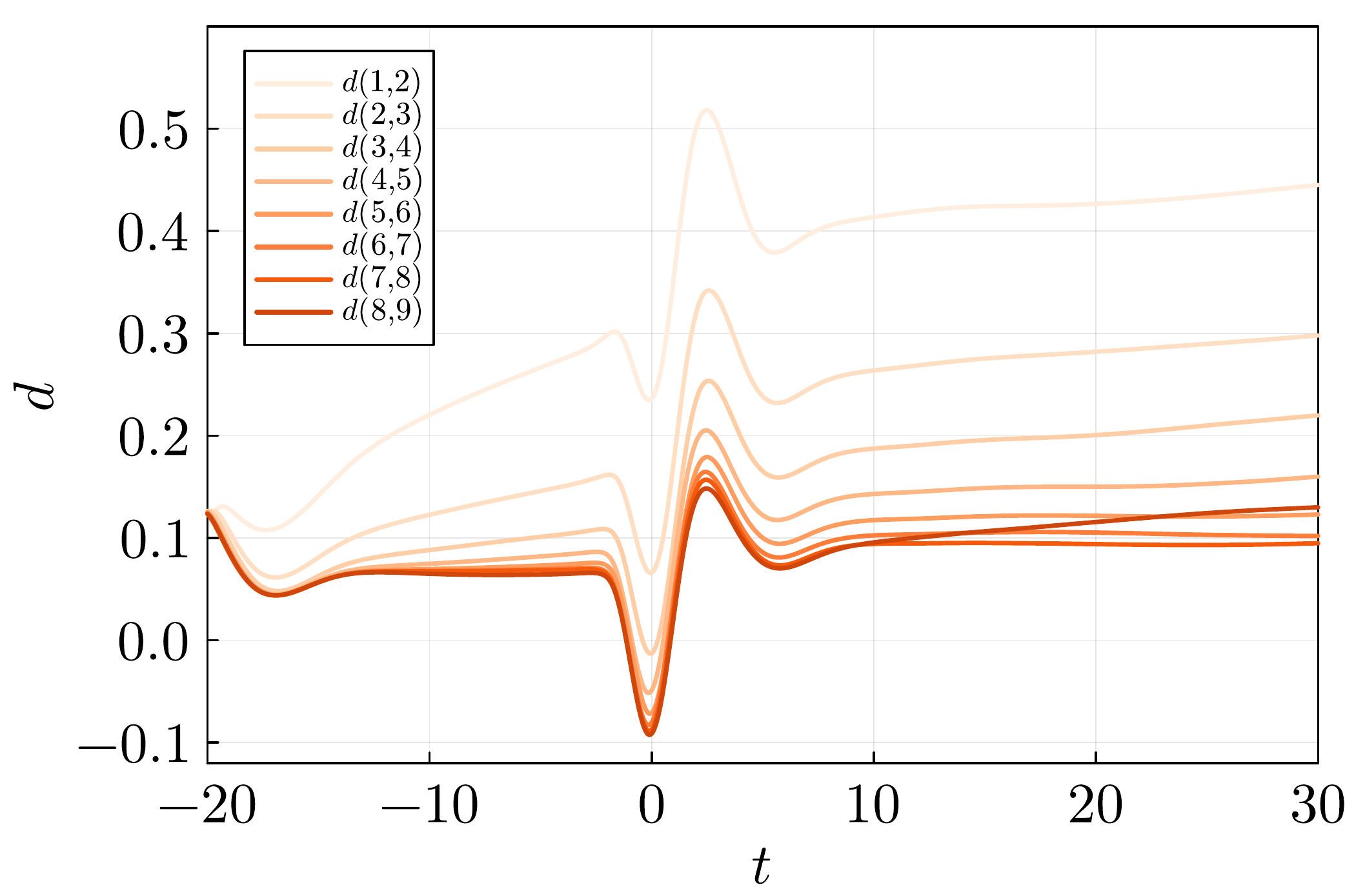}
        \caption{$Ca = 0.25$}
    \end{subfigure}
    \begin{subfigure}{\textwidth}
        \centering
        \includegraphics[width=.4\textwidth]{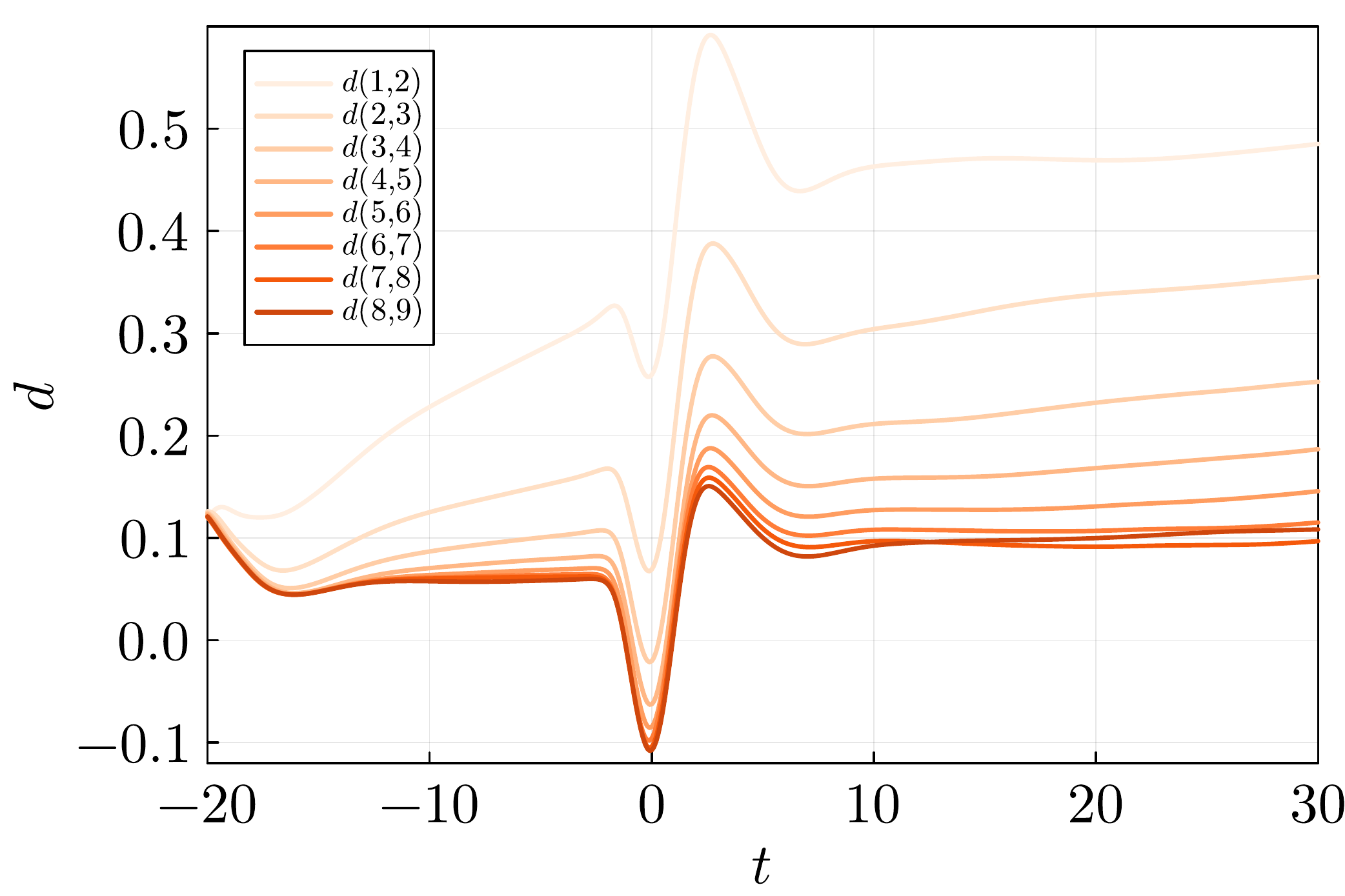}\\
        \caption{$Ca = 0.35$}
    \end{subfigure}
    \caption{Temporal evolution of $d$ for a train of 10 capsules at $Re = 50$ and $d_0 = 0.125$ for (a) $Ca = 0.15$, (b) $Ca = 0.25$ and (c) $Ca = 0.35$.}
    \label{fig:train_interdistance}
\end{figure}

\begin{figure}
    \centering
      \includegraphics[width=.45\textwidth]{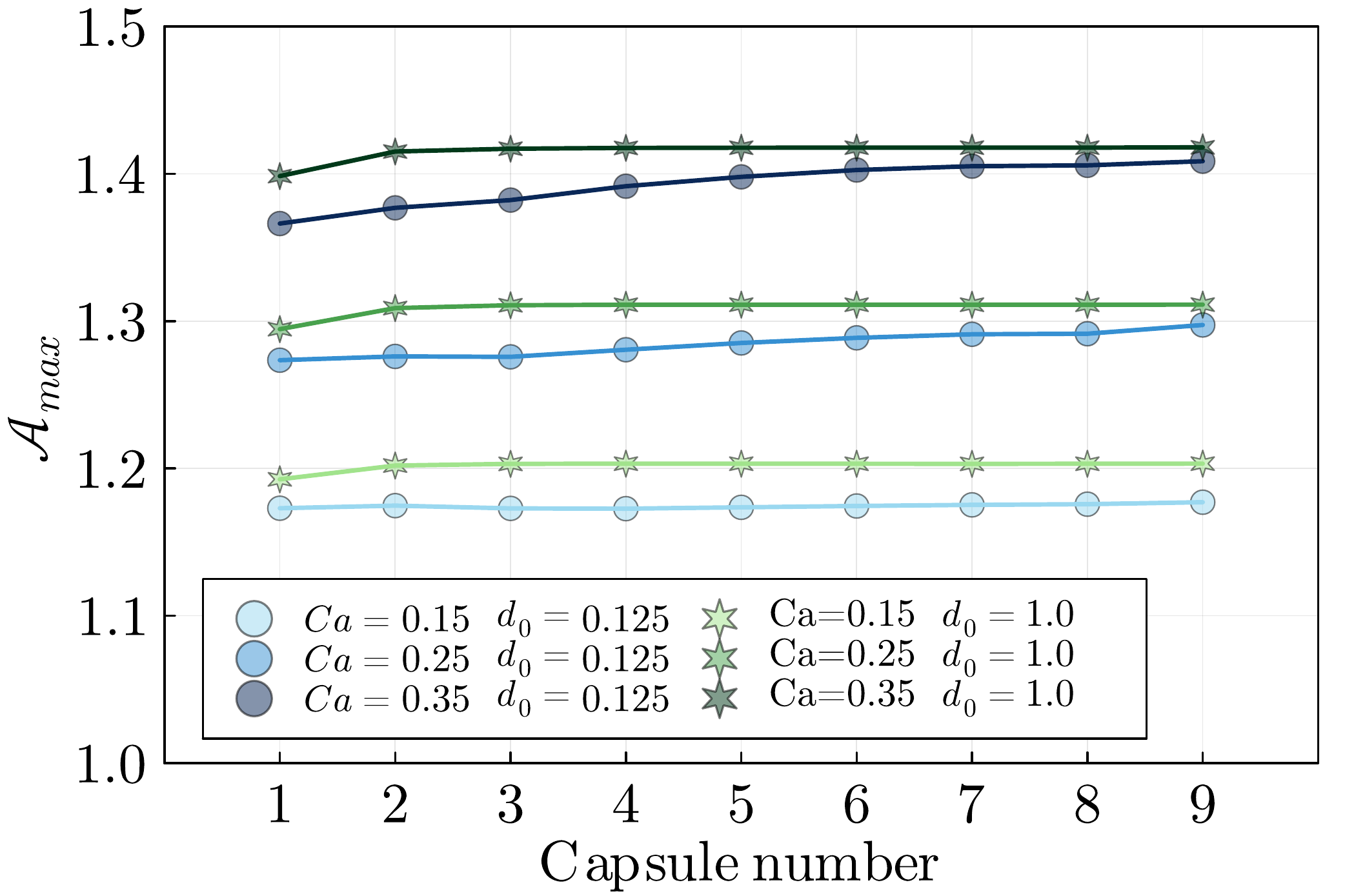}
      \caption{$\mathcal{A}_{max}$ as a function of the capsule number.}
    \label{fig:train_Amax}
\end{figure}

\begin{figure}
    \centering
    \begin{subfigure}{.48\textwidth}
        \includegraphics[width=.9\textwidth]{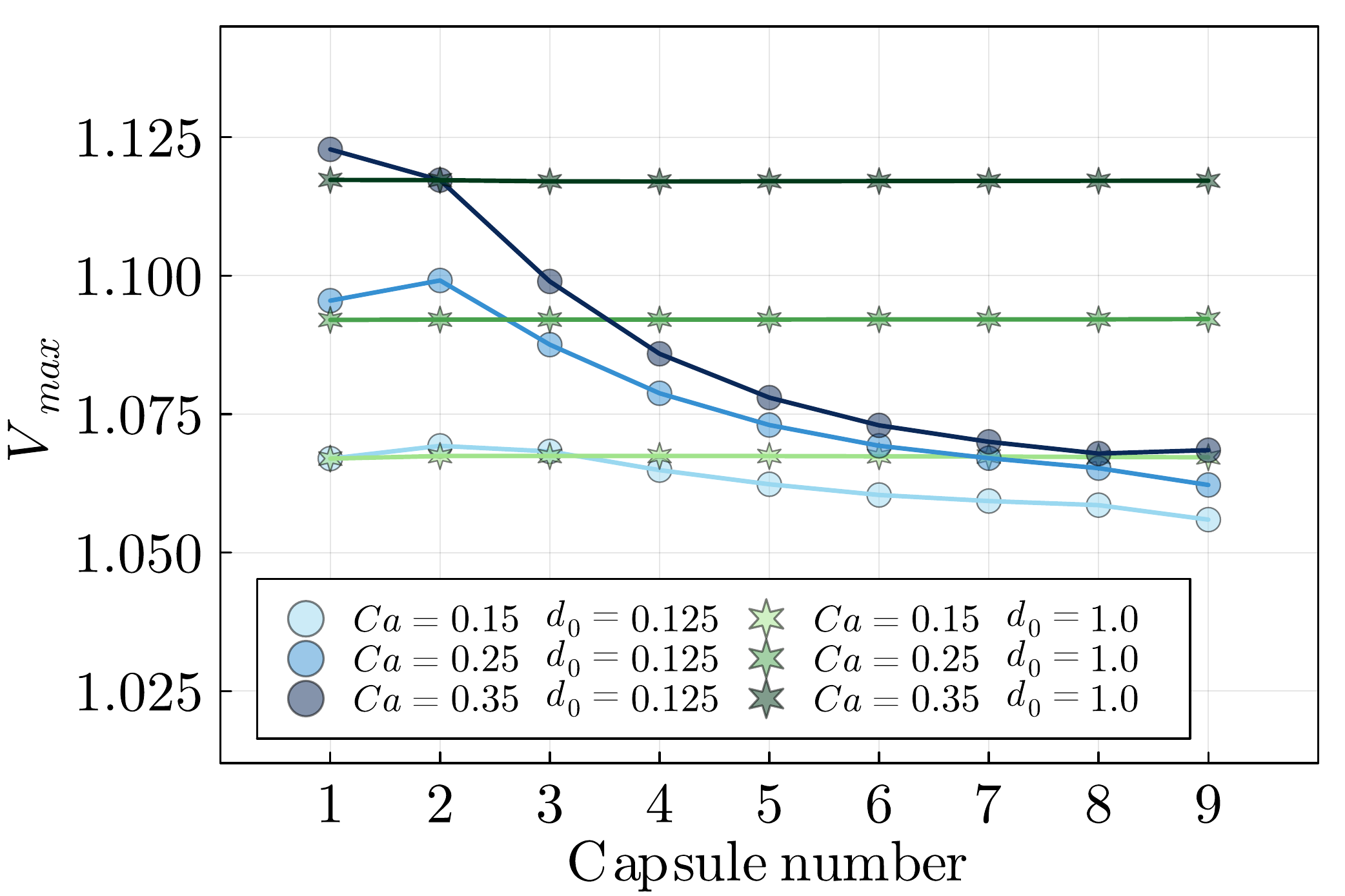}
        \caption{}
\label{fig:train_vmax}
    \end{subfigure}
    \hfill
    \begin{subfigure}{.48\textwidth}
        \includegraphics[width=.9\textwidth]{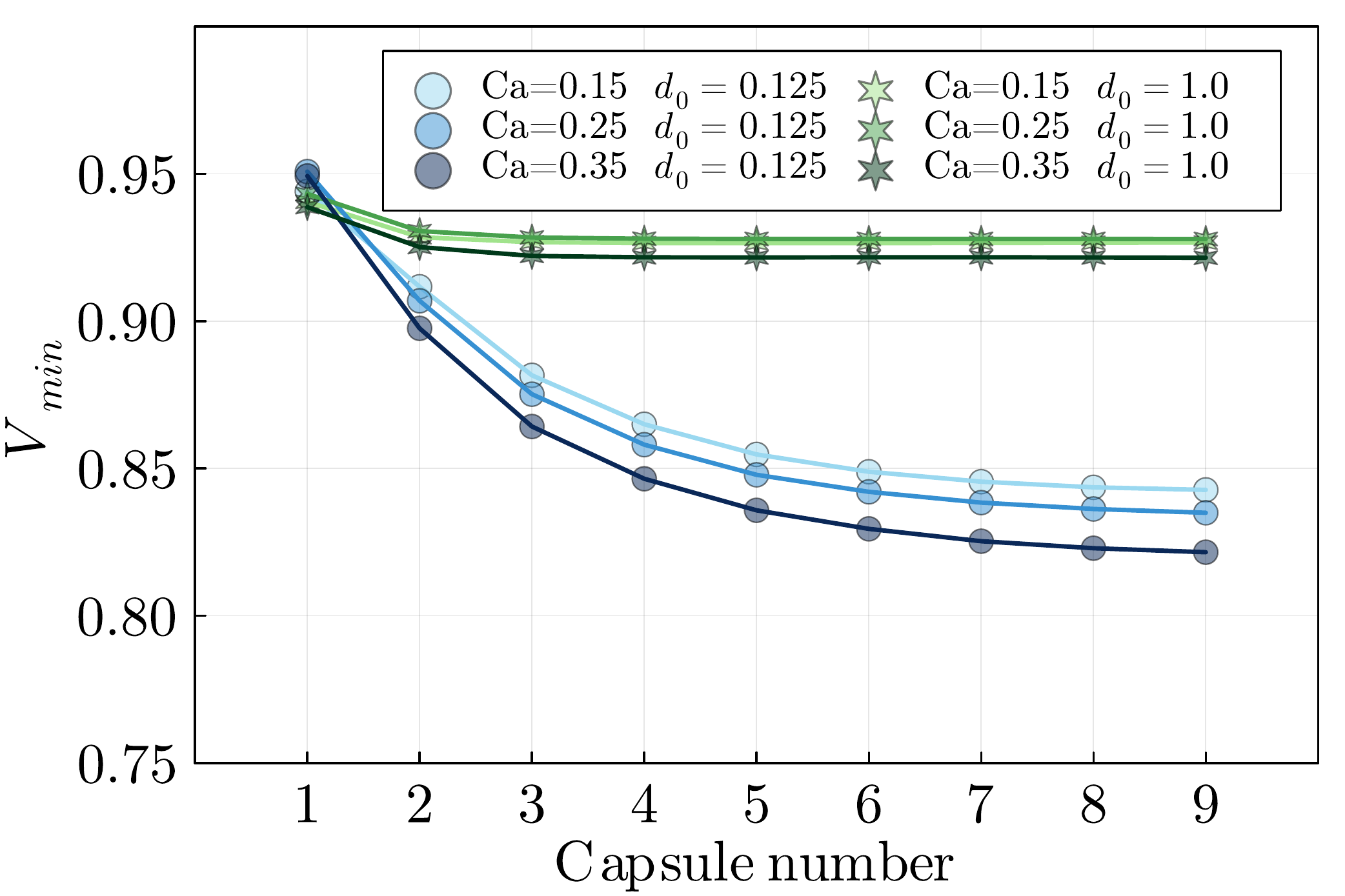}
        \caption{}
        \label{fig:train_vmin}
    \end{subfigure}
    \caption{(a) $V_{max}$ and (b) $V_{min}$ as a function of the capsule number.}
    \label{fig:train_vminmax}
\end{figure}

In this last section, we investigate the behavior of a train of ten capsules flowing through the corner. We insert each capsule using the same procedure employed in the previous section: a new initially spherical capsule appears at a distance $D_c = 30$ radii from the corner as soon as the preceding capsule has advanced by a reduced distance $\tilde{d} = 2\tilde{a}(1 + d_0)$. The capsules are removed from the computational domain when they are less than one initial diameter away from the outflow boundary. Our goal is to determine if the findings of the previous binary capsule analysis accumulate when more than two capsules are considered, especially with regards to the increased surface area of the capsules and the separating effect reported in \refsec{sec:corner_double}. As such, we plot in \reffig{fig:train_time} the normalized surface area and velocity of each capsule of the train at $Re = 50$, $d_0 = 0.125$ and $Ca$ ranging from 0.15 to 0.35. The same figure obtained in the case of $d_0 = 1$ is provided in Appendix \ref{app:train_d1}. In \reffig{fig:train_time}, the darkness of the color corresponds to the position of the capsule in the train: darker means increasing capsule number i.e. further downstream along the capsule train. As mentioned in \refsec{sec:corner_double}, the initial peaks in the surface area and velocity of the capsule are insertion artifacts and do not contribute to the physics that is the focus of this section.
We observe in \reffig{fig:train_time} that the behavior of the last capsule is significantly different than that of the rest of the train. In \refsec{sec:corner_double} we hypothesized that the difference in surface areas of the leading and the trailing capsules is due to the fact that the wake of the leading capsule is significantly affected by the presence of the trailing capsule. The present observation in \reffig{fig:train_time} corroborates this statement: all of the capsules in the train see their wake affected by a trailing capsule, except in the case of the last capsule. As a result, its deformation is greater and extends closer to the channel walls, thus decreasing its velocity. We also remark in \reffig{fig:train_time} that this effect is enhanced with increasing $Ca$. While noteworthy in the case of a pair of capsules, this effect is less pertinent to the study of a train of capsules, as only the core of the capsule train is relevant to typical microfluidic applications. As such, in the remainder of this section our analysis is focused on the first ninth capsules of the train.

As expected, a steady state is reached in the straight channel prior to the corner for each capsule and for all $Ca$. While the steady surface area remains constant with increasing capsule number, i.e. as we move further downstream in the train of capsules, we observe that the velocity of the capsules decreases. In particular the difference between the steady velocity of the first and ninth capsules increases with increasing $Ca$. As the capsules enter the corner region, they display the familiar pattern previously described in \refsec{sec:corner_single} and \refsec{sec:corner_double}, before relaxing to steady values. The shape of the deviation pattern is strikingly similar across different capsules of the train, regarding both the velocity and the surface area of the capsules, except that they are shifted in time and magnitude. More precisely, the surface area curves are shifted upwards with increasing capsule number while the velocity curves are shifted downwards with increasing capsule number. As a result, the maximum surface area of the capsule increases and the velocity extrema decrease with increasing capsule number. This behavior is more pronounced as $Ca$ increases. 
Additionally, we compare in \reffig{fig:train_interdistance} the normalized interspacing distance $d$ between each pair of capsules in the train. In \reffig{fig:train_interdistance}, each curve is shifted in time such that $t = 0$ corresponds to $d_{min}$ inside the corner. For all $Ca$, we observe that the interspacing distance $d(1,2)$ between the first and the second capsules increases to a steady value close to 0.5, and that the corner has marginal effects on the downstream evolution of $d(1,2)$: this behavior is identical to the case of two capsules studied in the previous section. However, as we move downsteam in the train of capsules, $d$ increases slower and slower prior to the corner until it remains constant for capsule numbers greater than 7, at a steady value $d \approx 0.7$ that decreases only marginally with increasing $Ca$. After the transient regime due to the corner, $d(i, i+1)$ for capsule numbers $i$ greater than 7 reaches a steady state that is slighly higher than prior to entering the corner. In other words, the corner tends to increase the interspacing distance, and therefore exhibits a separating effect. This seperating effect is observed regardless of the initial interspacing distance $d_0$, as was the case in the previous section when only two capsules were considered.


Finally, in order to investigate further the influence of the capsule number on the capsule dynamics, we plot in figures \ref{fig:train_Amax}-\ref{fig:train_vminmax} the maximum surface area as well as the maximum and minimum velocities of each capsule of the train for varying Capillary numbers and interspacing distances. The difference in minimum velocity (respectively, maximum velocity) between the first and the ninth capsule is about 15\% (respectively, about 7\%) at $Ca = 0.35$ while it is about 11\% (respectively, 2\%) at $Ca = 0.15$.
Similarly, the difference in maximum surface area between the first and the ninth capsule is about 4\% at $Ca = 0.35$ and less than 1\% at $Ca = 0.15$. These results correspond to $d_0 = 0.125$, while in the case of $d_0 = 1$ only deviations lower than 1\% are observed in the extrema of the capsule surface area and velocity (except in the case of $Ca = 0.35$ for which velocity deviations of 2\% are observed). The very small deviations observed in the case $d_0 = 1$ indicates that for this interspacing distance the capsules interact very weakly. As such, there exist a critical interspacing distance $d_c$ below which capsule interactions are observed, with $0.125 < d_c < 1$.

The fact that $d_c$ is less than 1 can be surprising, as a normalized interspacing distance of $d_0 = 1$ would typically be classified as a strongly interacting regime in other geometries, e.g. in the T-junction investigated by Lu et al. \cite{lu2021path}. The main reason for the low interaction we observed is likely due to the short residence time of the capsules in the corner region. Indeed, Lu et al. showed that the residence time is determinant in the path selected by the capsules in a T-junction geometry. Another reason for such a low critical interspacing distance is related to the very confined configuration we study: the capsule shape and behavior is primarily due to the presence of the walls, while the small disturbances of the flow field due to the other capsules only marginally contribute to each capsule dynamics. Future studies could explore the dynamics of a train of capsules in a wider channel, i.e. in a less confined configuration, where each capsule could be more influenced by the wake disturbances of their preceding neighbor.

\section{Conclusion\label{sec:corner_conclusion}}
In the present work, the inertial and non-inertial dynamics of three-dimen\-sional elastic capsules flowing through a sharp corner are investigated. The capsule trajectory, surface area, velocity and membrane stress are analyzed in the cases of one, two and a train of ten capsules released upstream of the corner. The channel Reynolds number ranges from 0.01 to 50, the Capillary number representing the ratio of viscous stresses over elastic stresses ranges from 0.075 to 0.35 and the initial normalized interspacing distance between two capsules is varied from 1 to 0.125. The goal of this study is to help provide practical guidelines in order to anticipate capsule breakup and estimate throughput in inertial microchannels.

The case of a single capsule with no inertia was previously studied by Zhu \& Brandt \cite{zhu2015motion}, who reported that the capsule follows the flow streamlines closely regardless of the Capillary number. In inertial flows, we found that this statement is still valid for all considered Reynolds and Capillary numbers. As the streamlines of the inertial flow cross the centerline of the secondary channel $-$ the horizontal channel downstream of the corner $-$, the capsule position is increasingly close to the top wall for increasing Reynolds number, especially in the case of large Capillary numbers. However no collision between the capsule and the wall of the secondary channel was observed thanks to strong lubrication forces.
In their study, Zhu \& Brandt also analyzed the velocity of the capsule centroid and the surface area of the capsule membrane: they found that the capsule velocity decreases in the corner and increases immediately after the corner, with an overshoot increasing with membrane deformability. The surface area of the capsule was also found to reach a maximum slightly shifted in time with respect to the minimum of velocity. In the inertial regime, we observed that this behavior is enhanced as the Reynolds number increases. However our results at $Re = 1$ do not differ significantly from results obtained in the non-inertial regime, which corroborates the same observation made by Wang et al. \cite{wang2016motion, wang2018path}. Moreover, at sufficiently high inertia, capsule surface areas lower to equilibrium surface areas are observed as the capsule relaxes to its steady state. In other words, immediately after the corner the capsule oscillates around its steady shape. This phenomenon is enhanced as the Capillary number increases. Additionally, we reported that the relationship between the maximum surface area $\mathcal{A}_{max}$ of the capsule and the Reynolds number is linear as long as the Capillary number is kept below 0.35. At $Ca = 0.35$, the relationship between $\mathcal{A}_{max}$ and $Re$ is not perfectly linear and the curve $\mathcal{A}_{max}(Re)$ is slightly concave. Moreover, from $Re = 1$ to $Re = 50$, the maximum surface area increases nearly linearly over the full range of $Ca$.
At $Ca = 0.35$, we compared the membrane stress to the capsule surface area and found that (i) the time evolution of the average stress presents a strong correlation to that of the membrane surface area, and (ii) in our configuration, the value of the maximum stress is double that of the average stress. As a result, observing the capsule surface area experimentally can provide reliable insight into the average stress as well as an estimate of the capsule maximum stress.
This finding is of primary importance in the design of microfluidic devices where capsule breakup is to be avoided, as well as in the development of targeted drug delivery methods for which a controlled capsule breakup is sought.

We then investigated the interaction of several capsules in the corner geometry. First, two capsules are considered with varying initial interspacing distances. Similar to the case of a single capsule, neither the trajectory of the leading nor of the trailing capsule is observed to significantly deviate from the flow streamlines. In the range of initial interspacing distance considered, the velocity of the trailing capsule is found to be generally lower than that of the leading capsule as well as that of a single capsule at the same Reynolds and Capillary numbers. Similarly, the velocity of the leading capsule is greater than that of a single capsule in the same conditions. This velocity difference is also visible in the time evolution of the interspacing distance $d$ between the pair of capsules. In particular, we found that capsules initially located at $d_0 \leq 0.5$ tend to separate. This suggests that there exists a minimum stable gap $d_{min}>0.5$ between two confined capsules. A systematic analysis of this effect is left for future studies. In contrast, inside the corner the surface area of the trailing capsule is found to be larger than that of the leading capsule and of the single capsule in the same conditions. However, in the configuration we consider where confinement is strong, the magnitude of these effects is small even for capsules located very close to each other: the velocity of the leading and trailing capsules only deviates by a few percents from that of a single capsule. 
Next, we examined the case of a train of capsules and sought to determine whether the effects observed with a pair of capsule accumulate. While no interaction occurs for a large initial interspacing distance $d_0 = 1$, we found that in the case $d_0 = 1/8$, the steady and extremum surface areas of the trailing capsules increase by up to 5\% and eventually saturate at the tail of the train, around the ninth capsule. In all cases, the corner is found to separate the pair of capsules as well as the capsule train, which can be further evidenced from the analysis of the time evolution of the capsule velocity inside the corner region.

We believe that the present work is a step forward towards providing practical guidelines to avoid capsule breakup in inertial and non-inertial microfluidic experiments. Future works could study capsule membranes exhibiting a strain-hardening elastic behavior, e.g. as described by the Skalak law \cite{skalak1973strain}, as well as vary the confinement ratio $\beta = 2\tilde{a}/\tilde{W}$ in order to consider high-throughput microfluidic devices. In the case of lower confinement ratios in particular, we expect to see stronger capsule interactions along with cross-stream capsule migration inside and downstream of the corner. Finally, the present work could also be useful to develop membrane characterization techniques, where viscoelastic membrane properties could be inferred from the time-dependant evolution of a capsule of interest through a corner.

\appendix
\section{Time evolution of capsule surface areas in the case of two interacting capsules\label{app:Dca_area}}

\begin{figure}[H]
    \centering
      \begin{subfigure}{.48\textwidth}
        \centering
        \includegraphics[width=.8\columnwidth, trim=0 0 0 0, clip]{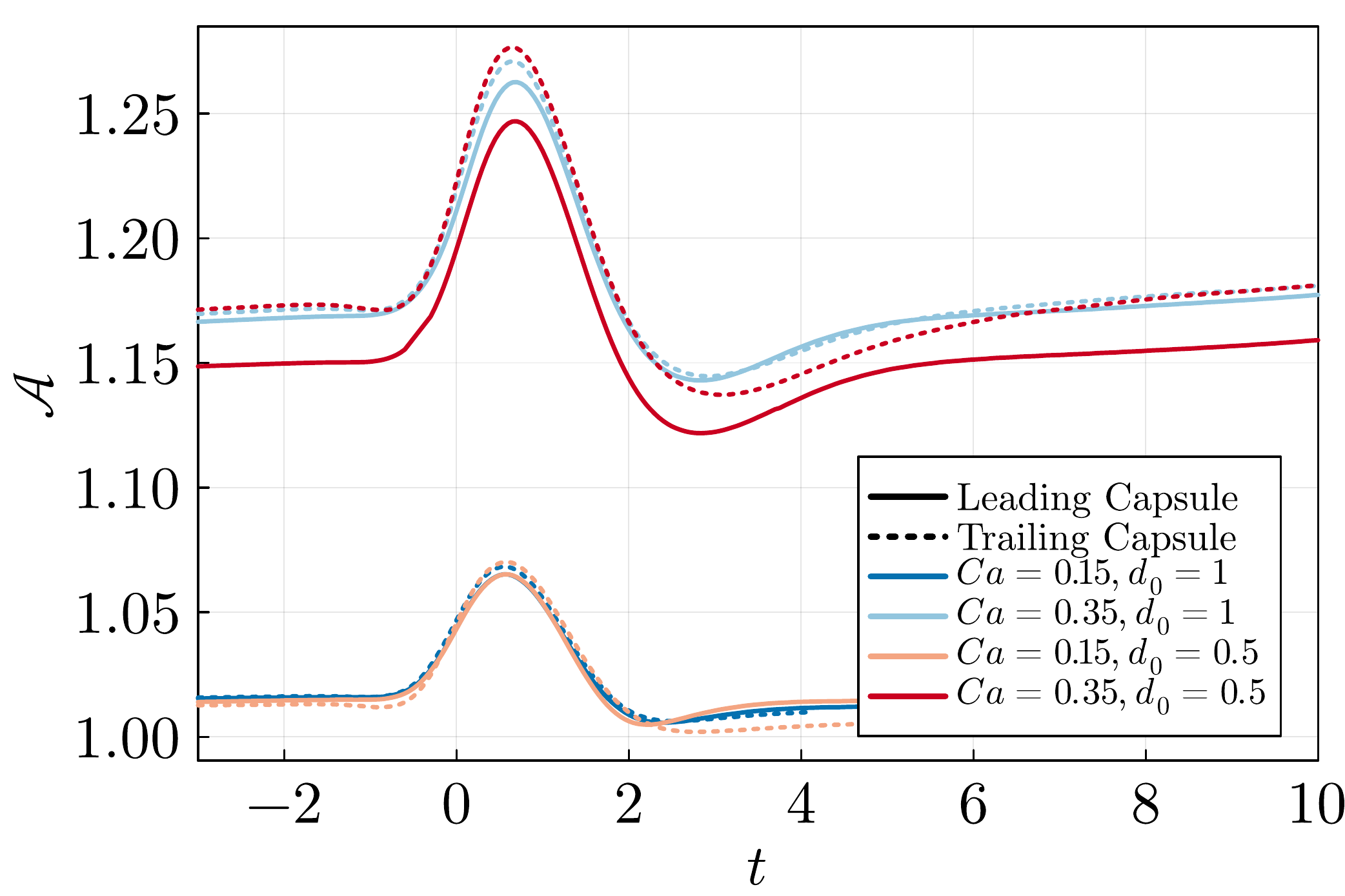}
        \caption{$Re=0.01$}
        \label{Dca_Area_NIn}
    \end{subfigure}\\
    \begin{subfigure}{.48\textwidth}
        \centering
        \includegraphics[width=.8\columnwidth, trim=0 0 0 0, clip]{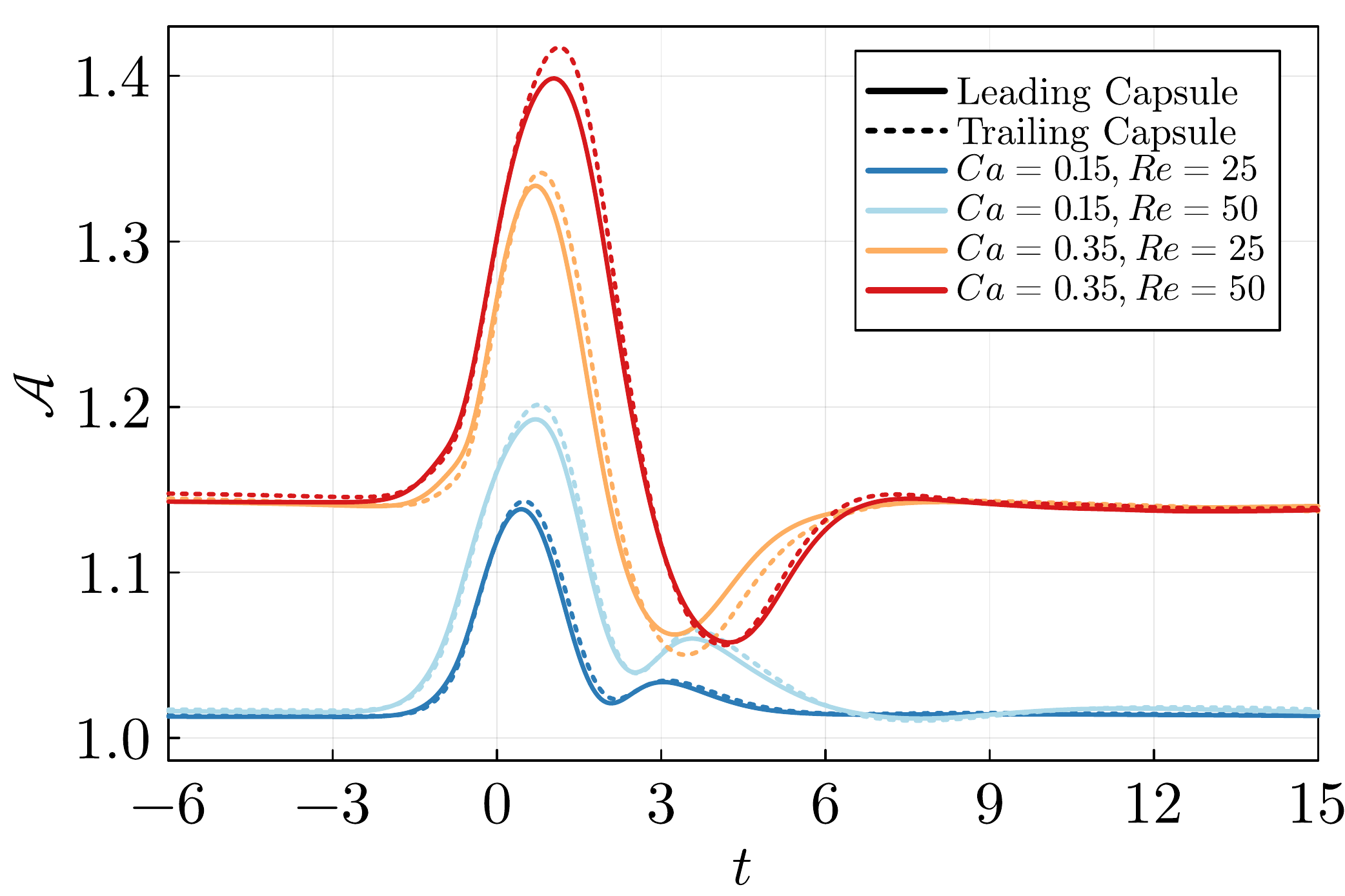}
        \caption{$d_0=1$}
        \label{Dca_Area_In_d1}
    \end{subfigure}
    \begin{subfigure}{.48\textwidth}
        \centering
        \includegraphics[width=.8\columnwidth, trim=0 0 0 0, clip]{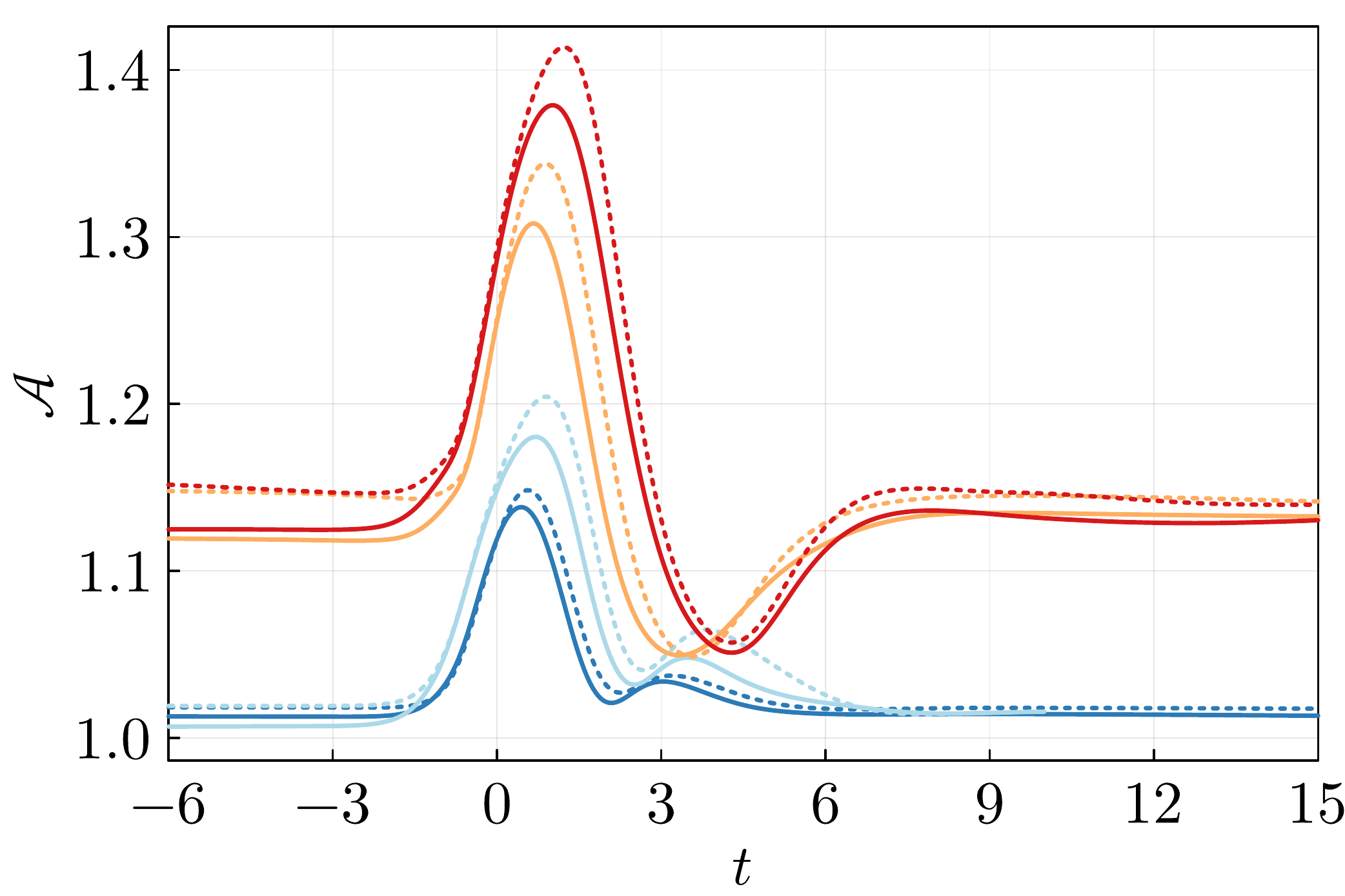}
        \caption{$d_0=0.25$}
        \label{Dca_Area_In_d0.25}
    \end{subfigure}

    \caption{Temporal evolution of the reduced surface area $\mathcal{A}$ of the leading and the trailing capsules at various interspacings $d_0$ and various Capillary and Reynolds numbers $Ca$ and $Re$.}
    \label{Dca_Area}
\end{figure}

\Reffig{Dca_Area} shows the evolution of the surface area of the leading and the trailing capsules in the non-inertial regime (\reffig{Dca_Area_NIn}), as well as at $Re = 25$ and $Re = 50$ where the initial interspacing $d_0$ is 1 (\reffig{Dca_Area_In_d1}) and 0.25 (\reffig{Dca_Area_In_d0.25}).

\section{Train of capsules at large initial interspacings\label{app:train_d1}}

\begin{figure}
  \centering
\begin{subfigure}{.48\textwidth}
      \centering
      \includegraphics[width=.8\columnwidth, trim=0 0 0 0, clip]{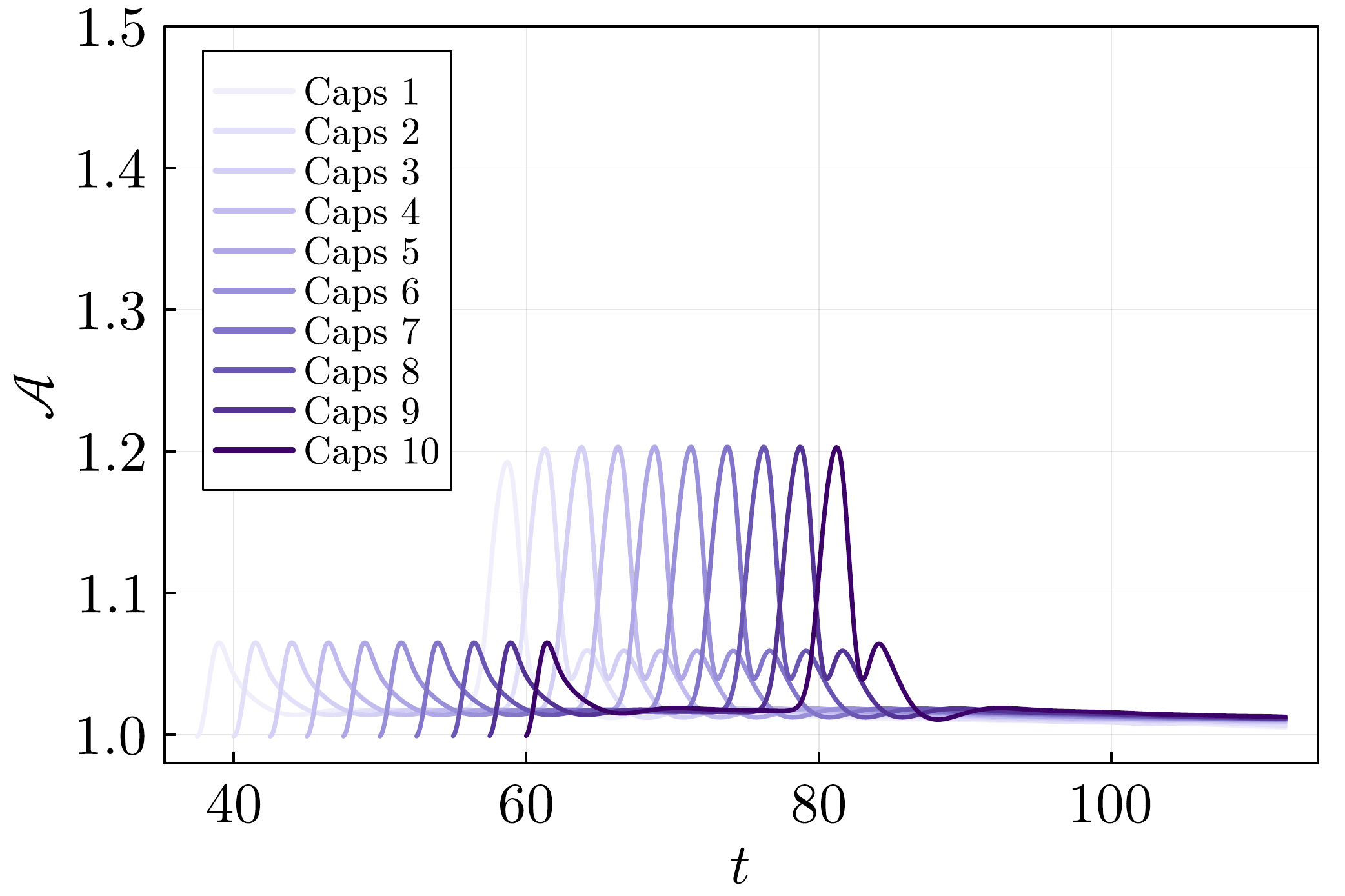}
      \caption{$Ca = 0.15$}
  \end{subfigure}
  \begin{subfigure}{.48\textwidth}
      \centering
      \includegraphics[width=.8\columnwidth, trim=0 0 0 0, clip]{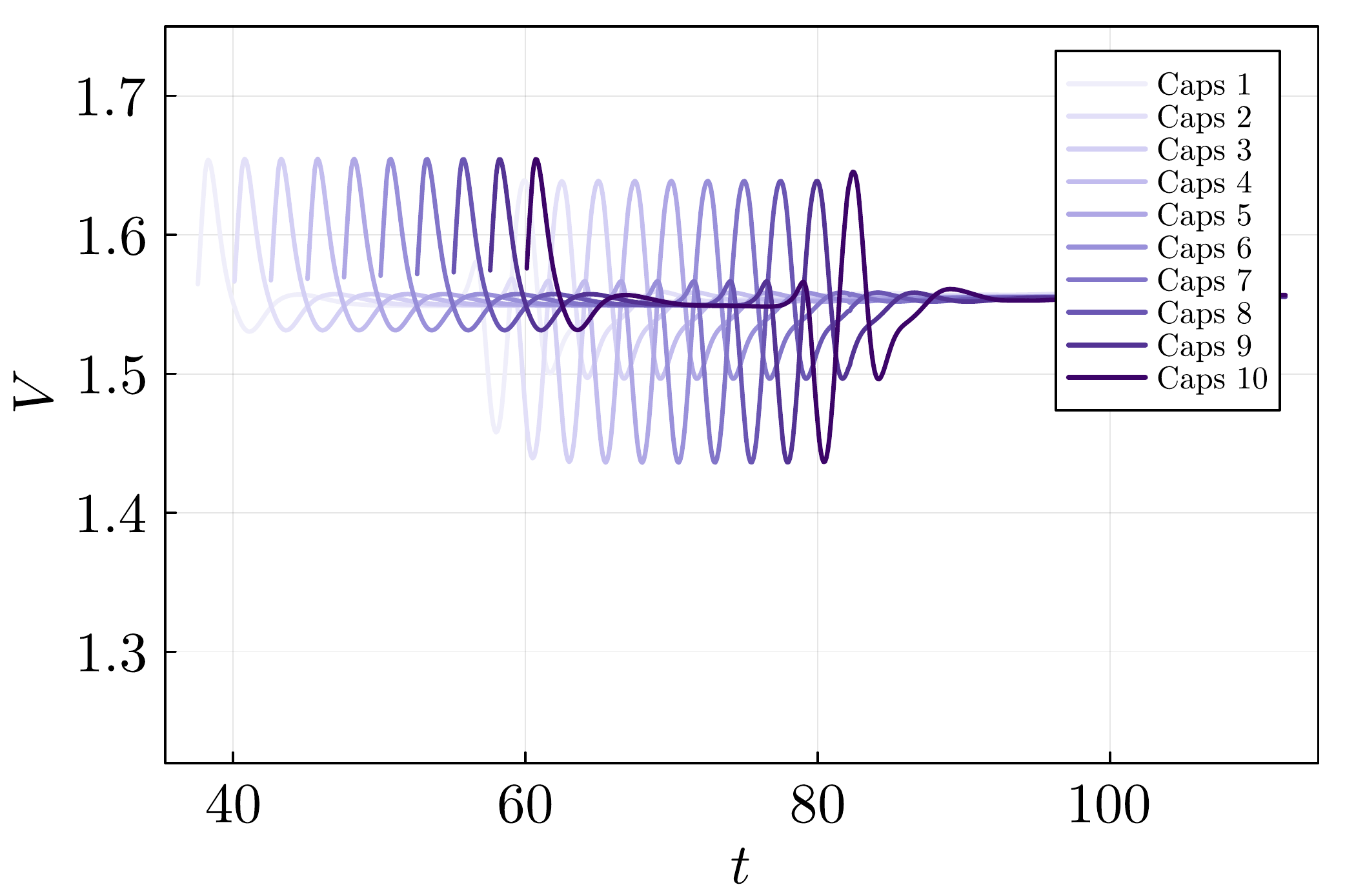}
      \caption{$Ca = 0.15$}
  \end{subfigure}\\

  \begin{subfigure}{.48\textwidth}
      \centering
      \includegraphics[width=.8\columnwidth, trim=0 0 0 0, clip]{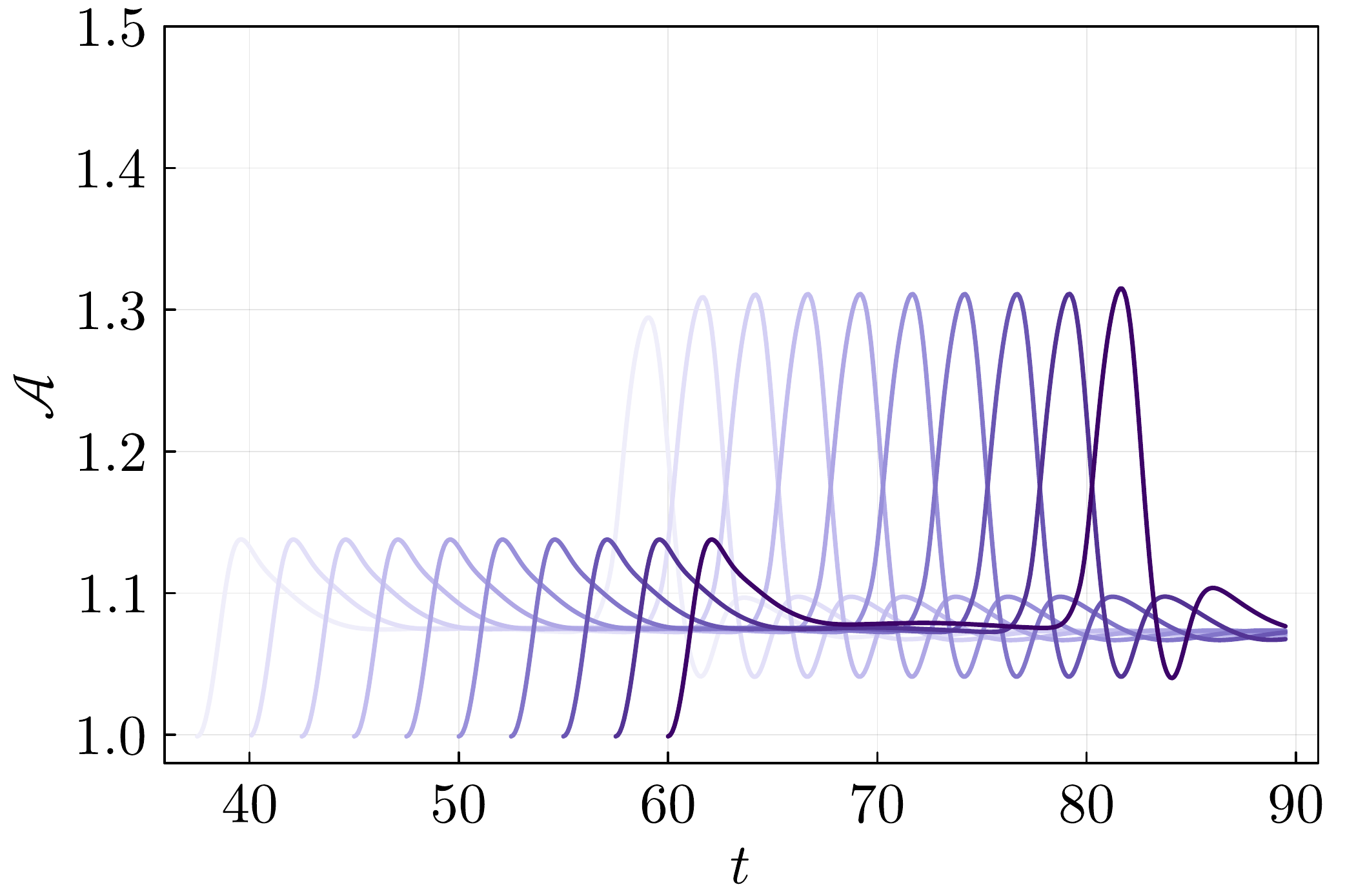}
      \caption{$Ca = 0.25$}
  \end{subfigure}
  \begin{subfigure}{.48\textwidth}
      \centering
      \includegraphics[width=.8\columnwidth, trim=0 0 0 0, clip]{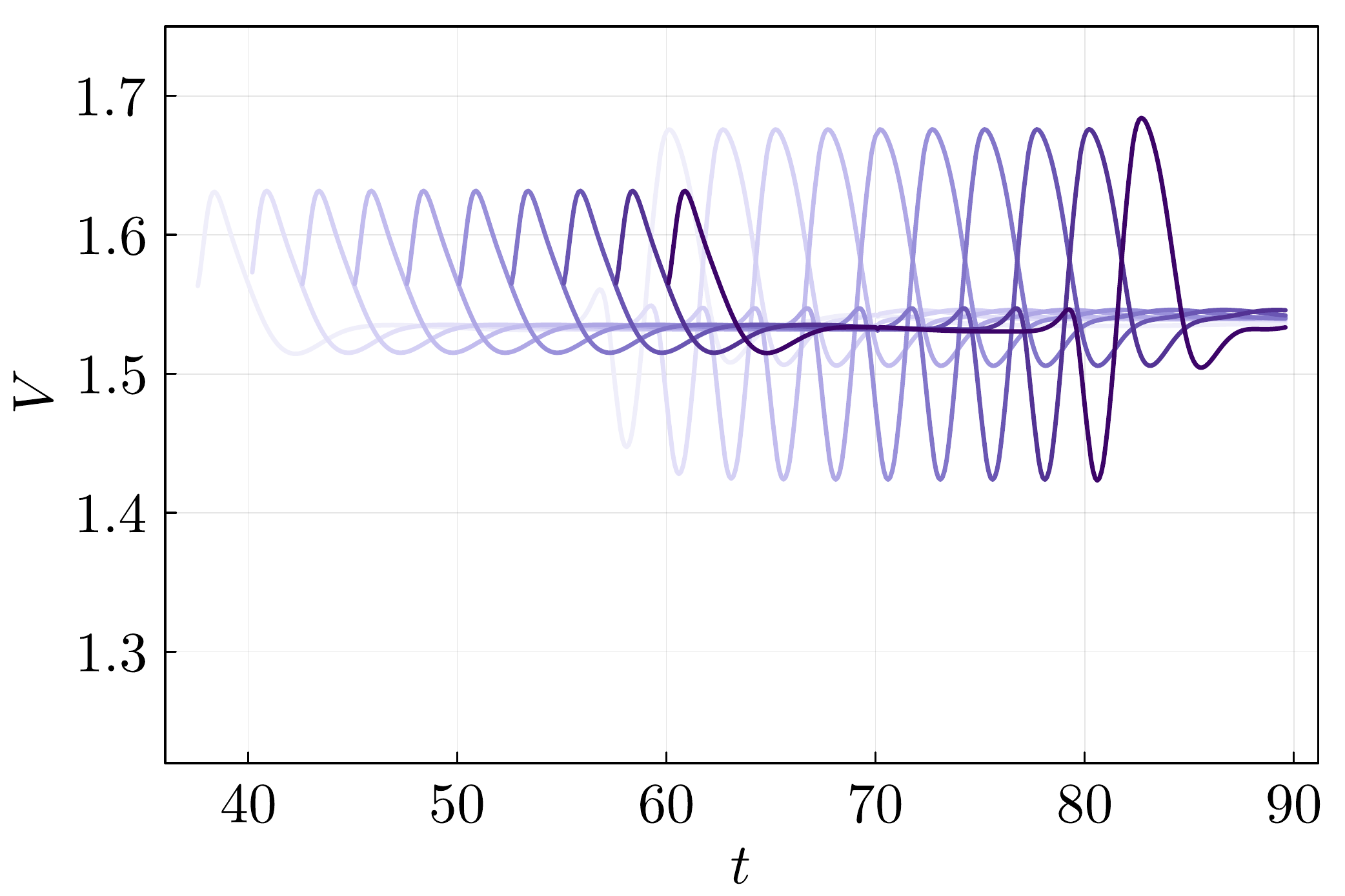}
      \caption{$Ca = 0.25$}
  \end{subfigure}  \\

  \begin{subfigure}{.48\textwidth}
      \centering
      \includegraphics[width=.8\columnwidth, trim=0 0 0 0, clip]{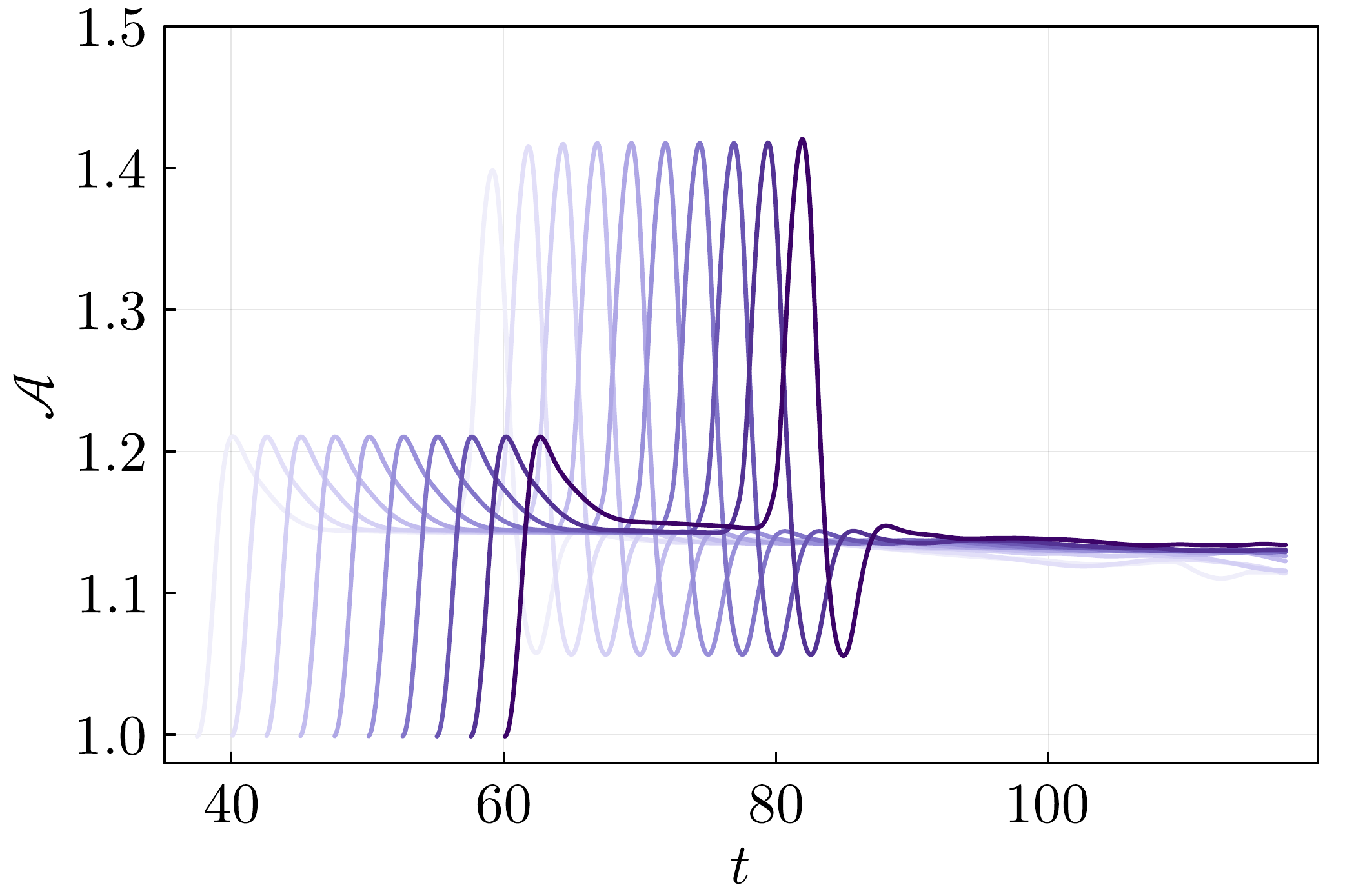}
      \caption{$Ca = 0.35$}
  \end{subfigure}
  \begin{subfigure}{.48\textwidth}
      \centering
      \includegraphics[width=.8\columnwidth, trim=0 0 0 0, clip]{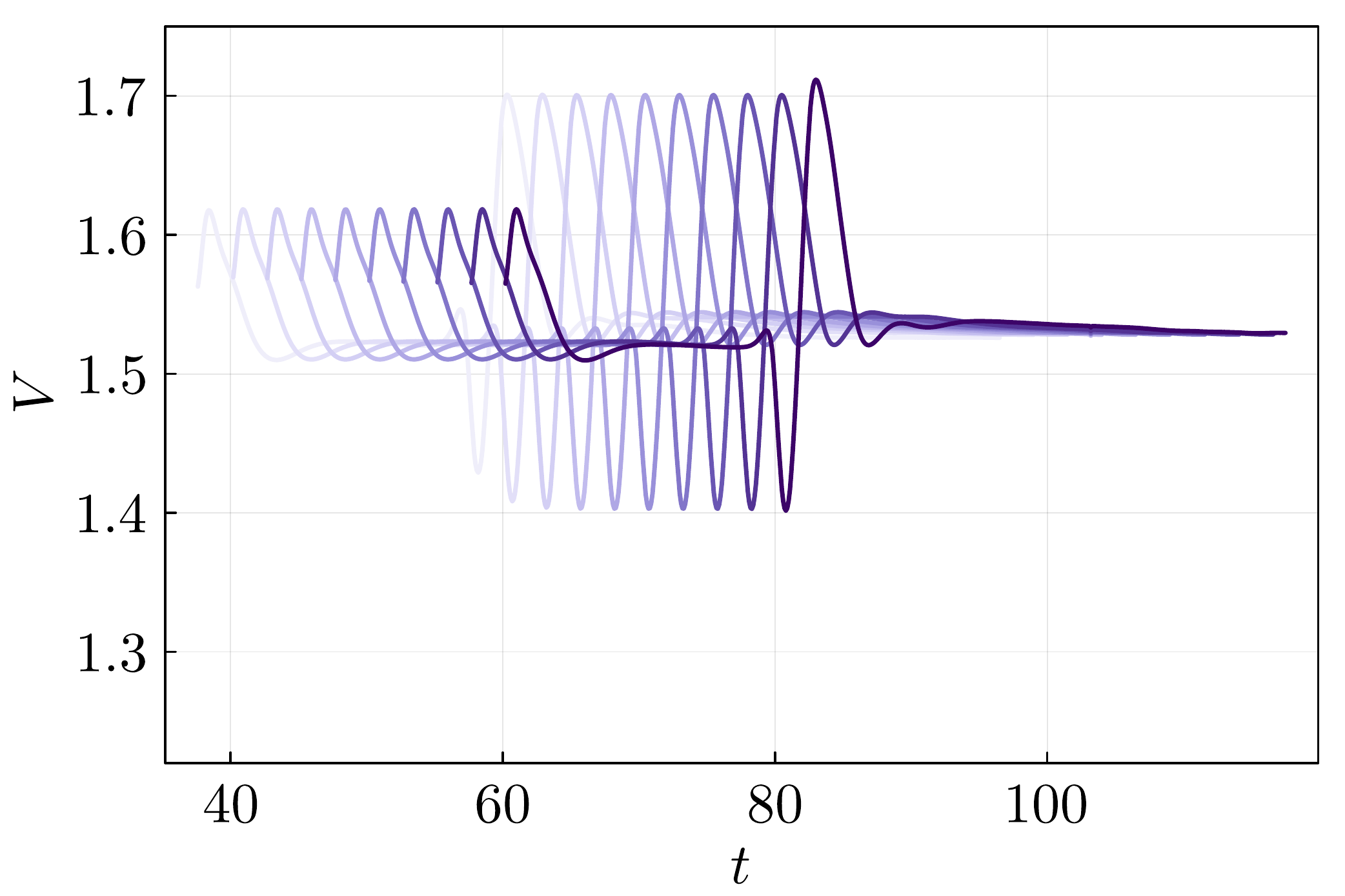}
      \caption{$Ca = 0.35$}
  \end{subfigure}
  \caption{Time evolution of the reduced surface areas and velocities of ten capsules at $Re = 50$ and $d_0 = 1$ for $Ca = 0.15$, 0.25 and 0.35.}
  \label{fig:corner_train_d1}
\end{figure}

We provide in \reffig{fig:corner_train_d1} the time evolution of the surface area and velocity of each capsule in a train of 10 capsules flowing through a corner at $Re = 50$, $Ca = 0.35$ and a reduced initial interspacing distance between each capsule $d_0 = 0.125$. As can be noted in this figure, the capsules in this regime do not interact as the surface area and velocity evolution of each capsule is almost identical.

\bibliographystyle{ieeetr}
\bibliography{MyReferences}

\end{document}